# VICIOUS CYCLE OF POVERTY IN HAOR REGION OF BANGLADESH: IMPACT OF FORMAL AND INFORMAL CREDITS

**Professor Dr. Md. Nazrul Islam**
**Department of Business Administration**
**Shahjalal University of Science and Technology**

**Sponsored by**
**Grants for Advanced Research in Education (GARE) Program**
**Ministry of Education, Government of Bangladesh**

# VICIOUS CYCLE OF POVERTY IN HAOR REGION OF BANGLADESH: IMPACT OF FORMAL AND INFORMAL CREDITS[1]

**Abstract**

This research attempts to explore the key research questions of: what are the different microcredit programs in Haor area in Bangladesh? And do microcredit programs have a positive impact on livelihoods of the clients in terms of selected social indicators viz. income, consumption, assets, net worth, education, access to finance, social capacity, food security and handling socks etc. in Haor area in Bangladesh? Utilizing difference-in-difference and factor analysis, we explore the nature and terms of conditions of available formal and informal micro-creditss in Haor region of Bangladesh; and investigate the impact of micro-creditss on the poverty condition of Haor people in Bangladesh. The findings showed that total income of borrowers has been increased over non-borrowers (z=6.75) significantly. Among the components of income, non-agricultural income has been increased significantly on the other hand income from labor sale has been decreased significantly. Total consumption expenditure with its heads of food and non-food consumption of both formal borrowers and informal borrowers have been increased over the period 2016-2019 significantly. Most of the key informants agreed that the findings are very much consistent with prevailing condition of micro-credits in Haor region. However, some of them raised question about the impacts of micro-credits. They argued that there is no straightforward positive impact of micro-credits on poverty condition of the households.

[1] Sponsored by the Grants for Advanced Research in Education (GARE) Program, Ministry of Education, Government of Bangladesh.

## Principal Investigator
**Professor Dr. Md. Nazrul Islam**
University Research Centre
Shahjalal University of Science and Technology

## Co-Investigator
**Professor Dr. Sabina Islam**
Department of Statistics
Shahjalal University of Science and Technology

## Other Contributors

Associate Investigator

**Professor Dr. Md Zakir Hossain**

Department of Statistics

Shahjalal University of Science and Technology

Research Assistant

**Fazlur Rahman**

M.S. Graduate Student, Department of Statistics

Shahjalal University of Science and Technology

**July 2021**

# TABLE OF CONTENTS

## *List of Tables*

## *List of Figures*

## List of Acronyms

| | |
|---|---|
| BBS | Bangladesh Bureau of Statistics |
| BHWDB | Bangladesh *Haor* and Wetland Development Board |
| CBN | Cost of Basic Needs |
| CEGIS | Center for Environment and Geographic Information Services |
| DCI | Direct Calorie Intake |
| DID | Difference-in-difference |
| FEI | Food Energy Intake |
| FGD | Focus Group Discussions |
| GoB | Government of Bangladesh |
| HH | Household |
| HIES | Household Income and Expenditure Survey |
| LPL | Lower Poverty Line |
| LPL | Lower Poverty Line |
| MAS | Multi-agent systems |
| MC | Micro-credits |
| MDGs | Millennium Development Goals |
| MFI | Micro Finance Institution |
| MPI | Multidimensional Poverty Index |
| NGO | Non-Government Organization |
| PCA | Principal Component Analysis |
| SACCOS | Savings and Credits Co-operative Societies |
| SFYP | Seventh Five Year Plan |
| SMCP | Savings and Micro Credits Program |
| SPSS | Statistical Package for Social Sciences |
| UP | Union Parishad |

## *Acknowledgement*

The project was funded in the window of Grant of Advanced Research in Education (GARE) for three years period (July 2018 to June 2021) by the Bangladesh Bureau of Educational Information and Statistics (BANBEIS) under the Ministry of Education Government of the People's Republic of Bangladesh. So, at first, we are grateful to the authorities of BANBEIS for giving opportunity in conducting the research project. Then we are also grateful to the data collectors for collecting data from the respondents of survey households. The contribution of Key Informant Interviewees and In-Depth Interviewees are also mentionable. Professor Dr. Md. Zakir Hossain, Department of Statistics (SUST) has contributed in designing questionnaire and sample size. So, we are thankful to him. The research assistant Mr. Fazlur Rahman has contributed throughout the study in respect of supervision of data collection, compilation and analysis of data. So, we are indebted to him. Mr. Minhaz Hasan has also contributed for mapping the project. So, his contribution is not least though at last.

**Professor Dr. Md. Nazrul Islam** (Principal Investigator)
and
**Professor Dr. Sabina Islam** (Co-investigator)

# *Executive Summary*

Though *Haor* districts of Bangladesh occupying 13.56% of total area of the country are though resourceful with water, fishing, mineral and biodiversity and *boro*-rice cultivation yet the basic avenues for life and livelihoods are largely absent for the people. The people of the regions are poorer than the other parts of the country where about one-third people are extreme poor as they lie below the lower poverty line and only about 30% of them lie above the upper poverty line. Thus, the lion part of them suffers from food insecurity and others basic needs. Due to the poverty, most of hoar people depend on borrowings from money lenders and micro-credits institutions for their livelihood mainly in crisis period and about 80% of them borrowed money from different formal and informal sources where 59% of them borrowed for purchasing food. The rate of interest of different types of micro credits varied from 12.5% to 43% and in some cases, it rose to 110%. The type and nature of both money and food borrowings in the *Haor* region are still be unexplored and need an in-depth investigation to make proper policy for the socioeconomic development of *Haor* people. Anecdotal evidences showed that most of the *Haor* households (HHs) could not overcome the borrowing cycle and as a result they have become handicapped to the lenders. Conventional literatures evidenced that most of the marginalized households adopted credits from NGOs and local money lenders at a high interest with rigid terms and conditions during the shocking/lean period. Most of the times, they could not comply the rigid terms and conditions and sell their assets to repay the loan. Finally, the victim HHs falls in credits trap and they could not increase income using the credits as they desired and as a result, one credits creates another credit persistently. Till today, no rigorous study was found on this issue and urgent research is essential to achieve the targets of the concerned SDGs (Sustainable Development Goals # 1, 2 and 10). Considering these the present study has been conducted to explore the nature of micro-credits in *Haor* region and evaluate the impact of micro-credits on the poverty condition using solid ground level data. In conducting the study, we have adopted cluster-sampling design where *Haor*-attached mouzas/unions have been treated as clusters and a total 30 clusters have been covered in the survey. The study has covered 52 borrowers and 26 non-borrowers as control (50% of the cases) respondents from each cluster and thus, the total sample size of the study stood at 2340 (78×30). The clusters have been selected using systematic probability proportionate to size (PPS) sampling procedure from the *Haor* prone districts. In collecting the household-level data from both micro-credits recipient and control households separate semi-structured questionnaire but in the same form was used where some parts were applicable for borrowers, some for non-borrowers and some were common for both. The Key Informant Interviews (KII) has been conducted with the people who have knowledge on *Haor* economy and impacts of micro- credits on livelihoods. The key informants were the community leaders, professionals, and other stakeholders including the NGO delegates working in *Haor* areas. The participants of In-depth Interviews were the selected victims of poverty cycle, local money-lenders and delegates of existing MFIs working in the *Haor* region. The study has conducted 30 KIIs and 30 IDIs for having in-depth understanding of the credits-trap and its impact on poverty situation. In analyzing and presenting the collected data, both descriptive and inferential statistical tools have been used. The descriptive statistics are frequency distribution, tabulation, mean standard deviation etc. while inferential statistical tools are, Z-test, t-test, before after comparison based, Difference-in-Difference (DID) method, factor analysis and discriminant analysis.

The real data showed that out of total 2340 respondents 69% were borrowers (treatment group) and 31% were non-borrowers (control group). In terms of relationship of respondents with household heads lion portion (81%) respondents were the household heads, in terms of gender 76.8% were male and 23.2% were female, in respect of age 62% were between 31-50, in regard to marital status 92.2% was married and 5% was widowed, in respect literacy more than half of the respondents attended primary level education, in respect occupational characteristics about 20% engaged in agriculture, 23% in labor selling, 13% in off-farm activities, 19% in business/service and 18.9% in household work, in respect of income earning nearly half are found as full-time income earner and about 29% were part-time income earner and 21% had no earning source while 3% of the respondents had disability.

The distribution of HHs' population by sex indicated that 51.17% were male and 48.83% female and from the age perspective 35.2% male and 37.9% female are up to 15 years, 26.7% male and 28.3% female between 16-30 years and 25.5% male 25.3% female are between 31-50 years 6.8% male, 4.9% female are in between 51-60 age and 5.7% male and 3.7% female are above 60. The marital status of the population 16 and above showed that the lion part of population is married (66.4% for male and 73% for female) and female population is being getting early married than their male counterpart, in terms of literacy near half of them has attended primary education and only about 2% completed graduate and the illiteracy rate was higher of women than men, in respect of occupation of male aged between 16 and 60 years is 8.9% farming, 13.5% day laborer, 8.4% off-firm activities 13.7% service/business 11% student, 34.6% household work and 9.9% others while 71% women is in household work, from earning capacity members aged 16 and above showed that 28% were full-time earner and about 22% were part-time earner and about one-quarter male and three-quarters female are found to no work at the survey point.

The ownership pattern of housing showed that about 95% owns house for living and 20% uses one room, 42% of uses 2 rooms and 10% four and more rooms for living. The data on landholdings showed that 8.5% HHs had no homestead land, 79.2% owned homestead land between 1 to 15 decimals and only 12.3% owned homestead land more than 15 decimals, over two-thirds have no agricultural land and 7% owned agricultural land between 1 to 15 decimals, 8% owned agricultural land between 16 to 50 decimals, and only 17% over 50 decimals.

The data possession of durable assets by HHs showed that 14.8% owned boats 1.2% owned deep tube well, 20.6% owned tube well and 8.1% owned bicycle, 16.2% owned television, 6.2% owned drawing room furniture, 28.5% owned watch 69.7% owned electric fan and 89.9% owned mobile phone. The data on possession of productive assets showed that 664 HHs of borrowers and 311 of non-borrowers possessed cultivating instruments, in respect possession of livestock 694 HHs of borrowers and 287 HHs of non-borrowers possessed on average 2.39 and 2.73 respectively showing significant difference between the groups. The average possession of cultivating instruments, livestock is 3.20 and 2.49 respectively. Fifteen causes were given to the eligible but not receivers of micro-credits to find out the causes of exclusion from micro-credits and through PCA major four factors are identified as (i) availability of micro-credits, (ii) institution and lenders, (iii) administration of loan and (iv) governance of loans.

Data revealed that in terms of loan receivers 1595 (99.3%) HHs borrowed cash loan and only 12(0.7%) in kinds and in terms of total amount the percentage is about same (99.29% cash and 0.71% in kinds). In terms of sources 1158 HHs (72.1%) borrowed from formal sources and 449

HHs (27.1%) from informal and in terms of amount of loan 66% from formal and 34% from informal sources. More in sight it is found that there are sources of formal credits viz. (i) government (Banks/Co-operatives; and (ii) nongovernment (MFI/NGO/Insurance) and 88 HHs (5.5%) borrowed from the former and the 1070 HHs (66.6%) from the later and there are three informal sources viz.(i) local money lender ; (ii) non-interest loan (Relatives/ friends/ neighbors); and (iii) more than one sources and the data revealed that 393 HHs (24.5%) has taken loans from the first, 53 HHs(3.3%) from the second and only 3 HHs(0.2%) from the third and in terms of amount of loan 89.84% from the first, 9.88% from the second and 0.28% from the third. Among the informal sources 408 HHs (90.9%) have taken from interest bearing sources and HHs 41(9.1%) from non-interest bearing and respect of amount loan 95.3% from the former and 4.70% from the latter. The data revealed that minimum amount of all types of loans is Tk. 2000, maximum amount of formal loan, informal loan, in total are Tk. 550,000 and Tk. 11, 00,000 and 11, 00,000 respectively and mean is Tk. 34596 for formal and 45475 for informal and 37636 in total. Among the informal loans maximum amount of interest bearing and non-interest bearing are Tk. 1100000 and 150000 respectively. The mean of interest bearing and non-interest bearing are loans are Tk. 47696 and Tk. 233656 respectively. In respect of interest rate for formal credits 525 HHs got loan at 11% to 15% following 211 HHs 16% to 20%, 171 HHs between 21% and 25%, 141 HHs up to 10% and only 110 HHs at above 25%. In case of informal credits 185 HHs borrowed at more than 25% following 106 HHs 1% to 10%, 55 HHs 11% to 15%, 21 HHs between 21% and 25% and 41 HHs at interest free. The data on type of installment revealed that in case of formal credits most of the HHs borrowed weekly and monthly basis while in case of informal credits most of them monthly and annual basis. In overall the lion parts of HHs borrowed on weekly, monthly and annual (634, 681 and 238 respectively) basis. The number of installments revealed that in case of formal credits most of HHs borrowed on 24 and 2 to 12 installment bases while incase informal credits most of them borrowed on 2 to 12 and one time installment basis and in overall most of them on 2 to 12 and more than 24 basis. The data revealed that most of the HHs (1096) borrowed informal loans for one year and formal loans (317 HHs) also for one year duration and in overall most of them (1413 HHs) for one year duration. In fact, all the loans of both formal and informal (1512 HHs) are non-collateral basis. The payment status of loans showed that 86% loan is recovered in time. The analysis of unpaid loan explored that 946 HHs of formal credits in which 944 principal and 945interest failed to pay loan in time and 402 informal borrowers in which 401 of principal and 326 of interests are in unpaid. In overall, 1348 HHs are in unpaid in which 1345 principal and 1271 interest. There were seventeen purposes of credits receiving and through PCA these dimensions are reduced into five factors as (i) daily life and livelihoods, (ii) adaptability of natural shocks and farming (iii) cropping and raring cattle the elements, (iv) business and marriage and (v) sending family member to abroad. The respondents were given a list of 24 expenditures and investments items of their loans and among those 14 items were marked by them. The items marked by 10% or above borrowers in case of formal credits are family enterprise (15.76%), agricultural inputs (15.09%), food consumption (15.05%), others (13%) and payment of loan (12.03%) similarly in case of informal micro-credits the these are food consumption (23.24%), others (16.86%), payment of loan (12.97%) and healthcare (11.68%). In total the top five sectors of expenditure and investment are food consumption (17.34%), others (14.08%) family enterprise (13.60%), agricultural inputs (13.03%) payment of loan (12.29%).

The annual income of borrowers depicted that income of majority 1053 (66%) HHs comes from labor sale and in total 60 million (36%) following from agriculture 830 (56%) and in total 40 million (24%) non-agriculture 83 (52%) HHs and in total 36 million (22%), from business 400 (25%) HHs and in total 28 million (17%), from donations 208 (13%) and in total 2 million (1%). The annual income of majority 464(63%) non-borrowers comes from business and in total income 25millions (34%) following from agriculture 333(45%) HHs and in total 15 million (20%), non-agriculture 331 (45%) HHs' and in total 18 million (25%), from business 197(27%) HHs and in total 15 million (20%) and from donations 81 (11%) HHs. It is seen that 22% of non-borrowers HHs under the pressure of debt. There are significant differences between income of borrowers and non-borrowers in respect of non-agriculture and debt.

Data on expenditure of both borrowers and non-borrowers HHs depicted that both the borrowers and non-borrowers HHs incurred expenditure for consumption in respect of both food and non-food but borrowers HHs incurred 80% for food and 20% for non-food while non-borrowers HHs incurred 84% for food and 16% for non-food and there is significant difference in respect non-food consumption as well as total consumption between borrowers and non-borrowers HHs. There are twelve items of investment expenditure and the items above 5% of total expenditure of borrowers HHs are agriculture (20%), healthcare (20%) family business (16%), house repair (9%) and poultry /livestock (7%) while in respect of non –borrowers the investment items above 5% of in terms of percentage of total expenditure are healthcare (27%), agriculture (23%), education (18%), family business 9% and house repair (9%). There is significant difference between borrowers and non-borrowers in the items of education, healthcare, poultry/livestock, productive assets, durable goods, house repair, others investment and also in total investment expenditure as well as total expenditure (consumption plus investment).

In analyzing the causes non-payment of loan in time showed that the top five for non-payment of loans agreed by formal borrowers are installment period is very short (80.9%) following rate of interest is very high (70.6%), natural calamities (67.9%) medical treatment/medicine (67.7%), acute food problem (63.7%). Similarly, the top five causes agreed by informal borrowers are rate of interest is very high (92.8%) following misappropriation of loan (82.8%), medical treatment/medicine (80.5%) installment period is very short (72.2%) and natural calamities (71.1%). In total top five causes of nonpayment of loans are installment period is very short (78.1%), rate of interest is very high (77.8%), medical treatment/medicine (71.9%), misappropriation of loan (69.2%) and natural calamities (68.9%). The PCA reduced the fourteen causes into four factors as related to food and medicine following marriage and legal problems, terms of loan and investment and lastly related to cost of loan.

Data portrayed that among the sixteen dimensions on attitudes of borrowers towards microcredits the top five positive dimensions of formal borrowers are -(i) by micro-finance your food security has increased (54.3%); (ii) micro-finance is helping you in better access to healthcare(53.8%); (iii) micro-finance is helping you in better financial situation of your family(50.7%); (iv) due to micro-finance, employment opportunities have been increased(45.6%) and (v) by micro-finance your income has increased (44%). In case of informal borrowers the top five positive dimensions are- (i) local loans are easier to get than MFIs (76.8%); (ii) micro-finance is helping you in better access to healthcare (61.2%); (iii) by micro-finance your food security has increased (56.1%); (iv) micro-finance is helping you in better financial situation of your family (52.3%); and (v) Operational assistance received from MFIs was helpful to run the business (49.4%). In total top five negative dimensions are (i) cost of local loans is lower than MFIs (73.6); (ii) local lenders are friendly than MFIs (58.4%); (iii) terms and conditions of local loans are easier than MFIs (57.9); (iv) duration of credits is sufficient (56.4%); and (v) by micro-finance your savings has increased (56.1%). The PCA reduced the sixteen causes into four factors as related to income and savings following terms of loans, cost of loans and security of food and health.

In analyzing the role of micro-credits on graduation of selected socioeconomic status of the borrowers showed that in graduation of food security status the borrowers are found more food secured than the non-borrower households. The proportion of households with moderate food insecurity is significantly ($p<0.01$) reduced to 14.5% from 19.0% due to micro-credits program. The incidence of normal food insecurity also significantly ($p<0.01$) reduced to 9.3% from 16.2%. On the other hand, such type of changes is also observed among non-borrower households but the decrease was found notably lower and statistically insignificant. In graduating socio-economic status data showed that the percentage households having extreme poor and moderately poor condition are estimated lower after receiving micro-credits. The micro-credits have significant influence for reducing extreme poverty condition among credits receiving households though the proportion also reduces to the non-borrower households but the decrement was statistically insignificant. In respect of graduation of investment on education the data showed the percentage of households with the increase of educational expenditure are estimated higher in case of borrower households (63.8%) than that of non-borrower households (51.0%) and the average educational expenditure increases significantly among borrower households as well as among non-borrower households. In respect of and healthcare investment data revealed that in case-control comparison healthcare expenditure has increased more among borrower household (70.1%) than non-borrower household (60.3%) and healthcare expenditures among borrower households has increased after inclusion in microcredits than the before inclusion significantly but it is insignificant in case of non-borrower households.

The analysis to find out the causes of not graduating from poverty in spite of using micro-credits among both case and control HHs fourteen dimensions were listed to the respondents and the data showed that the principal cause was found as natural calamity among non-borrower households (75.3%) and pressure of the payment of loan among borrower households (83.9%).

The second prime reason was loss of investment among non-borrower households (72%) and high interest rate among borrower households (73.1%). The PCA reduced these dimensions into five factors as (i) imbalance between amount and time span of loan, (ii) imbalance between dependent member and earning member (iii) incidence of investment loss (iv) pressure created by the loan provider; and (v) unable to utilize and diverse the loan.

In measuring the impact of micro-credits at the household level by comparing between 2016 and 2019 in terms of income, expenditure and investment the analysis of data revealed that in respect of income of borrowing HHs as well as non-borrowing HHs has been increased significantly and the income from agriculture, non-agriculture and labor sale have imbued this increase in total income in spite of insignificant increase in income from business and donations. The total income of borrowing HHs has been increased at higher rate than that of non-borrowing HHs significantly. This is permeated by the significant difference in increase of non-agricultural income and labor sale though the increase in agricultural income, business and donation are insignificant. In respect of expenditure and investment data showed that total consumption expenditure with food and non-food expenditures of both borrowers and non-borrowers has been increased significantly. The increase in total consumption of borrowers and non-borrowers did not significantly differ though non-food expenditure differs significantly. Total investment expenditure of both borrowers and non –borrowers has been increased significantly. All the items of investment have been increased in case of borrowers except land purchase but in case of non- borrowers only land agriculture, family business, house repair and others investment are significant. The increase of total investment expenditure of borrowers has been significantly increased than that of non-borrowers and in item-wise all the investment items except education. The net impact of the micro-credits on income, expenditure and investment showed that total income of borrowers has been increased over non-borrowers ($z=6.75$) significantly. Among the components of income, non-agricultural income has been increased significantly on the other hand income from labor sale has been decreased significantly. The incomes from agriculture, business and donations have been increased but insignificantly. The non-food expenditure has been increased significantly but food expenditure has been decreased insignificantly while total food consumption has been increased ($z=1.06$) significantly of borrowers over non-borrowers. The investment expenditure of borrowers has been increased over the non-borrowers significantly. Among the heads of investment expenditure healthcare, poultry livestock and productive assets have been increased significantly of the borrowers over non-borrowers while investment on education, agriculture, family business, house repair, land purchase and others investment have been increased insignificantly. Therefore, total expenditure, total investment and savings of borrowers have been increased significantly over the non-borrowers.

The more insight a comparative impact of formal and informal micro-credits in terms income, expenditure and investment showed that the total income and its heads of income from agriculture, non-agriculture and labor sale have been increased significantly of formal borrowers over the study period while in case of informal credits total income and its heads of income from agriculture and labor sale have been increased significantly. The heads of income from business and donations in case of formal credits and incase of informal credits the heads of income from non-agriculture, business and donations have been increased insignificantly. Data on total consumption expenditure with its heads of food and non-food consumption of both formal borrowers and informal borrowers have been increased significantly. In case of formal borrowers' total investment expenditure and its entire heads except land purchase and house repairing have been increased significantly. The increase in total consumption of borrowers and non-borrowers did not significantly differ though non-food expenditure differs significantly. Total investment expenditure of both borrowers and non –borrowers has been increased significantly. All the items of investment have been increased in case of borrowers except land purchase but in case of non- borrowers only land agriculture, family business, house repair and others investment are significant. Thus, the increase of total investment expenditure of borrowers has been significantly increased than that of non-borrowers and in item-wise all the investment items except education. The net impact on income and its heads showed that total income and its entire heads do not differ significantly among the formal and informal sources except donation income which has been decreased ($z=-20.6$) of formal sources over informal sources significantly($p=.001$).The data on total consumption showed that incase of formal borrowers the total consumption has two heads and total consumption and food consumption did not change significantly over informal borrowers ($z=-13.68$) but non-food expenditure has been decreased of formal borrowers than informal borrowers significantly (($p=<0.01$). Total investment has been increased ($z=18.53$) of formal borrowers than that of informal significantly ($p=<0.01$) among the items of investment, education, agriculture, poultry–livestock, productive assets and durable goods have been increased significantly of formal borrowers over informal borrowers on the other hand investment in health care, family business, house repair, land purchase and others investment have been decreased significantly. The total expenditure and savings have been increased of formal borrowers over informal significantly. The discriminant analysis between formal and informal borrowers was done in respect of following fifteen predictors(variables): gender, age, loan occupation, earning status, number of earning members, dependency ratio, regularity of income, assets index, types of toilets, quantity of land, total annual income, amount of loan, types of installments, duration of loan and interest rate. The analysis of data showed that there is significant discriminant between the two groups of borrowers in respect all predicators. Based on the estimates of the discriminant analysis such as Un-standardized coefficients, Standardized coefficients and Structure Matrix we found the magnetite of discriminating power of the predictors. The top most ten magnitude predictors are types of installments followed by duration of loan, interest rate, gender, earning status, regularity of income, occupation, number of earning members, age and amount of loan. We have established the Fisher's linear discriminant function for discriminating the formal and informal source of credits by the fifteen stated variables above to predict or classify a new individual with the respective characteristics who are supposed to borrow from formal source or informal source. The average of the functions at group centroid is estimated at 0.44. Therefore, it is obvious that new members with discriminant scores above 0.44 supposed to borrow from the informal sources, otherwise he/she would be preferred to borrow from formal sources. The qualitative survey showed that though there are numerous formal sources yet, the *Haor* people still depends on informal sources of micro-credits for covering the emergency situation like treatment of household members, covering the cost of food items during severe food shortages. The interest rate of micro-credits from informal sources is sometimes rose to 60% in some season, particularly before harvesting the crops. On the other hand, the interest rate of micro-credits is remains constant throughout the year for institutional sources like banks, NGOs etc. The qualitative survey showed that the main purpose of getting loans is for food expenditure, repairmen of living house, investment in agriculture farming, investment in purchasing livestock, and expenditure for treatment of household members are the main reasons for receiving micro-credits. The causes of not getting formal and informal micro-credits are they do not know about some micro-credits programs, nepotism, misappropriation of credits, lack of entry fee/bribe and bureaucratic problem. The heads of utilization of micro-credits are to meet-up the food expenditure, repairing the living house, investment in agriculture for harvesting crop along with loan repayment. The borrowers told that the rate of interest is very high, credits period is insufficient and terms are rigid. The causes of ineffectiveness of micro-credits for poverty reduction perspective are the lack of policy, lack of skills, and investment in non-productive heads appeared as the hindrances for effectiveness of the micro-credits on the perspective of poverty reduction. The main suggestions for maximization of benefits from micro-credits are that in general, the micro-credits from informal sources should be stopped, the volume of loan should be increased at lower rate, the micro-credits from formal sources should be encouraged for the households whose members acquired capability of start any economic activity through skill-development program. The government could undertake a plan of action to review the existing micro-credit programs. Skill-development training programs should be strengthening covering different investment areas for the adult members of credit-receiver households and finally new research may be undertaken to find out different pathways for eradication of poverty of *Haor* people in order to achieve the targets of SDGs.

# CHAPTER ONE
# INTRODUCTORY ASPECTS OF THE STUDY

**1.1 Introduction**

The *Haor* region is an economic lacked area of Bangladesh due to geographical remoteness and persistent negligence in development plan. As a result, the socioeconomic conditions of the *Haor* people are not up to the mark as others parts of the country. Though Government of Bangladesh (GoB) and many foreign and Non-Government Organizations (NGOs) have taken many development projects to improve their socioeconomic situations yet they are sufferings from unavailability of many socioeconomic benefits specially food insecurity, water and sanitation etc. Many of them are living on credit from formal and informal sources and persistently they are failing to come out from the credit. The first chapter of the study covered the followings:

- Background of the study;
- Poverty in Bangladesh;
- Measurement of incidence of poverty in Bangladesh;
- Trends of poverty incidence of Bangladesh;
- Statement of the problem
- Literature review;
- Key research questions
- Objectives of the study
- Conceptual framework of the study
- Rationality of the study;
- Scope and limitations of the study; and
- Conclusion of this chapter.

**1.2 Background of the study**

*Haors* are bowl-shaped large tectonic depressions which receive surface runoff water and consequently become a very expensive water bodies in the monsoon and mostly dries up during post-monsoon period. *Haor* districts (Sunamganj, Sylhet, Habiganj, Maulvibazar, Netrakona, Kishoreganj and Brahmanbaria) of Bangladesh cover 19,998 sq. km land, which is 13.56% of total area of the country. According to Center for Environment and Geographic Information Services (CEGIS) the total land of the *Haor* districts is about 43% (8585 sq. km) is under wetland (*Haor*) where 373 *Haors* are exists (CEGIS, 2012). The statistics of *Haors* in Bangladesh is given in Table-1.1.

**Table 1.1** Distribution of *Haor* areas of Bangladesh by districts

| Districts | Total area in ha (% of total) | *Haor* area in ha (% of total) | No. of *Haor* (% of total) |
|---|---|---|---|
| Sunamganj | 367,000 (18) | 268,531 (31) | 95 (25) |
| Sylhet | 349,000 (17) | 189,909 (22) | 105 (28) |
| Habiganj | 263,700 (13) | 109,514 (13) | 14 (4) |
| Maulvibazar | 279,900 (14) | 47,602 (6) | 3 (1) |
| Netrakona | 274,400 (14) | 79,345 (9) | 52 (14) |
| Kishoreganj | 273,100 (14) | 133,943 (16) | 97 (26) |
| Brahmanbaria | 192,700 (10) | 29,616 (3) | 7 (2) |
| Total | 1,999,800 (100) | 858,460 (100) | 373 |

*Source:* Master Plan of *Haor* Area, Volume,1, Summary Report, Government of the People's Republic of Bangladesh Ministry of Water Resources Bangladesh *Haor* and Wetland Development Board. p.1.

The *Haor* region has long been lagging behind the main stream of the national development due to geographical location and considering this, in 1974, then the Government of Bangladesh (GoB) established an independent Board for the development of the *Haor* region (CEGIS, 2012). Following this establishment of the Board, the GoB established the Bangladesh *Haor* and Wetland Development Board (BHWDB) through a resolution approved by the Cabinet in 2000. The mandate of the BHWDB is to coordinate the activities and formulate projects relating to a holistic development of the *Haor* and wetlands of the country (CEGIS, 2012). The map of *Haor* regions is given in Figure-1.1 (page 25).

### 1.3 Poverty in Bangladesh

Poverty alleviation is considered to be one of the most important indicators of the socioeconomic development of a state and society. Bangladesh has achieved remarkable development in poverty alleviation during the last few decades as a result of the combined efforts of both the Government and non-government sectors. According to the 'Millennium Development Goals**:** End period Stocktaking and Final Evaluations Report' the incidence of poverty has declined 1.74 percentage points on an average in Bangladesh during 2000-2010 against the MDGs target of 1.20 percentage points. According to the recently published 'Household Income and Expenditure Survey 2016' the present poverty rate is 24.3 percent whereas it was 56.7 percent in 1991. The Government has set up a target to reduce the poverty to 18.6 percent at the end of the 7th Five Year Plan (2016-2020). Despite all these positive changes in poverty reduction, still one-fourth population of Bangladesh lives below the poverty line. It would not be possible to attain the desired level of socioeconomic development without emancipating this portion of population from poverty. For this reason, the Government still considers poverty alleviation as a major agenda on the policy and development issues of the country. Bangladesh has achieved a significant progress in the Human Development Indicators. According to the UNDP Development Report-2016 the position of Bangladesh has been ranked at 139-th among 187 countries. Furthermore, the report reveals that Bangladesh's Multidimensional Poverty Index (MPI) reduced to 0.188 in 2016 from 0.237 in 2007.

**Figure 1** *Haor* areas of Bangladesh

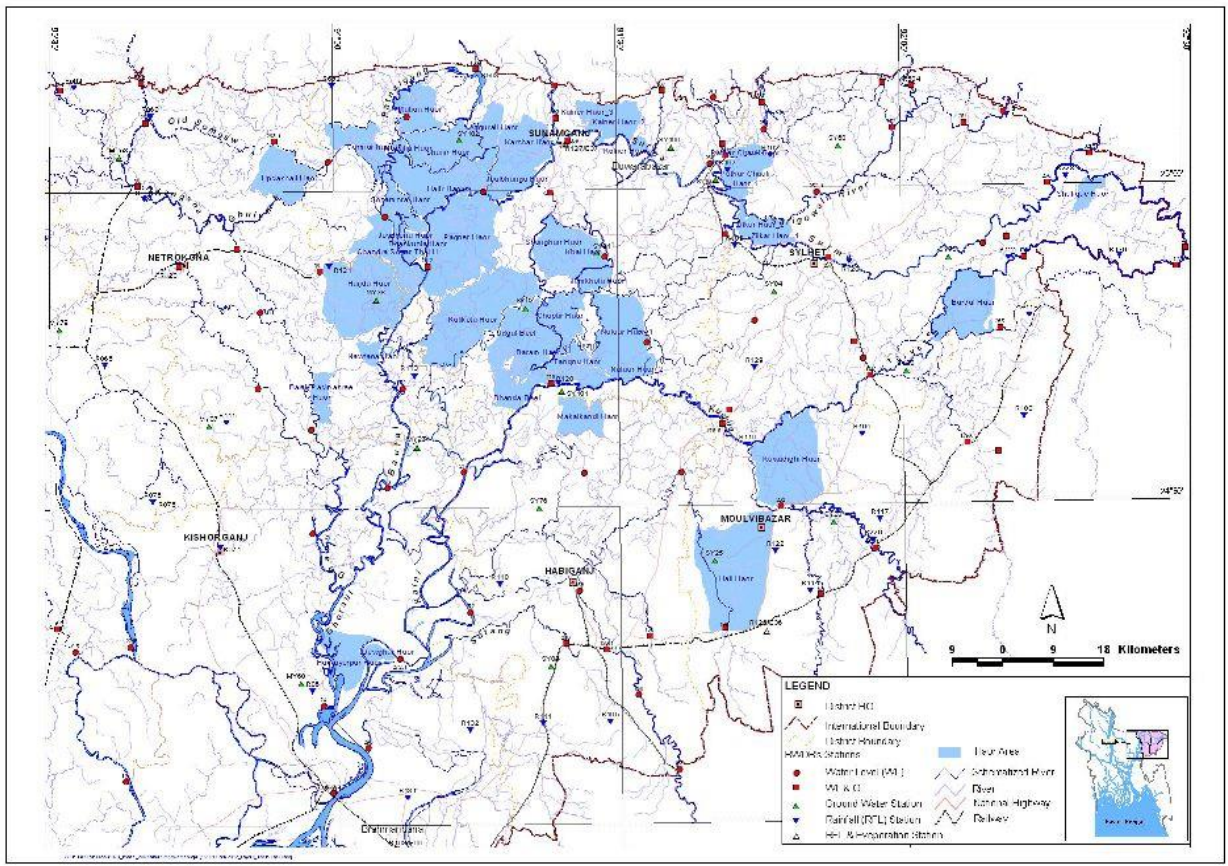

## 1.4 Measurement of the incidence of poverty in Bangladesh

The measurement of the incidence of poverty in Bangladesh is taken from the report of Household Income and Expenditure Survey (HIES). At present, the survey is renamed as Household Income and Expenditure Survey (BBS, 2016). The first HIES in Bangladesh was carried out in FY 1973-74 and after that up to FY 1991-92, few HIESs were conducted maintaining the same strategies of the first one. HESs were accomplished by Food Energy Intake (FEI) and Direct Calorie Intake (DCI) method. According to this survey, a man having calorie intake of less than 2,122 kilo-calories daily to be considered as absolute poor. On the other hand, a man having an intake of below 1,805 kilo-calories is considered as hard-core poor. The Bangladesh Bureau of Statistics (BBS) has adopted 'Cost of Basic Needs (CBN)' for HIES for the first time in 1995-96. The same method applied in the HIES in 2000, 2005 and 2010. This method also considered non-food consumption items for compiling poverty index. The latest HIES has been undertaken in 2016 and recently its result has been published.

## 1.5 Trends of poverty in Bangladesh

The incidence of income poverty (measured by CBN considering the upper poverty line) declined nearly 7 percent (from 31.5 percent to 24.3 percent) over the period in 2010-2016. During this period, the compound poverty declined 4.23 percent annually. On the other hand, the rate of income poverty declined from 40.0 percent to 31.5 percent from 2005 to 2010. At that time, compound poverty got reduced by 4.67 percent each year. Therefore, it is evident that though poverty is decreasing gradually, the pace of reduction rate declined during the period of 2010-2016 compared to the period of 2005-2010. In urban areas poverty reduction rate is higher (4.68 %) than rural areas (1.97%). During 2010 to 2016, the reduction rate of the depth of poverty (measured by poverty gap) was 4.28 percent. It has also been observed that income poverty reduction rate in urban areas is lower than that of rural areas (1.61% and 5.12% respectively). Moreover, the reduction rate of the depth of severity of poverty (measured by squared poverty gap) was also lower in urban areas compared to rural areas. The trends of poverty are depicted in table 1.2. The HIES-2016 reveals that on the basis of lower poverty line poverty rate in 31 districts is above national average. On the other hand, using the upper poverty line poverty rate in 36 districts is above national average.

**Table 1.2** Incidence of poverty in Bangladesh

| Index | 2016 | 2010 | Annual Change (%) (2010 to 2016) | 2005 | Annual Change (%) (2005 to 2010) |
|---|---|---|---|---|---|
| **Head Count Index** | | | | | |
| National | 24.3 | 31.5 | -4.23 | 40.0 | -4.67 |
| Urban | 18.9 | 21.3 | -4.68 | 28.4 | -5.59 |
| Rural | 26.4 | 35.2 | -1.97 | 43.8 | -4.28 |
| **Poverty Gap** | | | | | |
| National | 5.0 | 6.5 | -4.28 | 9.0 | -6.3 |
| Urban | 3.9 | 4.3 | -1.61 | 6.5 | -7.93 |
| Rural | 5.4 | 7.4 | -5.12 | 9.8 | -5.46 |
| **Squared Poverty Gap** | | | | | |
| National | 1.5 | 2.0 | -4.68 | 2.9 | -8.81 |
| Urban | 1.2 | 1.3 | -1.33 | 2.1 | -8.64 |
| Rural | 1.7 | 2.2 | -4.21 | 3.1 | -8.75 |

*Source*: BBS-2019 (Report of HIES 2016)

**1.6 Statement of the problem**

As stated earlier, *haors* are bowl-shaped large tectonic depressions which receive surface runoff water and consequently become a very expensive water bodies in the monsoon and mostly dries up during post-monsoon period. Most of the *Haors* are in the districts of Sunamganj, Sylhet, Habiganj, Maulvibazar, Netrakona, Kishoreganj and Brahmanbaria as given in table-1.1 of Bangladesh cover 19,998 sq. km land, which is 13.56% of total area of the country. Among the total land of the *Haor* districts, about 43% (8585 sq. km) area is wetland where 373 *Haors* are exists (CEGIS, 2012). Though the *Haor* regions are resourceful with water, fishing, mineral and biodiversity and *Boro*-rice cultivation, but the basic avenues for life and livelihoods are largely absent. The people of the *Haor* regions are poorer than the other parts of the country. The long seasonality of wet monsoon (6-7 months from May to October) forces the *Haor* people to remain out of work for most of the times (HILIP, 2011). It is documented that nearly one-third of the *Haor* people are extreme poor as they lie below the lower poverty line and only about 30% of the *Haor* people lie above the upper poverty line (Chowdhury, 2014; HILIP, 2011). As a consequence, the lion part of *Haor*-people suffers from food insecurity and others basic needs (Kazal *et al.*, 2017). Due to the poverty, most of *Haor* people depend on borrowings from money lenders and micro-credits institutions for their livelihood, mainly in crisis period. A study documented that about 80% of *Haor* people borrowed money from different formal and informal sources and 59% of them borrowed for purchasing food (Kazal et al*.,* 2010).

The rate of interest of different types of micro credits in Bangladesh varied from 12.5% to 43% and in some cases, it rose to 110% (Institute of Microfinance, 2015). The rate of interest for informal sector might be higher than the formal sectors. The type and nature of both money and food borrowings in the *Haor* region are still unexplored and needs an in-depth investigation to make proper policy for the socioeconomic development of *Haor* people of the country. Anecdotal evidences showed that most of the *Haor* households could not overcome the borrowing cycle and as a result they become handicapped to the lenders. Conventional literatures evidenced that most of the marginalized households adopted credits from NGOs and local money lenders at a high interest with rigid terms and conditions during the shocking/lean period. Most of the times, they could not comply the rigid terms and conditions and sell their assets to repay the loan. Finally, the victim households' fall in credits trap and they could not increase income using the credits as they desired; and as a result, one credit creates another credit persistently. Till today, no rigorous study was found on this issue and urgent research is essential to achieve the targets of the concerned SDGs (Sustainable Development Goals # 1, 2 and 10).

The study has been designed to explore the nature of micro-credits in *Haor* region of Bangladesh and evaluate the impact of micro-credits on the poverty condition using solid ground level data. The study has adopted both quantitative and qualitative approaches to collect the necessary data and information. In particular, household-level data has been gathered from both micro-credit recipient and control households. The qualitative data has been gathered through Key-Informant Interviews (KIIs) and In-depth Interviews (IDIs) from stakeholders and micro-credits recipients (including money lenders) respectively. The study has employed different descriptive statistical tools and techniques including mean, SD, factor analysis, discriminant analysis etc. in analyzing the data.

### 1.7 Literature review

There are a number of studies (Schreiner Mark, 2001; Mazumder Mohummed Shofi Ullah and Wencong Lu, 2013; Bangladesh Bank, 2015; Rahman Sayma, 2007; Habib & Jubb, 2015;

Khandker Shahidur R. 2003, 2005; CGAP, 2010; Chowdhury M. Jahangir Alam *et al*., 2002 ; Rabby Talukder Golam, 2012; Khanom Nilufa A., 2014; Zama Hassan,2001; Ahmed Salehuddin,2017; Uddin Mohammed Salim, 2011; Shoji Masahiro, 2008; Fant Elijah Kombian, 2010; Lawrence Peter,2004) recognized that benefits of micro credits exceeds cost and it instilled borrowers income, assets endowment, standard of living and poverty reduction, productivity of business and agriculture, increase wealth, savings, mobilize local economy, boost consumption, ensure sustainable finance for major expenses and increase capacity to cope with shocks, improve food intake , create sustainable community and increase acquisition of asset. Anthony Denise (2005) found that micro-credits as competing mechanisms has differential effects in group identity sanctions and help in borrowing more of the group. Alam Saad, (n.d) and Rankin Katharine N. (2001) opined that micro-credits amplify self-help approach and enhance self-employment profit by 50% to 81%. Micro credits enhances credits facilities and heighten repayment performance of borrowers (Ogaboh *et al*., 2014; Barboza and Barreto, 2006; Giné and Karlan , 2007 & 2009).The long- run effects of micro-credits reduce inequality in the society (Ahlin and Jiang, 2005) and make free poor people from the curse of poverty (Khandakar Shahidur R., 1998). The borrowers of micro-creditss get training facilities on different soft and hard skills (Karlan and Valdivia, 2006) and in household level play positive role in risk management (Karlan and Zinman,2009). Midgle James (n.d.) found that micro-credits and microfinance programs helped either to create small businesses or strengthened existing small businesses in low-income communities. Angelucci *et al*. (2013) found the average effects of micro credits on a rich set of outcomes and found that no transformative impacts, but more positive than negative impacts. Banerjee *et al*. (2011) showed the randomized impact of a micro-credits program to reach the poorest of the poor and elevate them out of extreme poverty and revealed that the program results in a 15% increase in household consumption. Moreover, a microcredit program targeted ultra-poor in haor areas has proven a short-run positive impacts on other measures of household wealth and welfare, such as the efficient utilization of scarce resources and promoting entrepreneurial activities (Alam and Hossain, 2018). Morduch Jonathan (1999) experienced that many relatively poor households can save in quantity when given attractive saving vehicle with micro-credits. A variant strand of literature found that micro finance is used as an instrument of crisis and risk management and limits the migration from rural to urban and help in creating and sustaining informal employment (Hossain and Ahamed, 2015; Shahnaz et al. 2018). Choudhury Haque Ariful *et al. (*2017) found that micro-credits have a significant positive impact on household income, expenditures and savings and also the level of education plays an important statistically significant role in increasing the household income, expenditure and savings. Rika Terano *et al.* (2015) micro-credits programs have effectiveness on earning ability, payment scheme, member's cooperation, and well-being and improve significantly the total income after joining the program. Khanam Dilruba *et al.* (2018) established that the micro-loans have a statistically significant positive impact on the poverty alleviation index and consequently improve the living standard of borrowers by increasing their level of income and child nutritional outcomes (Islam, Alam, and Afzal, 2021). Islam Asadul *et al.* (2014) found that households' access to microfinance reduces the incidence of borrowing from informal sources, but not the amount of borrowing and less poor households benefit more in terms of reducing their reliance on informal borrowing and that the benefit accrues over time and also found that access to microfinance increases women's informal borrowing for small consumption usage, without facilitating access to new business opportunities. Ferdousi Farhana (2018) attempted to measure the effectiveness of microenterprise loans on increasing entrepreneurs' incomes and innovation and found the positive impact on entrepreneurs' incomes. Rana Muhammad Sohail Jafar *et al*. (2017) showed that microfinance is an innovative way to poverty reduction because microfinance services poor people who can utilize their skills for enhancement of household economic growth, educate their children and enhanced quality of life of the poor resulting the more contribution to economic development and enhance the capacity to fight against hunger and poverty alleviation.

Micro-credits play positive role in empowering women (Rahman Sayma, 2007; Khandker Shahidur R., 2005; Omorodion, 2007). Pitt Mark M. and Khandker Shahidur R. (1998) found micro credits program has a larger effect on the behavior of poor households in Bangladesh when women are the program participants then men. Ahmed Ferdoushi (2011)explored that, 'with credits' women have a much lower percentage of poverty in terms of its incidence (80%), intensity (28%) and severity (12%) compared to the 'without credits' respondents (99, 59 and 37% respectively) and also found that educational attainment of the respondents and income earners in the family contribute positively to reduce poverty situation among the 'with credits' households more, as compared to 'without credits' households and his final conclusion was that micro-credits program helps the rural women to reduce their poverty more effectively. Abdalla Nagwa Babiker (2009) found that majority of women in Sudan live with low or no income; economically they are dependent on their husbands' income; burdened with their household activities and responsibilities to feed; educate and take care of many children, encounter a core problem which is lack of access to credits and financial services to economically, socially and politically empower themselves and improve their status. Rahman Aminur (1998) indicated that worker and borrowing peer loan group members in centers press on clients for timely repayment, rather than working to raise collective consciousness and borrower empowerment as envisaged in the Bank's public transcript. Mwongera Rose K. (2014) identified that majority of the young women entrepreneurs had borrowed money from the nearby micro-finance institutions the loan accrues interest rate imposed by the financial institutions as well as demand for collateral security and they found it unreasonable he also found out that most of the entrepreneurs had attained secondary school certificate as their highest academic qualification and was a determinant for uptake of loans and established that licensing of more financial institutions would encourage uptake of loan among women entrepreneurs since uptake of loan is low and this would influence business performance. Noreen Sara (2011) has made an attempt to explore the socio-economic determinants of women empowerment in which microfinance is crucial economic determinant. The results showed that women empowerment is considerably influenced by age, education of husband, father inherited assets, marital status, number of sons alive and microfinance. Age, education of husband, no of live sons and father inherited assets are more statistically significant variables in this study. Bhattacharjee Priyanka (2016) unveiled that most women failed to understand the process and effectiveness of micro-credits programs and took loan for meeting the cash-demand of the male(s) within family as the feel micro-credits as a medium to fulfill their emergency requirements. She also found that lack of education, awareness, unwillingness to join other programs of microfinance institutions, pessimistic thinking about micro-credits programs, hostile family structure, negligent attitude towards repayment of loan(s), limited investing opportunities, etc. are the main causes that hinders the road to development. Afrin Sharmina *et al*. (2008) showed that the financial management skills and the group identity of the women borrowers have significant relationship with the development of rural women entrepreneurship in Bangladesh and the experience from the parent's family of the borrowers and the option limit may also lead to the rural women borrowers to be entrepreneurial. Mohamed Fauzia Mtei (2008) argued that women and micro credits agencies have divergent understandings of money and its investment and its role in poverty reduction, and even a long-run impact on their degree of environmental awareness (Alam and Zakaria, 2021). Simojoki Hanna-Kaisa (2003) examined the socio-economic impact of micro-finance on urban female micro-entrepreneurs in Nairobi, Kenya and found that micro-credits play crucial role in empowering women in business control and decision making. Adeyemo Comfort Wuraola (2014) examined the influence of vocational skills acquisition and micro-credits loans on widows" socio-economic and psychological adjustments in South-western Nigeria and found that vocational skill acquisition centers afforded widows the opportunities to share their pains and experiences, thus assisting them to reduce loneliness, frustration, health-related problems and adjust to the reality of spouse's death. Vocational skill acquisition and micro-credits loans considerably assisted widows in overcoming their socio-economic hardship and psychological challenges. Karubi Nwanesi Peter (2006) established that micro-credits provide finance to enhance market and rural women's participation in production and trade. The study further ascertains that woman have some control over their loans. Churk Josephine Philip (2015) examined the Contribution of Savings and Credits Co-operative Societies (SACCOS) on promoting rural livelihood revealed that SACCOS have played minimal role towards promoting rural livelihoods in the study area. Waliaula Rael Nasimiyu(2012) conducted the study to evaluate the relationship between micro-credits and the growth of SMEs in Kenya and found that there was a very strong positive relationship between the variables. Gomez Rafael (2001) examined the determinants of self-employment success for micro-credits borrowers and found that social capital is a positive determinant of self-employment earnings and neighborhoods play in fostering social capital and improving micro-entrepreneurial performance. Asgedom and Muturi (2014) analyzed the socio-economic factors that affect the institution's loan repayment performance and revealed that the level of education, loan amount and loan category have insignificant effect on the probability of the Savings and Micro Credits Program (SMCP) loan repayment. On the other hand, age, gender, type of business and credits experience are significant determinants where age and type of business have negative relationship and gender and credits experience have positive relationship with the loan repayment probability. Islam Khan Jahirul (2014) argued that to get positive impact from micro-credits in reduction of poverty the nature and term should be dynamic with the nature of poverty. Rana Muhammad Sohail Jafar *et al*. (2017) explored that microfinance empowers the women so women can also contribute in the economic growth as well.

On the other hand, there are many scholars (Rogaly Ben, 1996; I.G. Okafor, 2016) who have found that micro credits programs have negative and insignificant role on poverty reduction. Diagne Aliou (1999) found that a higher share of land and livestock in the total value of household assets is negatively correlated with access to formal credits. However, land remains a significant determinant of access to informal credits. Matsuyama Kiminori (2006) found that a movement in borrower net worth can shift the composition of the credits between projects with different productivity levels and showed how investment-specific technological change may occur endogenously through credits channels. Furthermore, such endogenous changes in investment technologies in turn affect borrower net worth. These interactions could lead to a variety of nonlinear phenomena, such as credits traps, credits collapse, leapfrogging, credits cycles, and growth miracles in the joint dynamics of the aggregate investment and borrower net worth. Osmani and Mahmud (2015) found that the impact of micro-credits on the lives and livelihoods of the poor remains a contentious issue, there is no disputing the fact that there is something novel, something special, about micro-credits that has allowed an altogether new mode of financial intermediation to emerge, providing credits to millions of hitherto 'un-bankable' poor without breaking the lender's back. That is an extraordinary achievement in itself. Nissank (2002) examined economics of microfinance as an instrument of microenterprise development and poverty reduction as well as its delivery mechanisms and also assessed empirical evidence of the performance of microfinance institutions and their impacts on poverty alleviation and microenterprise development. The donors target might have unintended negative effects on the very objectives of poverty reduction and microenterprise development. Ahlin Christian (2009) extended the theory of micro credits movement management to examine sorting when group size is larger than two and joint liability can take several forms and found that moderately homogeneous sorting by risk in support of Ghatak's theory and proved the evidences of risk anti-diversification within groups while the anti-diversification results revealed a potentially negative aspect of voluntary group formation and point to limitations of micro-credits groups as risk-sharing mechanisms. Fasorant M.M. (2010) showed that the incidence of poverty was high among the economically active age bracket as the mean age was 33 years of the borrowers and all respondents acquired formal education as 60% had above primary school education and 39.2% of total respondents had no specific occupation before the inception of micro-credits scheme.

Some scholars found both positive and negative impact of micro-credits on poverty. Bylander Maryann (2014) found that using micro credits in combination with migration allows households to immediately meet consumption goals and utilize the credits being actively promoted by microfinance institutions but also retreating from insecure and less profitable local livelihood strategies. Mwangi and Sichei (2018) made a comparative study between the period 2006 and 2009 on access to credits of Kenyans using multinomial probit models and found that increase in household size reduced access to bank loans and ASCAs while it promoted access to loans from buyers of harvest. Increase in distance to service provider led to a decline in access to credits even though the impact was marginal. On the other hand, increase in age; education and income tend to enhance access to credits but the probability of access drops as one draws close to retirement age. Mamun *et al.* (2013) argued that microfinance has developed innovative management and business strategies by mobilizing the savings, loan repayment and empowerment of women, but its impact on poverty reduction remains in doubt. It certainly plays an important role in providing safety-net and consumption smoothening but identifying how microfinance can be used as an important vehicle to make an even larger and more critical contribution to alleviating poverty is in need of more careful assessment. Banerjee *et al.* (2015) showed the causal evidence on micro-credits impacts informs theory, practice, and debates about its effectiveness as a development tool and found that a consistent pattern of modestly positive, but not transformative effects. Angelucci *et al.* (2013) showed that the average effects on a rich set of outcomes measured of micro-credits suggest some good and little harm but estimators identify heterogeneous treatment effects and effects on outcome distributions, but again yield little support for the hypothesis that micro-credits cause harm. Khandker Shahidur R. *et al.* (2014) examined the level of indebtedness of the micro credit borrowers in Bangladesh with assumption that some borrowers might be taking loans that they will not be able to repay. He found that 26% of micro-credits borrowers are over-indebted while 22% of non-micro-credits borrowers. Cons and Paprocki (2010) opined that micro-credits and other 'self-help' development strategies operate through idealized notions of poverty and rural life. The scholars explored the ongoing debate over micro-credits in his study area and reflect on how re-rooting debates over development in specific places might move such debates from questions of 'self-help' to grounded and historicized projects of self-determination. Selinger Evan (2008) remarked on the debates about the Grameen Bank's micro-lending practices depict participating female borrowers as having fundamentally empowering or disempowering experiences and found that this discursive framework may be too reductive. It can conceal how technique and technology simultaneously facilitate relations of dependence and independence and it can diminish our capacity to understand and assess innovative development initiatives. Babajide Abiola (2012) found that access to microfinance does not enhance growth of micro and small enterprises in Nigeria. However, other firm level characteristics such as business size and business location, are found to have positive effect on enterprise growth. Karlan and Zinman (2011) found that microloans increase ability to cope with risk, strengthen community ties, and increase access to informal credits and therefore micro-credits here may work, but through channels different from those often hypothesized by its proponents.

Few scholars have come to find the impact of micro credits on the livelihoods of hoar area of Bangladesh. Among them Kazal *et al.* (2010) found the impact of micro credits on reduction of poverty and found the 40% of micro-credits receivers could not repay loan and fall in vicious cycle of credits. According to Khondker Bazlul Haque and Mahzab Moogdho Mim (laaging district dev.) the monthly per capita income is the third lowest of Sunamganj district which is only 2156 while the per capita monthly expenditure is 1978 the 11$^{th}$ lowest among the 15 lagging districts in Bangladesh. According to Consulting Group to Assist the Poor (CGAP) in (2010) micro finance helps hundreds of millions of people keep their consumption stable, finance major expenses, and cope with shocks despite incomes that are low, irregular, and unreliable in hoar area of Bangladesh. Planning Commission (2016) recommended from the workshop on "Environment and Climate Change Policy Gap Analysis in Haor Areas" that Haor area is enriched with valuable aquatic flora and fauna including different species of fish, other natural resources and natural habitat. The *Haor* basin has its ecological and economic importance and plays a vital role in the growth of the country but it is urgent to incorporate the economic importance on the Hoar areas in the national curriculum, particularly for the pupils of the hoar districts, so that the new generation can understand economic and environmental significance.

There two types of micro credits such as formal and informal and there is difference of impact between these two types. Gichuki *et al.* (2014) found that the key challenges hindering micro and small enterprises from accessing credits facilities to be high cost of repayment, strict collateral requirements, unwillingness of people to act as guarantors, high credits facilities' processing fees and short repayment period in case of formal micro- credits hence they recommended that financial institutions(formal sources) set more flexible, affordable and attractive requirements in financing micro and small enterprise while, Kazal *et al.* (2010) found that in the *Haor* area informal credits is more stronger than formal sector. Barnaud *et al.* (2008) explored the Multi-agent systems (MAS) and the results of a series of simulations exploring the ecological, social and economic effects of various rules for formal and informal credits suggested by the villagers-participants. The study highlighted the ability of MAS to deal with interactions between social and ecological dynamics and to take into account social interactions, in particular the concept of social capital which is a determining factor when dealing with sustainability issues. Secondly, the study addressed the potential and limits of MAS models to support a bottom-up (or participatory) modeling approach. This experiment suggests that the usefulness of models relies much more on the modeling process than on the model itself, because a model is usually useless if it is misunderstood by its potential users, or if it does not respond to their current preoccupations. The intuitive representation of real systems provided by MAS and their high flexibility are the two underlined characteristics favoring their appropriation by local stakeholders. Phan Dinh Khoi (2012) in his PhD dissertation investigated the determinants of households' borrowing decisions in terms of formal and informal micro-credits and micro-credits accessibility and found that informal micro-credits alter the households' decisions to obtain a formal microcredit. Khan Mohammad Mohaiminuzzaman (2013) in his master's thesis found that green formal micro-credits is becoming more popular among the natural resource-dependent borrowers in Bangladesh in *haior* based area

**1.8 Key research questions**

The key research questions of the project are:
   i.   What are the different MC programs in *Haor* area in Bangladesh?
   ii.  What are the different MC programs run by the government in *Haor* area in Bangladesh?

iii. What are the different MC programs run by the non-government/donors /agencies in *Haor* area in Bangladesh?
iv. Do MC programs have a positive impact on livelihoods of the clients in terms of selected social indicators viz. income, consumption, assets, net worth, education, access to finance, social capacity, food security and handling socks etc. in *Haor* area in Bangladesh?
v. Whether the MF clients are improving more their socioeconomic condition than non-clients in *Haor* area in Bangladesh?
vi. Whether the cost of micro-credits is higher than the rate of internal return of it or not in *Haor* area in Bangladesh?
vii. Whether the volume of micro-credits is insufficient in terms of productive investment or not in *Haor* area in Bangladesh?
viii. Whether the duration of credits is shorter than the client's need for productive outcome or not in *Haor* area in Bangladesh?
ix. Whether the payment schedule of credits is tighter than the expectations of clients in *Haor* area in Bangladesh?
x. Whether the continuation of credits is enough or not for productive outcome in *Haor* area in Bangladesh?
xi. Whether there is any difference of impact of micro credits in terms of duration in *Haor* area in Bangladesh?
xii. Whether there is any difference of impact of micro credits in terms of volume of credits in *Haor* area in Bangladesh?
xiii. Which type of credits among formal and informal does have more impact on poverty reduction of the borrowers in *Haor* area in Bangladesh?
xiv. Do all the eligible borrowers get credits in *Haor* area in Bangladesh?
xv. How does productive outcome of credits vary from one to another MC program in *Haor* area in Bangladesh?

## 1.9 Objectives of the study

In response of the above questions the objectives of the project are designed to**:**

- explore the nature and terms of conditions of available formal and informal micro-creditss in *Haor* region of Bangladesh; and
- investigate the impact of micro-creditss on the poverty condition of *Haor* people in Bangladesh

**1.10 Conceptual framework of the Study**

Vicious cycle of poverty means a rotation of a sum of forces tending to act and re act one another in such a way as to keep a poor man in a state of poverty permanently. For example, when a man does not get enough food day by day, he became so weak and ultimately, he failed to earn and remains poor or weak for a long time which pushes him to death. Thus, who are poor remain poor because they are poor. Garcíay Carrasco (2004) point out that "poverty reproduces poverty" and define vicious circle of poverty "as a circular dynamic of factors and lacks which feedback each other". As stated earlier in the literature reviews there are many studies (Schreiner Mark, 2001; Mazumder Mohummed Shofi Ullah and Wencong Lu, 2013; Bangladesh Bank, 2015; Rahman Sayma, 2007; Habib & Jubb, 2015; Khandker Shahidur R. 2003, 2005; CGAP, 2010; Chowdhury M. Jahangir Alam *et al*., 2002 ; Rabby Talukder Golam, 2012; Khanom Nilufa A., 2014; Zama Hassan,2001; Ahmed Salehuddin,2017; Uddin Mohammed Salim, 2011; Shoji Masahiro, 2008; Fant Elijah Kombian, 2010; Lawrence Peter,2004) that micro-creditss reduces poverty and help the borrowers to come out from vicious cycle of poverty . Micro-creditss enhance the self -help approach and enhance self-employment (Rankin Katharine N. (2001) and boosts borrowing capacity of group (Anthony Denise, 2005). It improves repayment performance of borrowers (Ogaboh *et al*., 2014; Barboza and Barreto, 2006; Giné and Karlan, 2007& 2009), reduce inequality in the society (Ahlin and Jiang, 2005) and make free poor people from the curse of poverty (Khandakar Shahidur R., 1998). The borrowers of micro-creditss get training facilities on different soft and hard skills (Karlan and Valdivia, 2006) and in household level play positive role in risk management (Karlan and Zinman,2009). Midgle James (n.d.) Micro-credits helped either to create small businesses or strengthened existing small businesses in low-income communities. Banerjee *et al*. (2011) found that micro-credits elevate the poor out of extreme poverty Morduch Jonathan (1999) found that micro-credits act as a vehicle of savings. Shahnaz *et al.* (2018) explored that micro-credit is as an instrument of crisis management. Micro-credits fuels household income, expenditures and savings and also the level of education (Choudhury Haque Ariful *et al.*,2017) and boosts earning ability, payment scheme, member's cooperation, and well-being and improve significantly the total income after joining the program (Rika Terano *et al.*, 2015; Khanam Dilruba *et al.*,2018). Islam Asadul *et al.* (2014) found access to microfinance increases women's informal borrowing for small consumption usage. Ferdousi Farhana (2018) attempted to measure the effectiveness of microenterprise loans on increasing entrepreneurs' incomes and innovation and found the positive impact on entrepreneurs' incomes. Rana Muhammad Sohail Jafar *et al.* (2017) found that micro-credits help to gather skills, utilization of the skills of the members/borrowers, educate their children and enhanced quality of life of the poor results the more contribution to economic development and enhance the capacity to fight against hunger and poverty alleviation. Micro-creditss play positive role in empowering women (Rahman Sayma, 2007; Khandker Shahidur R., 2005; Omorodion, 2007; Pitt Mark M. and Khandker Shahidur R., 1998; Ahmed Ferdoushi, 2011). Abdalla Nagwa Babiker (2009) found that access to credits and financial services to women makes economically, socially and politically empower themselves and improve their status. Mwongera Rose K. (2014) identified that micro-credits helped entrepreneurs to attain highest academic qualification and helped licensing of more financial institutions for women entrepreneurs. Noreen Sara (2011) showed that women empowerment is considerably influenced by microfinance with others. Bhattacharjee Priyanka (2016) unveiled micro-credits as a medium to fulfill their emergency requirements of women. Afrin Sharmina *et al.* (2008) found micro-credits provides financial management skills and the group identity of the women borrowers and helps the development of rural women entrepreneurship in Bangladesh. Adeyemo Comfort Wuraola (2014) proved that micro-credits assisted in acquiring vocational skill to widows by which they overcome their socio-economic hardship and psychological challenges. Karubi Nwanesi Peter (2006) established that micro-credits provide finance to enhance market and rural women's participation in production and trade. Nasimiyu (2012) found that micro-credits significantly helped in the growth of SMEs. Rana Muhammad Sohail Jafar *et al.* (2017) explored that microfinance empowers the women so women can also contribute in the economic growth as well.

According to Khondker Bazlul Haque and Mahzab Moogdho Mim(laaging district dev.) the monthly per capita income is the third lowest of Sunamganj district which is only 2156 while the per capita monthly expenditure is 1978 the eleventh lowest among the 15 lagging districts in Bangladesh. Planning Commission (2016) recommended from the workshop on "Environment and Climate Change Policy Gap Analysis in *Haor* Areas" that *Haor* area of Bangladesh is enriched with valuable aquatic flora and fauna including different species of fish, other natural resources and natural habitat. It also asserted that the *Haor* basin has its ecological and economic importance and plays a vital role in the growth of the country. Furthermore, the Climate Fiscal Framework is developed by the Finance Division (Finance Division, 2014) under the support of the. Poverty Environment and Climate Mainstreaming (PECM) Project focuses on the eradicate poverty and reduce inequalities through sustainable development in haor-basin areas by designing a governance framework for climate change funds in haor areas under national fiscal policy. Kazal *et al.* (2010) found that 40% of micro-credits receivers could not repay loan and fall in vicious cycle of credits in the *Haor* region of Bangladesh. According to Consulting Group to Assist the Poor (CGAP) in (2010) micro finance helps hundreds of millions of people keep their consumption stable, finance major expenses, and cope with shocks despite incomes that are low, irregular, and unreliable in *Hoar* area of Bangladesh. Khan Mohammad Mohaiminuzzaman (2013) in his master's thesis found that green formal micro-credits is becoming more popular among the natural resource-dependent borrowers in Bangladesh in *Haor* based area. Kazal *et al.* (2010) found that in the *Haor* area informal credits is stronger than formal sector. Considering the above discussion, the conceptual framework of the study is given in the following diagram:

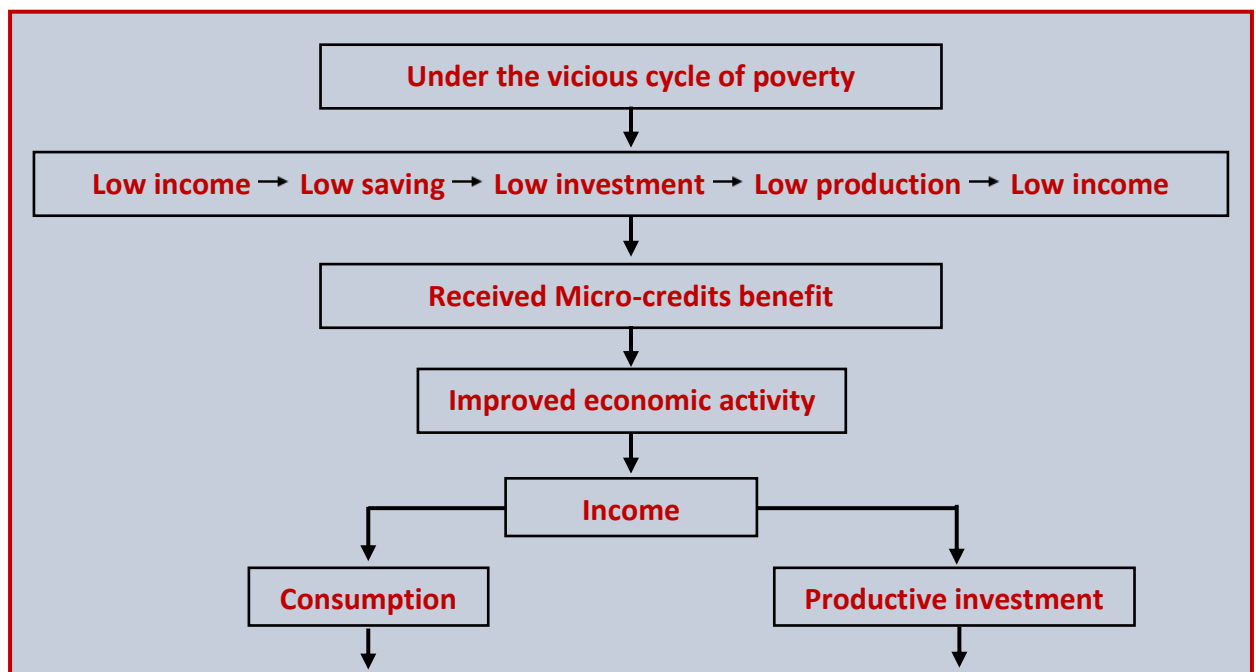

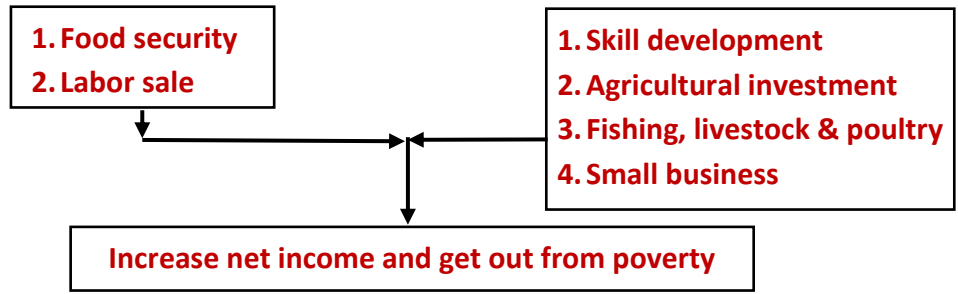

## 1.11 Rationality of the study

The review of literature revealed that micro credits have positive as well as a negative impact on of poverty alleviation on the other hand there are many studies showing both positive and negative and also neutral impact on poverty through different means or ways. Micro-creditss have a positive impact on a variety of micro-economic indicators including self-employment, business activities, household consumption and well-being (Bauchet *et al*., 2011). On the other hand, many studies showed that micro-credits programs have either a positive or limited impact on poverty reduction (Zeller& Meyer, 2003; Amin et al., 2001). About 33.33 million people included in micro credits services out which 91% women and total loans allocated about 1362750 million (MRA, Brucher 2020). However, the formal sectors through micro-finance institutions (MFIs) contribute nearly half of the credits and the rest are coming from informal sector. The formal sector loans need collateral. However, the credits services are even to render to the extreme poor households without any collateral and third-party guarantee in case of informal sectors (Mujeri, 2015). Generally, loans are provided under intensive monitoring mechanisms along with high interest rate in case of informal sectors. A number of studies in Bangladesh proved that the positive impact of micro-creditss on clients with regard to material well-being, reduction in exposure to seasonal vulnerability, contributions to consumption smoothing, and a better ability to deal with crises (Khalily 2004; Mustafa *et al.,* 1996). On the other hand, many witnesses also found against the micro credits. Islam (2014) showed that access of micro credits leads to an increase in vulnerability for those households which are chronically poor. Poor peoples are trapped in the vicious cycle of poverty since the emergence of micro-credits and are failing to come out from the cycle of poverty as well as micro credits (Pitt and Khandker, 1996; Khandker, 2001). Another study found that 26% of micro-credits borrowers are over indebted (Khandker *et al.*, 2014). Therefore, there are debates on the actual impact of micro-credits in improving the wellbeing of the poor (Montgomery and Weiss, 2005).

People of *Haor* areas in Bangladesh are regarded as backward section citizens as they are far behind from the main-stream population of the country in terms of education, per capita income, per capita consumption, roads and electricity facility, infant mortality and poverty (Khondker and Mahzab, 2015). It is documented that 56% of households of Bangladesh use micro-creditss (Khandker *et al.*, 2014). This rate is extremely high for *Haor* regions as it is documented that about 80% of households borrowed money from different sources (Kazal, *et al*., 2015). As 70% of *Haor* people are poor and suffer from food insecurity, most of them have no alternative other than micro-credits, either formal or informal sectors, for their survival. The terms and conditions of micro credits are not important issues at the time of receiving micro-credits for *Haor* people. It is observed that local money lenders lend with rigid terms of references, which are generally fail to comply by most of the borrowers and consequently, most of the borrowers take one credits to repay another credit, and finally they remain in debt persistently. As a result, most of them could not improve their poverty condition which is known as vicious cycle of poverty. A very few studies with limited coverage are found in *Haor* areas on micro credits (Khan, 2014; Uddin, 2011), which are not sufficient to understand the actual scenario of the nature and behavior of micro-credits on livelihoods. In addition, no study has been found to deal with the impact of micro-credits in details. From the view of policy perspective, the exploration of the nature and behavior of existing micro-credits is of utmost importance to remove the vicious cycle of poverty of *Haor* people. The quantification of the impact of these micro-credits on life and livelihood will guide for proper policy implementation. Hence an in-depth study on the nature and impact of micro-credits, both formal and informal, is very much important for this very region. For doing this, the study is planned to discover the existing nature and availability of micro credits in the *Haor* areas meticulously and find out the negative facets of micro credits which are especially accountable for vicious cycle of poverty.

**1.12 Scope and limitations of the study**

The study has designed to explore the nature of micro-credits in *Haor* region of Bangladesh and evaluate the impact of micro-credits on the poverty condition using solid ground level data. The study has adopted both quantitative and qualitative approaches to collect the necessary data and information. In particular, household-level data has been gathered from both micro-credits recipient and control households. The qualitative data has been gathered through Key-Informant Interviews (KIIs) and In-depth Interviews (IDIs) from stakeholders and micro-credits recipients (including money lenders) respectively. The study has used different descriptive statistical tools and techniques including instrumental variable regression and logistic regression to analyze the data. That the study has unveiled the existing credits-trap in *Haor* regions and has also identified the factors acting for vicious cycle of poverty for proper policy formulation. But in collecting data from the borrowers and non-borrowers it is observed that the respondents are not interested to provide actual data on their income, expenditure, assets and liabilities. As a result the study has failed to present actual data in the regards.

## 1.13 Conclusion

The present chapter has presented background of the study, poverty in Bangladesh and its measurement and trends, statement of the problem, literature review, research questions, objectives and conceptual framework, rationality and scope and limitations of the study. The methodology of the study is presented in the next chapter.

# CHAPTER TWO

## METHODOLOGY OF THE STUDY

### 2.1 Introduction

In the previous chapter we have presented the background, problem literature review, rationality objectives and scope of the study. Following the previous chapter, the study has adopted both quantitative and qualitative techniques to gather the data and information and, in this line, the present chapter aims to discuss the methodology followed in conducting the study

### 2.2 Survey area

The study has covered the six *Haor* prone districts of Bangladesh viz. Sunamgong, Sylhet, Moulvibazar, Habiganj, Kishoreganj and Netrokona. Among these districts *Haor* are mostly concentrated in Sunamgonj, Netrokona and Kishorgonj districts. According to the Haor Master Plan sources, there are 370 *Haor* in the aforesaid six districts and only three belong to Brahmanbaria district as stated in table 1.1 earlier. The selected clusters are given in table 2 (page 44).

### 2.3 Designing the sample

#### 2.3.1 Sample design for household-level survey

The study has adopted cluster-sampling design and *Haor*-attached mouzas/unions have been treated counted as clusters. A total of 30 clusters have been covered in the survey. Thirty clusters are regarded as statistically representative sample of a population by internationally recognized survey designs, such as WHO's EPI cluster sampling design (Turner *et al*., 1996). The recognized sample size determination formula[2] [on the basis of 70% indicator percentage. (proportion of households lie below the upper poverty line), 95% confidence interval, 0.04×p relative precision and highest response distribution with an assumed design effect 1.5] yields that at least 1544 targeted (micro-credits recipient) households are required to cover for the study. For rounding-up in clusters, the study increased the sample size from 1544 to 1560. Since one of the core objectives of the study is to evaluate the impact of micro-credits on poverty cycle, hence a control group must be selected to apply the sophisticated statistical techniques. The study has covered 780 households as control (50% of the cases). The characteristics of the control households are almost similar to the cases (micro-credits recipient) households. Thus, the total sample size of the proposed study was 2340 poor households.

---

[2] $n = \dfrac{p(1-p)Z^2}{(0.04p)^2} \times Deff$ where, $p$ is the indicator percentage, $Z$ is the value of normal variate with 95% confidence interval, $0.04p$ is the relative error margin and *Deff* is the design effect.

**Table 2** List of selected clusters (mouza/union) for primary data collection

| Cluster No | District name | Upazila name | Union name |
|---|---|---|---|
| 1 | Sylhet | Osmani nagar | Pachim Pailanpur |
| 2 | Habiganj | Ajmiriganj | Kakailseo |
| 3 | Sunamganj | Jamalganj | Beheli |
| 4 | Sunamganj | Jamalganj | Fenarbak |
| 5 | Sunamganj | Sunamgnaj sadar | Rangar Char |
| 6 | Sunamganj | Jagannathpur | Raniganj |
| 7 | Sunamganj | Jagannathpur | Raniganj |
| 8 | Sunamganj | Derai | Charnachar |
| 9 | Sunamganj | Derai | Taral |
| 10 | Sunamganj | Dakshin Sunamganj | Shimulbag |
| 11 | Sunamganj | Dakshin Sunamganj | Patharia |
| 12 | Sunamganj | Doarabazar | Surma |
| 13 | Sunamganj | Tahirpur | Tahirpur |
| 14 | Sunamganj | Tahirpur | Uttor Badaghat |
| 15 | Sunamganj | Tahirpur | Balijuri |
| 16 | Sunamganj | Bishwamvarpur | Dakshin Badaghat |
| 17 | Sunamganj | Dharmapasha | Chamardani |
| 18 | Kishorganj | Itna | Mriga |
| 19 | Kishorganj | Nikli | Jaraitala |
| 20 | Kishorganj | Nikli | Jaraitala |
| 21 | Kishorganj | Katiadi | Masua |
| 22 | Habiganj | Lakhai | Bulla |
| 23 | Habiganj | Baniachong | Kagapasha |
| 24 | Habiganj | Baniachong | Subidpur |
| 25 | Habiganj | Ajmiriganj | Shibpasha |
| 26 | Netrokona | Mohangonj | Suair |
| 27 | Netrokona | Mohangonj | Magan Siadhar |
| 28 | Netrokona | Mohangonj | Magan Siadhar |
| 29 | Sunamganj | Shalla | Bahara |
| 30 | Sunamganj | Shalla | Shalla |

The clusters have been selected using systematic probability proportionate to size (PPS) sampling procedure. Since the numbers of *Haor* are different in the six districts (Annex 1), a stratified random sampling with proportional allocation was adopted to estimate the number of *Haor* from each district (stratum).

Therefore, following steps have been adopted in order to select the clusters and sample households:

i. The number of *Haors* in each of the six strata was determined and defined.
ii. The *Haors* have been selected individually from the six strata using systematic PPS sampling procedure.
iii. A mouza/union has been selected randomly from each of the selected *Haors*.
iv. The UNICEF pencil-spin method was used to select the households randomly within the cluster.
v. Finally, 52 micro-credits recipient households and 26 control households have been selected for interview from each cluster, totaling 2340 households from 30 clusters.

### 2.3.2 Sample design for qualitative component

The Key Informant Interviews (KII) have been conducted with the people who have knowledge on *Haor* economy and impacts of micro credits on livelihoods. The key informants were the community leaders, professionals, and other stakeholders including the NGO delegates working in *Haor* areas. The participants of In-depth Interviews were the selected victims of poverty cycle, local money-lenders and delegates of existing MFIs working in the *Haor* region. The study has conducted 30 KIIs and 30 IDIs for having in-depth understanding of the credits-trap and its impact on poverty situation.

### 2.4 Survey instrument

A structured interview schedule (questionnaire) has been developed and administered for conducting household survey. The questionnaire contained several sections, viz., profile of household members, basic household information and possessions of assets, knowledge on microcredits, profile of micro-credits including expenditure and investment pattern, and impact of micro-credits on livelihood issues, causes of nonpayment of loan, causes of not overcoming from poverty etc. The detail interview was conducted with the borrowers about availability of borrowings, volume of borrowings, sources of borrowings, socioeconomic condition before and after borrowings, time period of borrowings, repayment of borrowings, cost of borrowings, terms of reference of borrowings, household's savings, behavior of lenders etc. Separate Checklists have been prepared and used to collect the data through IDIs and KIIs. The IDIs have been

47 | P a g e

conducted covering the availability of the micro-credits in the community; interest rate of micro-credits in the community; purposes receiving micro-credits; causes of not getting formal and informal micro-credits; heads of utilization heads of micro-credits; attitudes of borrowers on formal and informal micro-credits programs; causes of ineffectiveness of micro-credits for poverty reduction perspective; causes of remaining poor after using micro-credits and suggestions for maximization of benefits from micro-credits.

## 2.5 Data collection

A well-trained fieldwork team was employed and sent to the project locale for collecting data and information. The Field Assistants were the graduate level students of Shahjalal University of Science & Technology. They have been given 5-day long training on different aspects of the questionnaire and research objectives. It is to be noted that the Field Assistants were selected considering their competence on project objectives as well as communication with local dialects. The questionnaire and the checklists were designed in such a way as to cover maximum but relevant and useful information and data for the survey. The Supervisors was randomly re-interviewed some sample units to check and validate the already collected data by Field Assistants.

## 2.6 Analytical techniques of data

The study has used several descriptive tools and inferential techniques to achieve the project objectives. Among the descriptive statistical tools, percentage, average, cross-tabulation, Chi-square test, Z-test, t-test and F-test have been used to analyze the data. On the other hand, the study used case-control (borrowers vs. non-borrowers) comparison, before-after (2016 vs. 2019 assuming before borrowing and after borrowing) comparison, Diff-in-diff technique, factor analysis and discriminant analysis as inferential techniques. The brief description of the analytical techniques is given below:

### 2.6.1 Before-after comparison

This study would try to explore the productive outcomes of micro-credits programs through evaluating their explicit performances. Due to lack of baseline data of the study households, this study intends to assess the productive outcomes in three different ways:

i. comparing the socio-economic status of beneficiary (borrower) households with their non-beneficiary counterparts;

ii. comparing the current socio-economic status of beneficiary (borrower) with their status before entering the program (before-after comparison);

iii. comparing the current socio-economic status of study households, both beneficiary (borrower) and non-beneficiary (non-borrower), with their status before 3 years (before-after comparison).

**2.6.2 Difference-in-differences (DID) method**

Difference in differences is typically used by project management to evaluate the impact of program intervention over a specific time gap. It is a quasi-experimental research design using observational study data, by studying the differential effect of a treatment on a 'treatment group' versus a 'control group'. In reality, DID makes use of longitudinal data from treatment and control groups to obtain an appropriate counterfactual to estimate a causal effect.

To estimate the effect of benefits from micro-credits programs, this study employed DID to compare the changes in outcome variables over time between a population that is enrolled in a program (the borrower group) and a population that is not (the non-borrower group). Following steps have been followed:

i. Estimate the difference of the outcome variables related to socio-economic status by comparing the current status (at survey date) of beneficiary/borrower households (Treatment group) with their status before entering the program (the study considered before 3 years);

ii. Estimate the difference of the outcome variables related to socio-economic status by comparing the current status (at survey date) of non-beneficiary/non-borrower households (Control group) with their status before 3 years of the survey;

iii. Estimation of the difference of the above two differences [(ii)-(i)] will yield the net impact of the benefits from micro-credits.

### 2.6.3 Factor analysis

Factor analysis is a multivariate statistical technique that addresses itself to the study of interrelationships among a total set of observed variables (Manly, 2005; Rencher, 2002). The technique allows looking at groups of variables that tend to be correlated to one-another and identify underlying dimensions that explain these correlations. While in multiple regression model, one variable is explicitly considered as dependent variable and all the other variables as the predictors; in factor analysis all the variables are considered as dependent variables simultaneously. In a sense, each of the observed variables is considered as a dependent variable that is a function of some underlying, latent, and hypothetical set of factors. Conversely, one can look at each factor as dependent variable that is a function of the observed variables.

If $\{X_1, X_2, \ldots X_n\}$ be a set of n observed variables and $\{F_1, F_2, \ldots, F_m\}$ be a set of unobservable variables then the factor analysis model can be expressed as:

$$X - \mu = LF + \varepsilon$$

where $L_{n \times m}$ is the matrix of factor loadings (coefficient $l_{ij}$ is the loading of i-th variable on the j-th factor) and μ is a vector of the means of $X_i$.

Several methods are available in literature to estimate factor loadings and factor scores. The study considers principal component method to estimate the factor loadings and communalities [ $h_i^2 = \sum_{j=1}^{m} l_{ij}^2$ ], a measure of the variation of observed variables through factors. Several factor rotation methods like 'Varimax', 'Equamax', 'Quartimax' are adopted to find better estimates of factor loadings. Once the factors are identified and factor loading matrix is estimated then the estimated values of factors, factor scores, are calculated for each individual. The estimated values of factor scores are often used for diagnostic purposes as well as inputs to a subsequent analysis.

### 2.6.4 Discriminant analysis

The two-group discriminant analysis is used to identify the associated factors which are essential for discriminating the choice of micro-credits benefits from formal and informal source. Besides, the Fisher's linear discriminant function provides appropriate direction for classifying an individual who is supposed to receive credits either from formal source or form informal source having specific individual and household level characteristics.

Fisher's Linear Discriminant Function:

The discriminant analysis model involves linear combinations of the following form:

$$D = b_0 + b_1 \times X_1 + b_2 \times X_2 + \ldots + b_k \times X_k$$

were,

$D$ = Discriminant score

$b_0, b_1, b_2, \ldots, b_k$ = Discriminant coefficient or weight

$X_1, X_2, \ldots, X_k$ = Predictor or independent variables

The coefficients, or weights of the Fisher's linear discriminant function are estimated so that the groups of the dependent variable differ as much as possible on the values of the discriminant function.

## 2.7 Conclusion

The present chapter has explained the methodology used in the study the next chapter has attempted to explore the socio-economic features of the surveyed households in respect of micro-credits programs.

# CHAPTER THREE

# SOCIO-ECONOMIC FEATURES OF THE SURVEYED HOUSEHOLDS IN RESPECT OF MICRO-CREDITS PROGRAMS

## 3.1 Introduction

The analysis of profile of the households is essentail in describing the kind of subjects involved in the study. Their profile can provide the delimitation of the study so that what ever finidings brought out of the study can be described within the ambit only of this profile. So, to have insights into the micro-creditss outcomes, exploration of these characteristics is essential. To have the in-depth understanding, the profiles of the surveyed households in terms of socio-demographic characteristics, landholdings and housing conditions, household assets, and reasons for exclusion from the targeted micro-credits programs are discussed in this chapter. The major aspects of analysis are as follows:

- ❖ Socio-demographic characteristics of the households;
- ❖ Housing conditions and facilities of housing;
- ❖ landholdings of households;
- ❖ Possession of Assets by Household;
- ❖ Knowledge and perception of respondents on micro-credits benefits; and
- ❖ Reasons of not getting micro-credits from targeted source by eligible non-borrowers.

These aspects are discussed below:

## 3.2 Socio-demographic characteristics of the households

This section provides the characteristics of respondents and household members, which are helpful in identifying the socio-economic behavior of the households. The characteristics of respondents, viz., age, marital status, educational status, occupation, income earning status, SSNP benefit status, and disability status have been analyzed according to their micro-credits borrowing status (appendix table 3.1). On the other hand, the characteristics of the household members have been analyzed according to the gender as shown in appendix table 3.2.

### 3.2.1 Profile of respondents

The analysis of total respondents (2340) indicates that 69% respondents were borrowers and 31% were non-borrowers. The analyses of relationship of respondents with household heads indicate that about 81% respondents were the household heads. This proportion respondents as household head were found significantly ($p<0.01$) higher for non-borrower households (98.1%) than borrower households (73.8%). About 24% respondents were identified as husbands/wives for borrower households.

The distribution of respondents according to sex indicates that 76.8% were male and 23.2% were female. The age of the respondents has been classified into four classes as 16-30, 31-50, 51-60 and above 60. The age distribution of the respondents indicates that about 17% were of ages 16-30 years, about 62% were of ages 31-50 years, and only about 7% were of ages over 60 years. There is no significant variation in the age distribution of the respondents between borrower and non-borrower households.

The average of age of the respondents is estimated at 42.81 years with a standard deviation of 11.46 years. The average of age of the respondents was found slightly lower for borrower group of households (42.47 years) than that of borrower group of households (43.57 years). The analysis of marital status of the respondents indicated that 92.2% were married and 5% were found as widowed.

Data showed that about 27% of the respondents were found to have no education at all and they must have no ability to read or write too. A little more than half of the respondents attended primary level education and it can be inferred that they could hardly read or write. Only about 7.2 of the respondents were found who have completed secondary level of education. There is no significant variation in the education of the respondents according to the borrowing status of the households.

It is found out that about 20% the respondents were engaged in agriculture, about 23% were engaged in labor selling, about 13% in off-farm activities, about 19% engaged in business/service. Another 18.9% of the respondents are found to engage in household work, most of them were women. Regarding the income earning status of the respondents, nearly half are found as full-time income earner and about 29% were part-time income earner. About 21% of the respondents were found to have no earning source. Only about 3% of the respondents are found have any disability.

**3.2.2 Profile of the household population**

The socio-demographic characteristics of household population in terms of age group, marital status, educational status, occupation, income earner and disability status is shown in Appendix Table 3.2. The distribution of household population according to sex indicates that 51.17% were male and 48.83% were female (Total population was 11628, out of which 5950 male and 5678 female). For clear understanding of the age distribution of the household population, the age of the population has been classified into five classes. The first class (0 to 15 years) is found to have 35.2% male and 37.9% female population. The second class (16-30 year) is found to have 26.7% male and 28.3% female population. The third class (31-50 years) included 25.5% male and 25.3% female population. The fourth class was 51-60 age groups where 6.8% are found male and 4.9% female. The last and fifth class (ages above 60 years) were of 5.7% male and 3.7% female. The average age of the household population was found 27.52 years (with SD ±18.92 years) for male and 25.38 years (SD ±17.50 years) for female.

In terms of marital status of the population 16 years or more, the findings indicate that the lion part of population is married (66.4% for male and 73% for female). The proportion of unmarried people was found significantly low for female population in comparison with that of the male population, which indirectly indicates that female population is being getting early married than their male counterpart.

About 19% of the study population aged 7 years or older was found to have no education at all and they must have no ability to read or write too. Near half of the study population are found to attend primary level education and it can be inferred that they could hardly read or write. Only about 2% people were found who have attained/completed graduate level of education. The differences in educational attainment by sex showed that illiteracy was higher among women than men (appendix table 3.2). The proportion of attainment of secondary or higher level of education was found higher for men than women, partly due to gender discrimination against girl children. Parents favor boys' education than girls' with the expectation that the former would support them in the future. The practices of early marriage of girls, another consequence of discrimination, also contribute to high female illiteracy.

In respect of occupation the population aged in-between 16 years and 60 years, the occupation has been categorized into seven groups. The percentage of population in each group were found as farming - 8.9%, day laborer - 13.5%, off-firm activities -8.4%, service/business - 13.7%, student - 11%, household work - 34.6% and others - 9.9%. It is to be noted that about 71% women were engaged in household work, which may include vegetable cultivation, poultry rearing, weaving mattress, stitch *khatha*, fire-wood collection, feeding cattle, helping in agricultural work *etc.* (Kazal *et. al,* 2017). Therefore, women might have a significant contribution to household incomes, specially, income from vegetables, fruits, nursery, poultry, eggs, handicraft etc.

Regarding the distribution of earning of household members aged 16 years or more, it is found that about 28% were full-time earner and about 22% were part-time earner and about one-quarter male and three-quarters female are found to no work at the survey point. The disability status of the male household members aged 16 years and above indicated that about 4% of the household members (both male and female) were disabled.

**3.3 Housing conditions and facilities of housing**

The analysis of housing conditions and landholdings of a household plays an important role in determining its economic condition. The rural households are mostly depending on land for their livelihoods and it is considered as permanent asset of a household. The analysis of housing condition may have several implications in addition to assessing the economic ability. For example, households without proper sanitation facilities have a risk of incidence of diseases like diarrhea, dysentery and typhoid. The housing condition and related facilities of the study households is analyzed through Appendix Table 3.3 from different perspectives below:

### 3.3.1 Housing condition

The ownership pattern of the house indicates that about 95% of the households use their own house for living (appendix table 3.3). The number of sleeping-rooms gives an indication to what extent the households are crowded. Crowding in one sleeping room increases the risk of infection. It is found that about 20% of the households use one room for living and about 42% of the households use 2 rooms for the same purpose. Only about 10% households were identified who had four or more rooms for living. It is estimated that on average 2.48 persons live in a room in surveyed households, indicating a moderate congestion rate. The analyses indicate that there was no remarkable difference in household occupancy between borrowers and non-borrower households. The most common housing materials were found tin as about 86% houses were made of tin shed roof with different kinds of fence. Only 4.3% of the houses were made of straw roof with bamboo/muddy fence. The analysis reveals that the quality of houses was found inferior for borrower households in comparison to non-borrower households. Indoor pollution has important implications for the health of the household members. The type of fuel and the place where cooking is done are related to indoor air quality and the degree to which the household members are exposed to the risk of respiratory infections and other diseases. About 39% of the households reported that they did not have any separate kitchen room for cooking.

### 3.3.2 Facilities of housing

The facilities of housing are included sources of drinking water, cooking facilities, sanitation facilities, electricity coverage in the house, electricity coverage in the village. These are discussed below:

Sources of drinking water
Access to an improved source of drinking water is almost universal in Bangladesh. Nine-in-ten of the surveyed households are found to use tube-well as the source of their drinking water (appendix table 3.3). About 5% of the surveyed households reported that they used drinking water from pond or well.

Cooking facilities
The main sources of cooking fuel were straw/leaf/husk (43.6%) and wood/kerosene (30.1%) as near 74% of the households are found to use these ingredients for cooking.

Sanitation facilities

Most of the surveyed households (91.6%) reported that they have toilet facility of their own. Of the surveyed households, about 15% reported that they used hygienic latrines that include water-preventing capacity, about 28% households used pit latrine without water preventing capacity. The findings indicated that over half of the households used completely non-hygienic toilet that include *katcha* and open field/space

Electricity coverage in the house

The information about the sources of lighting of the households was also collected in the survey. The results indicated that about 81% of the HHs had improved lighting facility in terms of electricity and 19% has no access of electricity. The lighting of the rest of the HHs depends mainly on kerosene.

Electricity coverage in the village

The information about the sources of lighting of the village was also collected in the survey. The results indicated that about 87% of the community had improved lighting facility in terms of electricity and 13% has no access of electricity. The lighting of the rest of the community depends mainly on kerosene.

**3.4 Landholdings of households**

The landholdings pattern of the study households is presented in appendix table 3.4. The landholding of a household has been estimated by taking into account all types of land (homestead, agricultural and pond) that a household owns. Several studies categorized the landholding size considering its merit in production and income. Households with less than 2000 $m^2$ of land (<50 decimal) are commonly considered as 'Functionally Landless Households', households with 2000 $m^2$ up to 8000 $m^2$ (50-200 decimal) are considered as 'Marginal Households' and households with 8000 $m^2$ up to 30,000 $m^2$ are considered as 'Medium Households' (Hossain and Keus, 2004). In addition, households with only homestead land are considered as absolute landless (Hossain and Bayes, 2019). In this study households owning landholdings sized 15 decimals or less are considered as absolute landless.

The findings showed that 8.5% of the surveyed households had no homestead land of their own, 79.2% of the households owned homestead land between 1 to 15 decimals, and only 12.3% households owned homestead land more than 15 decimals. Among the surveyed households, over two-thirds have no agricultural land at all and about 7% owned agricultural land between 1 to 15 decimals, about 8% owned agricultural land between 16 to 50 decimals, and only about 17% were found to have agricultural land over 50 decimals. Among the land-owning households, the average cultivable agricultural land has been estimated at 105.85 decimals for non-borrower households and 96.96 decimals for borrower households. The findings indicate that there is no considerable variation in case of owning homestead or agricultural land according to the borrowing status of the surveyed households.

In rural Bangladesh, agricultural land can either be leased-in (not owned but cultivated) or leased-out (owned but not cultivated). It is found that about 27% of the households had taken some sort of land through leased-in, mainly as sharecropper. The average size of the leased-in or sharecropped land is found 102.66 decimals. In the study households, about 94% are found to have to pond in their possession. The size of the pond land is estimated at 8.65 decimals for the owning households.

Data regarding the total of all of types of land showed that about 5% households have no land of their own including homestead. The analysis shows that about 43% households own only 1 to 15 decimals of land; about 18% own land 16 to 50 decimals of land. The overall analysis of landholding indicates that about two-thirds of the surveyed households are functionally landless. Table 3.1 (page 56) depicts the household composition by micro-credits receiving status. The study found that the male-female ratio was 100:104 for borrower group while 100:107 for non-borrower group. The study revealed that female headed households are significantly ($p<0.05$) lower for borrower households (2.6%) than that of non-borrowers (3.5%). The study also found that the unemployment rate was 2.1% for borrower and 2.8% for non-borrower household. The study also revealed that the child dependency ratio was 54.35% for borrower households, whereas 50.86% for non-borrower households. Similar scenario revealed for aged dependency ratio. The overall dependency ratio was also differed significantly ($p<0.01$) for borrower (61.3%) and non-borrower (56.5%) households. The average family size was found significantly (p<0.01) higher for borrower (5.09) and non-borrower (4.71) households.

**Table 3.1** Household composition by micro-credits receiving status

| Characteristics | Borrower (N = 8172) | Non-borrower (N = 3451) | Z-statistic | P-value | Overall (N = 11623) |
|---|---|---|---|---|---|
| **Household composition (%)** | | | | | |
| Sex Ratio (Male per 100 female) | 104 | 107 | | | 105 |
| Female-headed household (in %) | 2.6 | 3.5 | -2.656 | 0.012 | 2.8 |
| Unemployment rate (Age 15-60) | 2.1 | 2.8 | -1.816 | 0.077 | 2.3 |
| Total household (Age 15-60 years) | 5023 | 2180 | | | 7203 |
| **Dependency Ratio (%)** | | | | | |
| Child (0–14) dependency ratio | 54.35 | 50.86 | 3.446 | 0.001 | 53.19 |
| Aged (60+) dependency ratio | 6.94 | 5.63 | 2.608 | 0.013 | 6.55 |
| Dependency ratio | 61.29 | 56.49 | 4.824 | <0.001 | 59.74 |
| **Family size (Mean ± SD)** | 5.09 ± 1.49 | 4.71 ± 1.52 | 5.644 | <0.001 | 4.97 ± 1.51 |

*Note:* HH=Households; SD=Standard Deviation.

## 3.5 Possession of assets by households

The possession of assets by study households has been divided into two parts viz. (i) durable assets and (ii) productive assets. These two are discussed below:

### 3.5.1 Possession of durable assets

The details of possession of durable assets have been given in appendix table 3.5. Among the work-related items, about 14.8% households are found to own boat, about 1.2% owned deep tube well, 20.6% owned tube well and 8.1% owned bicycle. A number of surveyed households owned luxury items: 16.2% of the households owned television, 6.2% owned drawing room furniture, 28.5% owned watch, and 69.7% of the households owned electric fan and 89.9% owned mobile phone. The findings indicate that in most of the cases possession of durable goods was significantly higher in the formal micro-credits receiving households than that of informal micro-credits receiving households when the differences between borrower and non-borrower households is not following any pattern.

**3.5.2 Possession of productive assets**

The average number of productive assets of the study households by micro-credits receiving status is shown in appendix table 3.6. Among the productive assets, are possessed and livestock are found to own by most of the study households Table showed that among the borrowers 664 HHs and 311 of non-borrowers HHs possessed cultivating instruments on average 3.27 and 3.05 respectively without any significant difference between the groups while livestock possessed by 694 HHs of borrowers and 287 HHs of non-borrowers on average 2.39 and 2.73 respectively showing significant difference between the groups(p=.003). On an average, 3.20 cultivating instruments are used by the households. On the other hand, the average number of livestock were 2.49. The ownership of others productive assets is not notable as a few households have owned those assets.

**3.6 Knowledge and perception of respondents on micro-credits benefits**

The study has collected the necessary data and information from 2340 sample households on perception and experience of micro-credits. The respondents were asked to provide their knowledge on micro-credits in terms of different aspects as shown in appendix table 3.7. The analysis revealed that about 95% respondents have the knowledge of micro credits benefits and about 86% has tried to get micro-credits. Among the credits-receiving households, 55.1% has tried to get credits from non-government sources (MFI/NGOs/Insurance) and 16% have tried to get credits from local moneylenders (*Mohajan*/ Private *Samittee*) in their first attempt. In addition, 4.1% has tried to get credits from government sources (banks/ cooperatives) and 2.4% from non-interest-bearing sources (relatives/friends/neighbors). About 8% credits-receiving households has mentioned that they tried to get credits from more than one sources. The findings indicate that 8.6% have succeed to get loan at their second attempt after failing in the first attempt and most of them received loan from local moneylenders (*Mohajan*/ Private *Samittee*). The data on respondents where/whom they communicated to get micro-credits revealed that a little more than one-third has reported that they went to UP office, followed by relatives/neighbors/friends (14.7%), government officers (4.8%), NGOs (3.3%). It is to be mentioned that about 4% of the respondents were demanded to provide money as bribe for getting micro-credits benefits. Six-in-ten of the total respondent believed that micro-credits benefits would help them to come out from poverty.

## 3.7 Reasons of not getting micro-credits from targeted source by eligible non-borrowers

The respondents of the eligible non-borrower households were asked to put their views on the reasons for exclusion from the targeted micro-credits programs. The literature suggests several reasons are responsible for exclusion from the targeted micro-credits programs. Total sixteen causes are listed as in appendix table 3.8 as the causes those are responsible for not getting micro-credits from targeted organizations by eligible non-borrowers. The descriptive statistics of the causes is explored under the subsection 3.6.1 and the inferential statistics i.e., Principal Component Analysis (PCA) is given in section 3.6.2.

### 3.7.1 Descriptive statistics on reasons for exclusion from the targeted programs

The respondents were asked to put their perception on five-point Likert scale (starting from strongly disagree=1 to strongly agree=5) on each of the pre-assigned reasons for not getting micro-credits from the targeted schemes. The percentage in each of the agreement level indicates that a significant number of respondents did not make any comments regarding all the reasons (appendix table 3.8). Among the strongly agreed responses, bureaucratic complexity was endorsed by 13.6% respondents, followed by nepotism (11.2%), non-cooperation from local lenders (11.2%) and lack of networking or lobbying (9.1%). The highest percentage of response against 'agree' was found for lack of networking or lobbying (29.2%), followed by non-cooperation from local lenders (28.6%), non-cooperation from public delegate of micro-credits institution (25.9%), distance of MFIs from villages (21.6%), misappropriation of credits (17.7%), no political exposure (16.9%). On the other hand, 39.4% of the respondents have put their response as strongly disagree on 'non-availability of NID. The findings revealed that the maximum of the observed respondents put their response on 'disagree' for reasons limitation of budget (according to selector), couldn't provide bribe or entry fee, didn't have any idea about such program, non-availability of NID, distance from micro-credits from the village, no micro-credits in the area, non-availability of collateral and biasness. On the other hand, majority of the total respondents respond to 'agree' for reasons non-cooperation from public delegate of micro-credits institution, non-cooperation from local lenders, lack of networking or lobbying, misappropriation of credits. The respondents had no idea about the reasons like no political exposure and nepotism for the exclusion from micro-credits program.

### 3.7.2 Factor analysis on reasons for exclusion from the targeted micro-credits programs

The study has employed factor analysis to identify the major dimensional reasons (to reduce a large number of variables into fewer numbers of factors) for exclusion from the expected micro-credits programs. The factor analysis has been performed using PCA with Varimax rotation technique. The summarized results of factor analysis are shown in table 3.2. The scree plot (Figure 3.1) explored that four factors are mainly responsible for not getting the micro-credits from the expected credits programs/schemes. That means the 15 reasons can be reduced into four major dimensional factors.

Thus, the factor analysis extracted four factors as the reasons for not being included in the micro-credit programs with a cumulative percentage of variance 64% by the 15 variables and the KMO value as 0.735. The test statistics including chi-square value of Bartlett's test of sphericity, KMO value indicates that the factor analysis is appropriate technique to explore the major dimensions of the underlying variables (Table 3.2). Based on the maximum variation of the factors, the study identified four main factors of reasons for not being included in micro-credits programs. These factors are:

**Figure 3** Choice of the number of factors for reasons of excluded from micro-credits program

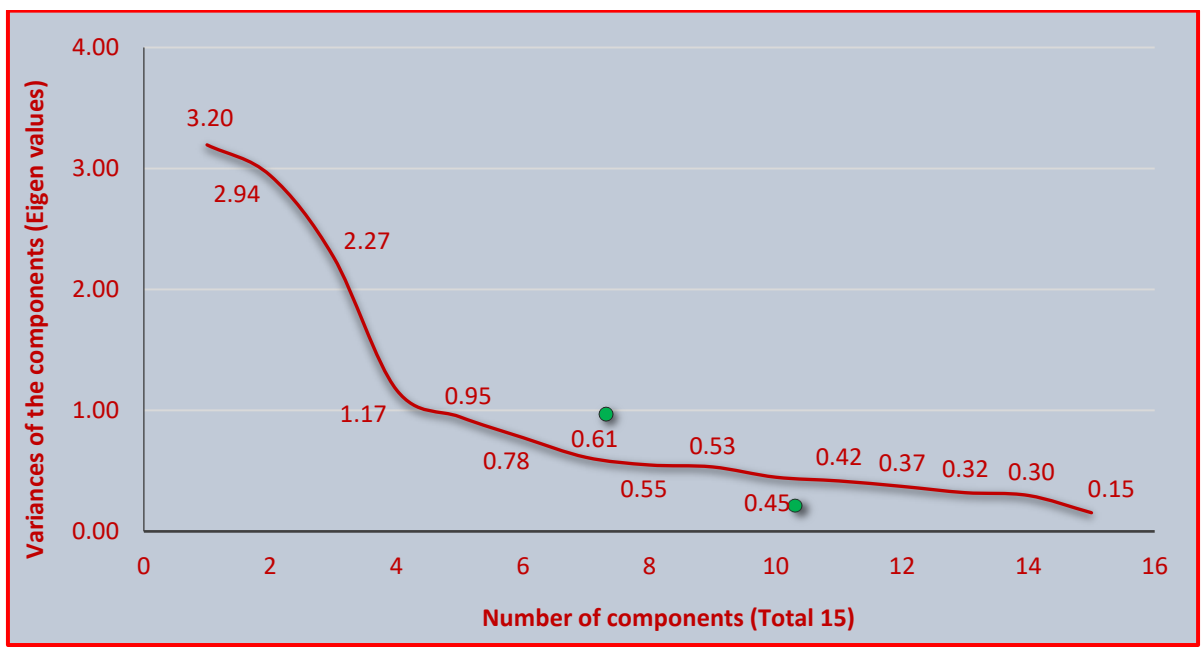

Factor-I related to *credits misappropriation and biasness* and composed of four reasons: (i) no micro-credits in the area; (ii) non-availability of collateral; (iii) misappropriation of credits and (iv) biasness. Factor-II related to *non-cooperation of local authority* and is includes four causes: (i) non-cooperation from public delegate of micro-credits institution; (ii) non-cooperation from local lenders (iii) lack of networking or lobbying and (iv) distance from micro-credits from the village. Factor-III regarding *bureaucracy and budget limitation* and included four causes :(i) bureaucratic complexity; (ii) limitation of budget (according to selector); (iii) no political exposure and (iv) nepotism. Factor-IV associated with *ignorance and corruption* and is composed of three causes: (i) couldn't provide bribe or entry fee; (ii) didn't have any idea about such program and (iii) non-availability of NID.

Table 3.2 Major dimensional factors for excluding from the micro-credit benefits

| Serial | Reasons for | Rotated Factor Loadings (Varimax Rotation) | | | | Communalities |
|---|---|---|---|---|---|---|
| | | F1 | F2 | F3 | F4 | |
| a. | Bureaucratic complexity | | | 0.611 | | 0.525 |
| b. | Limitation of budget (according to selector) | | | 0.759 | | 0.597 |
| c. | Couldn't provide bribe or entry fee | | | | 0.559 | 0.599 |
| d. | No political exposure | | | 0.623 | | 0.680 |
| e. | Didn't have any idea about such program | | | | 0.833 | 0.704 |
| f. | Nepotism | | | 0.599 | | 0.608 |
| g. | Non-cooperation from public delegate of Micro-credits institution | | 0.727 | | | 0.567 |
| h. | Non-cooperation from local lenders | | 0.818 | | | 0.673 |
| i. | Non-availability of NID | | | | 0.700 | 0.517 |
| j. | Lack of networking or lobbying | | 0.775 | | | 0.652 |
| k. | Distance from Micro-credits from the village | | 0.635 | | | 0.468 |
| l. | No Micro-credits in the area | 0.620 | | | | 0.591 |
| m. | Non availability of collateral | 0.862 | | | | 0.794 |
| n. | Misappropriation of credits | 0.896 | | | | 0.844 |
| o. | Biasness | 0.863 | | | | 0.755 |
| Percentage of Variation Explained | | 21.30 | 19.61 | 15.13 | 7.80 | |
| Total Variation explain by the extracted factors | | | | | | 63.84 |
| Kaiser-Meyer-Olkin Measure of Sampling Adequacy | | | | | | 0.735 |
| Bartlett's Test of Sphericity | | Chi-Square =3261.82; df =105 & P-value <0.001 | | | | |
| Extraction Method | | Principal Component Analysis | | | | |

## 3.8 Conclusion

The socioeconomic and -demographic profile of respondents has been presented in this chapter meticulously in the next chapter profile of formal and informal micro-credits will be discussed in details.

# CHAPTER FOUR
# PROFILE OF FORMAL AND INFORMAL MICRO-CREDITS

## 4.1 Introduction

The nature and characteristics of both formal and informal micro-credits used in the study area are explored in this chapter under the following aspects:

- Sources and types of micro-credits in *Haor* region;
- Terms and conditions of formal and informal micro-credits;
- Volume of over-all loans;
- Payment structure of loans;
- Purpose of micro- credits;
- Expenditure and investment pattern of both formal and informal credits;
- Comparative economic performance between borrowers and non-borrowers households;
- Causes of non-payment of loans; and
- Attitudes of borrowers towards micro-credits.

## 4.2 Sources and types of micro-credits in *Haor* region

There sources of micro-credits can be classified into formal and informal. The formal is regulated by country's banking and financial services acts/rules or regulation where informal sources are embracing all financial transactions taking place beyond various countries' regulations on banking and other financial sectors. The definition of informal finance includes such schemes as the operations of Savings and Credits Associations (SCA), known all over Africa, professional moneylenders, part-time moneylenders (estate owners, traders, grain millers, smallholder farmers, employers, relations and friends) mobile bankers, known as collectors in West Africa; credits unions; and cooperative societies and these exist in both urban and rural areas (Ernest, 2011). While savings collectors fall under the first category of deposit mobilizers, moneylenders - including relations and friends - do not generally accept deposits and may be assigned to a second category. SCAs (credits unions and credits cooperatives) take in deposits and also lend in varied forms. Most informal units deal with specific groups of people, ensuring that only those satisfy distinct selection criteria are able to either deposit with them or borrow from them (Ernest, 2011). In this study the formal sources included government and private banks and cooperatives, different micro-finance institutions NGOs and insurance companies which are regulated either by the government directly or by the affiliated body or authority such as, MRA (Micro-credits Regulatory Authority, Bangladesh Bank, Bureau of NGOs, Bangladesh Bank, Ministry of Finance or any other authorised body. The informal sources are (i) interest bearing –includes local money lender, mohajon/private somitee (ii) non–interest bearing includes relatives, friends/neighbour/land owner. The descriptive statistics are given below in table 4.1

**Table 4.1** Sources and types of micro-credits

| Characteristics | No. of HH | % of HH | Average amount | Total amount | % of total | SD amount |
|---|---|---|---|---|---|---|
| **Loan Type** | | | | | | |
| Cash | 1595 | 99.3 | 37651 | 60053106 | 99.29 | 53321 |
| Kinds (food items) | 12 | 0.7 | 35625 | 427500 | 0.71 | 52938 |
| Total | 1607 | 100.0 | | 60480606 | | |
| **Sources of loan** | | | | | | |
| Formal | 1158 | 72.1 | 34596 | 40062504 | 66.00 | 40826 |
| Informal | 449 | 27.9 | 45475 | 20418100 | 34.00 | 76126 |
| Total | 1607 | 100.0 | 37656 | 60480606 | | |
| **Formal sources** | | | | | | |
| Government (*Banks/Co-operatives*) | 88 | 5.5 | 52313 | 4603500 | 11.50 | 62822 |
| Nongovernment (*MFI/NGO/Insurance*) | 1070 | 66.6 | 33139 | 35458998 | 88.50 | 38142 |
| Total Formal | 1158 | 72.1 | 34596 | 40062504 | 100.00 | 40826 |
| **Informal sources** | | | | | | |
| Local money lender (*Mohajan/Private Samittee*) | 393 | 24.5 | 46675 | 18343098 | 89.84 | 74777 |
| Non-interest loan (*Relatives/friends/neighbors*) | 53 | 3.3 | 38076 | 2018000 | 9.88 | 87567 |
| More than one sources | 3 | 0.2 | 19000 | 57000 | 0.28 | 1732 |
| Total Informal | 449 | 27.9 | 45475 | 20418100 | 100.00 | 76126 |
| **Informal by interest** | | | | | | |
| Interest bearing | 408 | 90.9 | 47696 | 19460099 | 95.30 | 79057 |
| Non- interest bearing | 41 | 9.1 | 23366 | 957 1000 | 4.70 | 27571 |
| Total | 449 | 100.0 | 45475 | 20418100 | 100.00 | |

Table 4.1 showed that 1595 HHs (99.3) borrowed cash loan and only 12(0.7%) in kinds and in terms of total amount the percentage is about same (99.29% cash and .71% in kinds). In terms of sources 1158 HHs (72.1%) borrowed from formal sources and 449 HHs (27.1%) from informal sources and in terms of amount of loan 66% from formal sources and 34% from informal sources. There are two types of formal sources viz. (i) government (Banks/Co-operatives; and (ii) nongovernment (MFI/NGO/Insurance) and the data showed that 88 HHs (5.5%) borrowed from the former and the 1070 HHs (66.6%) from the later. There are three informal sources *viz.*, (i) Local money lender (Mahajan/Private Samittee); (ii) non-interest loan (Relatives/friends/ neighbors); and (iii) more than one sources and the data revealed that 393 HHs (24.5%) taken loans from the first, 53 HHs (3.3%) from the second and only 3 HHs (0.2%) from the third and in terms of amount of loan 89.84% from the first, 9.88% from the second and 0.28% from the third. Again, the informal sources have been divided into interest bearing and non-interest bearing and the data revealed that 408 HHs (90.9%) have taken from interest bearing sources and HHs 41 (9.1%) from non-interest bearing and respect of amount loan 95.3% from the former and 4.70 from the latter.

### 4.3 Terms and conditions of formal and informal micro-credits

Table 4.2 shows the profile of micro-credits benefits for the borrower households in the year 2019. The study found that the maximum average loan was Tk.56536.36 for formal micro-credits receiver at the rate of interest more than 25% while this amount was Tk.69853.66 for informal micro-credits receiver at the interest rate of 21-25%. There is highly significant (p<0.01) difference between formal and informal micro-credits receiver for all kind of interest rate. The study also found that the annually installment type had the maximum average amount of loan which was Tk.79105.26 for formal and Tk.64589.04 for informal micro-credits receiver. On the other hand, biweekly installment type had the minimum average amount of loan which were Tk.18846.15 and Tk.24657.14 for formal and informal micro-credits receiver respectively. There are significant (p<0.01) differences between formal and informal micro-credits receiver for weekly, monthly and annually installment types. The study revealed that the maximum average amount of loan was Tk.88400.00 for formal micro-credits receiver at installment type 'one time' whereas the maximum amount was Tk.52157.40 for informal micro-credits receiver at installment type 'up to 12 times. On the other hand, the loan amount was minimum at installment type 'more than 24 times' and 'one time' respectively for formal and informal micro-credits receiver. In addition, there was highly significant (p<0.01) difference between formal and informal micro-credits receiver for all kind of installment type. The study also revealed that the maximum average amount of loan was taken in two years of duration for both formal and informal micro-credits receiver though the minimum amount was Tk.33050.64 for formal receiver in one year of duration and Tk. 18790.48 for informal receiver in six months of duration.

**Table 4.2** Terms and conditions of micro-credits in *Haor* region in Bangladesh

| Profile of micro-credits | Formal credits (N = 1158) | | Informal credits (N = 449) | | P-value | Overall (N = 1607) | |
|---|---|---|---|---|---|---|---|
| | HHs | Average | HHs | Average | | HHs | Average |
| **Interest rate** | | | | | | | |
| No interest (0%) | - | - | 041 | 23365.85 | | 041 | 23365.85 |
| 1% to 10% | 141 | 39539.01 | 106 | 25576.42 | <0.001 | 247 | 33546.96 |
| 11% to 15% | 525 | 28456.19 | 055 | 25254.55 | <0.001 | 580 | 28152.59 |
| 16% to 20% | 211 | 39123.22 | 021 | 65095.24 | 0.001 | 232 | 41474.14 |
| 21% to 25% | 171 | 29672.51 | 041 | 69853.66 | 0.001 | 212 | 37443.40 |
| More than 25% | 110 | 56536.36 | 185 | 60156.76 | <0.001 | 295 | 58806.78 |
| **Installment type** | | | | | | | |
| Weekly | 619 | 30378.03 | 015 | 31133.33 | <0.001 | 634 | 30395.90 |
| Biweekly | 013 | 18846.15 | 007 | 24657.14 | 0.258 | 020 | 20880.00 |
| Monthly | 501 | 38613.77 | 180 | 26580.56 | <0.001 | 681 | 35433.19 |
| Quarterly | 006 | 27500.00 | 028 | 30321.43 | 0.176 | 034 | 29823.53 |
| Annually | 019 | 79105.26 | 219 | 64589.04 | 0.001 | 238 | 65747.90 |
| **Total installment** | | | | | | | |
| One time | 015 | 88400.00 | 180 | 37544.44 | <0.001 | 195 | 41456.41 |
| 02 to 12 times | 508 | 35810.04 | 223 | 52157.40 | <0.001 | 731 | 40796.99 |
| 13 to 24 times | 023 | 53652.17 | 030 | 46566.67 | 0.001 | 053 | 49641.51 |
| More than 24 times | 612 | 31553.92 | 016 | 39500.00 | <0.001 | 628 | 31756.37 |
| **Duration of loan** | | | | | | | |
| Six months | 039 | 33410.26 | 021 | 18790.48 | <0.001 | 060 | 28293.33 |
| One year | 1096 | 33050.64 | 317 | 33285.49 | <0.001 | 1413 | 33103.33 |
| Two years | 023 | 110260.87 | 111 | 85333.33 | <0.001 | 134 | 89611.94 |
| **Collateral type** | | | | | | | |
| Collateral | 067 | 56358.21 | 028 | 77807.14 | 0.003 | 095 | 62680.00 |
| Non collateral | 1091 | 33259.85 | 421 | 43324.23 | <0.001 | 1512 | 36062.17 |

*Note:* HHs=Households;

There had strong evidence that all duration of total installment showed significant ($p<0.01$) difference between formal and informal micro-credits receiver. It was also found in the study that the average amount of received loan was higher in 'collateral' type than 'non collateral' for both formal and informal micro-credits receiver and the difference of averages were also statistically significant ($p<0.01$) for both formal and informal receiver. The study further found that the formal micro-credits receiver took maximum average loan Tk.20969.69 at the condition of 'total paid' whereas the informal micro-credits receiver took maximum average loan Tk.20911.05 at the condition of 'principal' paid. All type of condition of loan pay between formal and informal micro-credits receiver was highly significant ($p<0.01$). The study again revealed that the total unpaid loan had maximum average for both formal (Tk.25786.90) and informal (Tk.54035.89) receiver and their differences also statistically significant ($p<0.01$).

### 4.4 Volume of over-all loans

Table 4.3 depicted that total formal loan receiver is 1158 and informal receiver is 449. The total amount of formal loan is about Tk. 40 million and informal loan is about Tk 20 million and in total is Tk 60 million. The average of formal loan is Tk. 34596 and informal loan is Tk. 45475 and in overall-all average is Tk. 37636. The maximum amount of formal loan is Tk 5.5 lac... and informal is Tk 11.00 lac and the minimum amount of both formal and informal loan are Tk. 2000. The over-all minimum and maximum amount of loan are Tk. 20000 and 11.00 lac with SD Tk. 53303.

**Table 4.3** Total, minimum, maximum, average, standard deviation, median and inter quartile range of micro-credits

| Statistics | Formal (N=1158) | Informal (N=449) | | | Overall (N=1607) |
| | | Interest bearing (N=408) | No interest bearing (N=41) | All informal (N=449) | |
|---|---|---|---|---|---|
| Minimum | 2000 | 2000 | 3000 | 2000 | 2000.00 |
| Maximum | 550000 | 1100000 | 150000.00 | 1100000 | 1100000 |
| Total | 40062504 | 19460099 | 958000 | 20418100 | 60480602 |
| Mean | 34596 | 47696 | 233656 | 45475 | 37636 |
| SD | 40826 | 79057 | 27571 | 76126 | 53303 |
| Median | 25000 | 30000 | 15000 | 25000 | 25000 |
| IQR | 25000 | 25000 | 15000 | 40000 | 25000 |

### 4.5 Payment structure of loans

The payment structure shows that during the period (2016-19) 1113 households of formal credits borrowers totally paid amounted to Tk. 20969.69 on average out of which principal loan paid by 1108 receivers and interest paid by 1092 receivers on average amount of Tk. 17679.33 and Tk. 3434.59 respectively while the 245 households of informal credits borrowers totally paid amounted to Tk. 20878.61 on average out of which principal loan paid by 144 receivers and interest paid by 233 receivers on average amount of Tk. 20911.05 and Tk. 9030.34 respectively (table 4.4). The table showed that there are significant differences between formal and informal credits in respect of total amount of loan paid, principal paid and interest paid on average. The unpaid loan analysis showed that 946 households of formal credits borrowers failed to pay credits amounted to Tk. 25786.90 on average out of which principal unpaid by 944 borrowers and interest unpaid by 945 receivers on average amount of Tk. 21688.35 and Tk. 4148.79 respectively while the 402 households of informal credits borrowers totally failed to pay principal amounted to Tk. 54035.89 on average out of which principal loan unpaid by 401 receivers and interest unpaid by 326 receivers on average amount of Tk. 20911.05 and Tk. 9030.34 respectively. The table showed that there are significant differences between formal and informal credits in respect of total amount of unpaid loan, unpaid principal and interest.

**Table 4.4** Total, average, paid and unpaid amount of loan in *Haor* regions of Bangladesh

| Profile of micro-credits | Formal credits (N = 1158) | | Informal credits (N = 449) | | P-value | Overall (N = 1607) | |
|---|---|---|---|---|---|---|---|
| | HHs | Average | HHs | Average | | HHs | Average |
| **Paid loan** | | | | | | | |
| Total paid | 1113 | 20969.69 | 245 | 20878.61 | <0.001 | 1358 | 20953.26 |
| Principal | 1108 | 17679.33 | 144 | 20911.05 | <0.001 | 1252 | 18051.03 |
| Interest | 1092 | 3434.59 | 233 | 9030.34 | 0.001 | 1325 | 4418.60 |
| **Unpaid loan** | | | | | | | |
| Total unpaid loan | 946 | 25786.90 | 402 | 54035.89 | <0.001 | 1348 | 34211.31 |
| Unpaid principal | 944 | 21688.35 | 401 | 43408.75 | <0.001 | 1345 | 28164.10 |
| Unpaid interest | 945 | 4148.79 | 326 | 13237.79 | 0.001 | 1271 | 6480.04 |

*Note:* HHs=Households;

## 4.6 Purpose of micro-credits

The purpose receivers of micro-credits of both formal and informal are analyzed in terms of descriptive statistics as well as inferential statistics under the subsections 4.6.1 and 4.6.2 below:

### 4.6.1 Descriptive statistics of purposes of micro-credits

The borrowers borrow money for different purposes and from pilot survey seventeen purposes of loan are identified and data table 4.5 showed that the top three items of purposes of credits in both formal and informal credits are purchasing food items (32.1%) for formal and 47.4% for informal credits and in total 36.4% following payment of previous loan 27.5% for formal and 34.1% for informal and in total 29.% and crop production 24.1% for formal and 32.4% for informal and in total 26.4%. It is found that there are significance differences between formal and informal sources in all the three purposes. The purpose healthcare expenditure in total is ranked fourth (19.4%) and also in informal loans 32.3% for formal but in case of formal loans it is ranked fifth (14.4%) after trade /business purpose (20.2%) showing significant differences between two types of loans. The fifth purpose is trade/business in total (17.8%) but in case formal its' position is fourth and ninth in informal with showing significant difference between them. In order to find the important purposes PCA analysis is done in subsection 4.5.2.

**Table 4.5** Descriptive statistics of purposes of credits receivers

| Purpose of taking loan | Formal sources (N = 1158) Yes (%) | Informal sources (N = 449) Yes (%) | P-value | Both sources (N = 1607) Yes (%) |
|---|---|---|---|---|
| Purchasing of food items | 32.1 | 47.4 | <0.001 | 36.4 |
| Crop production | 24.1 | 32.4 | 0.001 | 26.4 |
| Rearing cattle/poultry | 13.7 | 9.8 | 0.043 | 12.6 |
| Sending family member to abroad | 2.8 | 2.7 | 0.397 | 2.8 |
| Trade/business/industry | 20.2 | 11.6 | <0.001 | 17.8 |
| Fish farming/fishing | 4.6 | 2.4 | 0.052 | 4.0 |
| Daughter/son's marriage | 2.9 | 6.5 | 0.001 | 3.9 |
| Constructing housing | 7.7 | 14.7 | <0.001 | 9.6 |
| Tackling shocks of natural calamities | 4.0 | 15.4 | <0.001 | 7.2 |
| Tackling shocks of sudden death of HH head | 0.9 | 0.9 | 0.399 | 0.9 |
| Purchasing of livelihood equipment | 12.8 | 21.4 | <0.001 | 15.2 |
| Payment of loan | 27.5 | 34.1 | 0.013 | 29.3 |
| Repairing cost of houses | 12.5 | 22.7 | <0.001 | 15.4 |
| Healthcare expenditure | 14.4 | 32.3 | <0.001 | 19.4 |
| Education | 7.4 | 17.6 | <0.001 | 10.3 |
| Others | 13.3 | 9.1 | 0.027 | 12.1 |

### 4.6.2 Inferential statistics of purposes of micro-credits

The study has employed PCA to identify the major dimensional purposes (to reduce a large number of variables into fewer numbers of factors) of micro credits receivers. The PCA has been performed using Varimax rotation technique. The summarized results of factor analysis have shown in figure 4.1 and table 4.6.

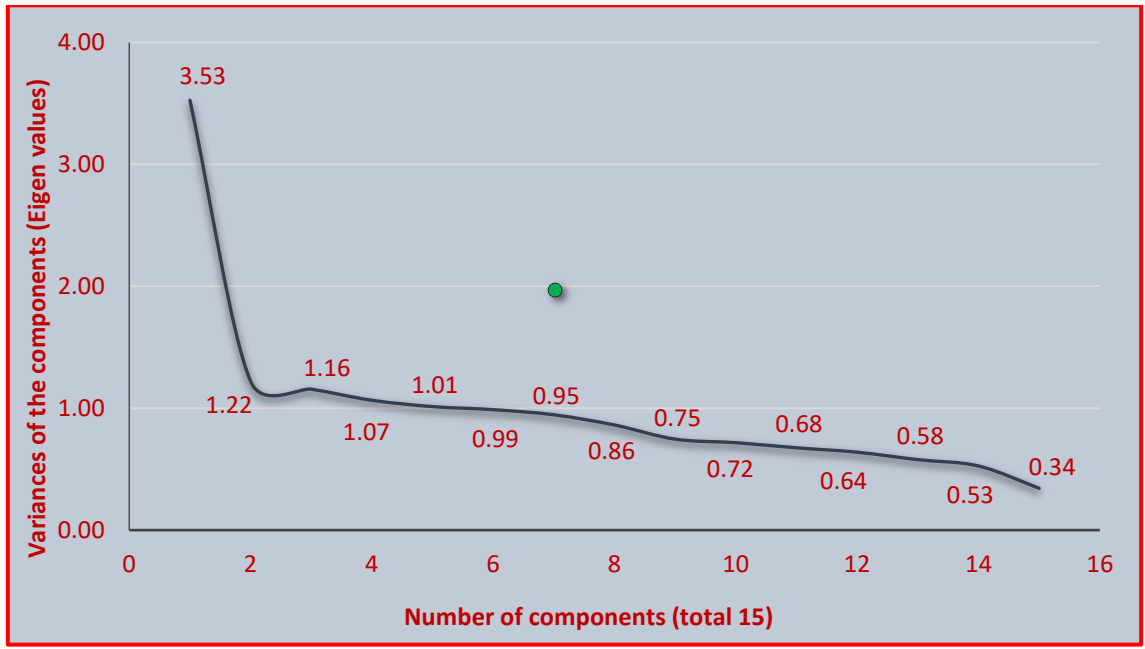

**Figure 4.1** Choice of the number of components for the purposes of micro-credits

The factor analysis extracted seventeen purposes into five factors as the purposes of taking micro-credits services with a cumulative percentage of variance 53%. The KMO value is 0.823 (table 4.6). The test statistics including chi-square value of Bartlett's test of Sphericity, KMO value indicates that the factor analysis is appropriate technique to explore the major dimensions of the underlying variables. Based on the maximum variation of the factors, the study identified five main factors of taking loan purposes. Factor- I related to daily life and livelihoods and is composed of more than 50% (nine) listed purposes - (i) purchasing of food items; (ii) constructing housing; (iii) Tackling shocks of natural calamities (iv)purchasing of livelihood equipment; (v)payment of loan; (vi) repairing cost of houses; (vii) healthcare expenditure; (viii) education; and (ix) others. Factor-II associated adaptability of natural shocks and farming and is included two purposes- (i) fish farming/fishing and (ii) tackling shocks of sudden death of HHs' head. Factor-III connected to cropping and raring cattle and is formed of two purposes- (i) crop production; (ii) rearing cattle/poultry. Factor-IV is linked with business and marriage and is blended with two causes: (i) trade/business/industry; (ii) daughter/son's marriage. Factor-V is associated a unique and single purpose and that is (i) sending family member to abroad.

**Table 4.6** Major dimensional factors behind the purpose of receiving loan

| Serial | Reasons for receiving loan | Rotated factor loadings (Varimax) | | | | | Comunalities |
|---|---|---|---|---|---|---|---|
| | | F1 | F2 | F3 | F4 | F5 | |
| a. | Purchasing of food items | 0.552 | | | | | 0.478 |
| b. | Crop production | | | 0.614 | | | 0.478 |
| c. | Rearing cattle/poultry | | | 0.623 | | | 0.445 |
| d. | Sending family member to abroad | | | | | 0.915 | 0.866 |
| e. | Trade/Business/Industry | | | | 0.806 | | 0.675 |
| f. | Fish farming/Fishing | | 0.761 | | | | 0.610 |
| g. | Daughter/son's marriage | | | | 0.613 | | 0.413 |
| h. | Constructing housing | 0.638 | | | | | 0.447 |
| i. | Tackling shocks of natural calamities | 0.801 | | | | | 0.699 |
| j. | Tackling shocks of sudden death of HH head | | 0.677 | | | | 0.489 |
| k. | Purchasing of livelihood equipment | 0.652 | | | | | 0.435 |
| l. | Payment of loan | 0.493 | | | | | 0.509 |
| m. | Repairing cost of houses | 0.691 | | | | | 0.488 |
| n. | Healthcare expenditure | 0.646 | | | | | 0.454 |
| o. | Education | 0.675 | | | | | 0.493 |
| p. | Others | 0.552 | | | | | 0.478 |
| Percentage of variation explained | | 23.51 | 8.13 | 7.71 | 7.10 | 6.76 | |
| Total Variation explain by the extracted factors | | | | | | | 53.20 |
| Kaiser-Meyer-Olkin measure of sampling adequacy | | | | | | | 0.823 |
| Bartlett's test of sphericity | | Chi-square = 3622.224; df = 105 & P-value < 0.001 | | | | | |
| Extraction method | | Principal Component Analysis | | | | | |

*Note:* HHs = Households; df = degrees of freedom.

## 4.7 Expenditure and investment pattern of both formal and informal micro-credits

The respondents were given a list of 24 expenditures and investments items of their loans and among those 14 items were marked by them as shown in table 4.7. The study revealed that formal micro-credits receiver spent 15.05% of total loan in food consumption whereas this percentage increased to 23.24% for informal micro-credits receiver. There existed significant (p=0.02) difference in food expenditure between formal and informal micro-credits receiver.

The study found that the expenditure in clothing and essential sector of received loan were 1.55% for formal and 3.36% for informal micro-credits receiver. Formal micro-credits receiver spent 15.09% of total loan in agricultural inputs. On the other hand, this rate decreased to almost half (7.70%) for informal micro-credits receiver. About 1.40% of formal and 0.23% of informal micro-credits receivers expended in purchasing durables. Almost 5% of total loan used in housing for both formal and informal micro-credits receiver. The study also found that the share of total loan expended in purchasing land was quite similar for both formal (2.23%) and informal (2.59%) micro-credits receiver. The formal micro-credits receiver spent their loan more than double in purchasing animals than informal micro-credits receiver. The formal as well as informal micro-credits receiver shared almost equal amount of received money in payment of their previous loan. In family enterprise sector, the formal micro-credits receiver spent highest (15.76%) amount of their loan whereas the informal micro-credits receiver spent 8.02%.

Table 4.7 Expenditure and investment pattern of both formal and informal micro-credits

| Expenditure Heads | Formal (N = 1158) | | | Informal (N = 449) | | | P-value (Share) | Overall (N = 1607) | |
|---|---|---|---|---|---|---|---|---|---|
| | HHs | Average | Share (%) | HHs | Average | Share (%) | | Average | Share (%) |
| Food consumption | 384 | 12287.24 | 15.05 | 199 | 13908.04 | 23.24 | 0.020 | 12840.48 | 17.34 |
| Clothing & essentials | 078 | 6365.38 | 1.55 | 058 | 9965.52 | 3.36 | 0.314 | 7900.74 | 2.06 |
| Agricultural inputs | 285 | 16686.32 | 15.09 | 076 | 15927.63 | 7.70 | 0.098 | 16526.59 | 13.03 |
| Purchasing durables | 031 | 13193.55 | 1.40 | 003 | 7666.67 | 0.23 | 0.393 | 12705.88 | 1.07 |
| Housing | 153 | 12939.22 | 5.27 | 064 | 16968.75 | 4.89 | 0.396 | 14127.65 | 5.16 |
| Purchasing land | 039 | 18448.72 | 2.23 | 018 | 35333.33 | 2.59 | 0.398 | 23780.70 | 2.33 |
| Purchasing animals | 168 | 16041.67 | 9.24 | 037 | 18000.00 | 4.39 | 0.251 | 16395.12 | 7.88 |
| Payment of loan | 294 | 15964.29 | 12.03 | 117 | 21469.23 | 12.97 | 0.385 | 17531.39 | 12.29 |
| Family enterprises | 234 | 31602.56 | 15.76 | 045 | 47644.44 | 8.02 | 0.160 | 34189.96 | 13.60 |
| Health care | 138 | 13060.87 | 4.49 | 110 | 24063.64 | 11.68 | 0.043 | 17941.13 | 6.50 |
| Human capital | 039 | 14989.74 | 1.34 | 018 | 9833.33 | 1.26 | 0.399 | 13361.40 | 1.32 |
| Going member abroad | 027 | 112777.8 | 2.19 | 004 | 312500.0 | 0.83 | 0.392 | 138548.4 | 1.81 |
| Children marriage | 037 | 30270.27 | 2.63 | 029 | 49948.28 | 5.19 | 0.344 | 38916.67 | 3.35 |
| Others | 240 | 23039.17 | 13.00 | 113 | 27420.35 | 16.86 | 0.250 | 24441.64 | 14.08 |

*Note:* HHs = Households.

Informal micro-credits receiver spent almost three times more money of received loan than formal micro-credits receiver in health care sector. The study further found that the share of total loan expended in human capital was quite similar for both formal (1.34%) and informal (1.26%) micro-credits receiver. To send family member abroad the formal micro-credits receiver spent more than 2% whereas informal micro-credits receiver spent less than 1% of total loan. Informal micro-credits receiver spent 5.19% of total loan in children marriage but this rate decreased to half (2.63%) for informal micro-credits receiver.

## 4.8 Comparative economic performance between borrower and non-borrower households:

The economic performance of micro-credits borrowers and non-borrowers are discussed under two heads *viz.*, income and expenditures under the sub sections 4.8.1 and 4.8.2 below:

### 4.8.1 Comparison of income between borrower and non-borrower households

Table 4.8 showed that annual income of borrowers' majority 1053 (66%) HHs depend on income from labor sale with average income 57 thousand and in total 60 million (36% of total) with SD 39 thousand following from agriculture 830 (56%) HHs on average 45 thousands in total 40 million (24% of total) with SD 45 thousand, from non-agriculture 83 (52%) HHs on average 43 thousand and in total 36 million (22% of total) with SD with average, from business 400 (25%) HHs on average 69 thousand and in total 28) million (17% of total with SD 58 thousand and 208 (13%) HHs' income from donations/begging on average and in total 2 million (1% of total) with SD 12 thousand.

**Table 4.8** Sources and status of income between micro-credits receiving and non-receiving households in *Haor* region of Bangladesh

| Major income source | Borrower (N = 1607) | | | | Total in Million (% of total) | Non-borrower (N = 733) | | | | | t-statistic |
| --- | --- | --- | --- | --- | --- | --- | --- | --- | --- | --- | --- |
| | HHs | % of HHs | Average (000) | SD (000) | | HHs | % of HHs | Average (000) | Total in millions (% of total) | SD (000) | |
| Agricultural | 900 | 56% | 45 | 325 | 40 (24) | 333 | 45% | 46 | 15 (20) | 42 | -0.51 |
| Non-agricultural | 830 | 52% | 43 | 48 | 36 (22) | 331 | 45% | 56 | 18 (25) | 65 | -3.63*** |
| Labor sale | 1053 | 66% | 57 | 39 | 60 (36) | 464 | 63% | 54 | 25 (34) | 28 | 1.06 |
| Business | 400 | 25% | 69 | 58 | 28 (17) | 197 | 27% | 74 | 15 (20) | 60 | -1.02 |
| Donation/begging | 208 | 13% | 98 | 12 | 2 (1) | 81 | 11% | 11 | 0.90 (1) | 16 | -0.71 |
| **Total income excluding debt** | 1607 | 100% | 103 | 59 | 165 (100) | 733 | 100% | 102 | 74 (100) | 64 | 0.44 |
| Debt | 1253 | 78% | 29 | 36 | 36 | 158 | 22% | 12 | 2 | 14 | 5.82*** |
| **Total income including debt** | 1607 | 100% | 70 | 49 | 109 | 733 | 100 | 86 | 62 | 53 | |

*Note:* HHs = Households; SD = Standard Deviation

The table showed that 78% of HHs has debt on average 29 thousand and in total 359 million with SD 36 thousand. The annual income of majority 464 (63%) non-borrowers comes from business on average 54 thousands with total income 25millions (34%) income following 333 (45%) HHs' income come from agriculture with 46 thousands in total 15 million (20%) with SD 42 thousands, 331 (45%) HHs' income from non-agriculture and on average 56 thousands and in total 18 million (25%), 197(27%) HHs' come from business on average 74 million and in total 15 (20%) million and 81 (11%) HHs' income comes from donations/begging on average 11 thousand and in total .90 million (1%) with SD 16 thousand. It is seen that 22% HHs of non-borrowers under the pressure of debt on average 12 thousand and in total million. Finally, we observed that there are significant differences between income of borrowers and non-borrowers in respect of non-agriculture and debt.

### 4.8.2 Comparison of expenditure between borrowers and non-borrower households:

Table 4.9 (page 74) depicted that both the borrowers and non-borrowers HHs incurred expenditure for consumption of both food and non-food but borrowers HHs incurred 80% for food and 20% for non-food while non-borrowers HHs 84% for food and 16% for non-food and there is significant difference in respect non- food consumption as well as total consumption between borrowers and non-borrowers HHs. There are twelve items of investment expenditure and the items above 5% of in terms of percentage of total expenditure of borrowers HHs are agriculture (20%), healthcare (20%) family business (16%), house repair (9%) and poultry /livestock (7%) while in respect of non –borrowers the investment items above 5% of in terms of percentage of total expenditure are healthcare (27%), agriculture (23%), education (18%), family business 9% and house repair (9%). There is significant difference between borrowers and non-borrowers in the items of education, healthcare, poultry/livestock, productive assets, durable goods, house repair, others investment and also in total investment expenditure as well as total expenditure (consumption plus investment).

### 4.9 Causes of non–payment of loan

For identifying the reasons behind the non-payment of loans in time the borrowers were given list of 15 causes. The collected data is measured in terms of percentage of the respondents with respect to disagree, neutral and agree. The causes are discussed showing descriptive as well as inferential statistics under the subsections 4.9.1 and 4.9.2 as follows:

**Table 4.9** Expenditures between borrower and non-borrower households in *Haor* region

| Major expenditure heads | Borrower (*N* =1607) | | | | | Non-borrower (*N* = 733) | | | | | t-statistic |
|---|---|---|---|---|---|---|---|---|---|---|---|
| | HHs | %of HHs | Average (000) | Total in million (% of total) | SD (000) | HHs | % of HHs | Average (000) | Total (in millions (% total) | SD (000) | |
| 1. Food | 1607 | 100 | 563 | 90 (80) | 24 | 733 | 100 | 55 | 41 (84) | 25 | 1.25 |
| 2. Non-food | 1607 | 100 | 14 | 22 (20) | 95 | 733 | 100 | 11 | 8 (16) | 83 | 5.11*** |
| A.Total consumption (1+2) | 1607 | 100 | 70 | 112 (100) | 27 | 733 | 100 | 66 | 49 (100) | 28 | 2.79*** |
| 3. Education | 1190 | 74 | 10 | 12 (16) | 12 | 481 | 66 | 9 | 4 (18) | 12 | 1.79* |
| 4.Healthcare | 1539 | 96 | 10 | 15 (20) | 12 | 686 | 94 | 8 | 6 (27) | 10 | 3.53*** |
| 5.Agriculture | 811 | 50 | 19 | 15 (20) | 12 | 279 | 38 | 19 | 5 (23) | 18 | -0.14 |
| 6.Poultry-livestock | 541 | 34 | 10 | 5 (7) | 11 | 202 | 28 | 5 | 1 (5) | 6 | 5.58*** |
| 7.Family business | 220 | 14 | 25 | 6 (8) | 16 | 79 | 11 | 22 | 2 (9) | 12 | 1.17 |
| 8.Productive asset | 233 | 14 | 12 | 3 (4) | 14 | 54 | 7 | 6 | 0 | 7 | 2.74*** |
| 9.Durable goods | 146 | 9 | 9 | 1 (1) | 11 | 84 | 11 | 4 | 0 | 4 | 3.78*** |
| 10.House repair | 574 | 36 | 11 | 7 (9) | 21 | 235 | 32 | 7 | 2 (9) | 9 | 3.11*** |
| 11.Land purchase | 38 | 2 | 30 | 1 (1) | 35 | 13 | 2 | 25 | 0 | 23 | 0.43 |
| 12.Others | 561 | 35 | 16 | 9 (12) | 14 | 158 | 22 | 11 | 2 (9) | 12 | 4.03*** |
| B. Total investment (3-12) | 1606 | 100 | 46 | 74 (100) | 36 | 730 | 100 | 31 | 22 (100) | 29 | 10.29*** |
| C. Total expenditure (A+B) | 1607 | 100 | 116 | 186 | 51 | 733 | 100 | 97 | 71 | 47 | 8.50*** |
| Savings | 507 | 32 | 4 | 59 | 4 | 111 | 15 | 19 | 2 | 24 | -13.59*** |

*Note:* HHs = Households; SD = Standard Deviation

### 4.9.1 Descriptive statistics of causes of non-payment of loan in time

The descriptive statistics of the cause of non-payment of loan timely by formal and informal borrowers are given in table 4.10. As stated earlier in our study there is 1607 borrowers' respondents among them 1158 borrowed from formal sources and 449 from informal sources. Among the causes top five causes are agreed by formal borrowers are installment period is very short (80.9%) following rate of interest is very high (70.6%), natural calamities (67.9%) medical treatment/medicine (67.7%), acute food problem (63.7%). On the other hand, the top five causes agreed by informal borrowers are rate of interest is very high (92.8%) following misappropriation of loan (82.8%), medical treatment/medicine (80.5%) installment period is very short (72.2%) and natural calamities (71.1%). In total top five causes of nonpayment of loans in time are installment period is very short (78.1%), rate of interest is very high (77.8%), medical treatment/medicine (71.9%), misappropriation of loan (69.2%) and natural calamities (68.9%). It is observed that the top five causes are among the top five causes marked either by formal or by informal borrowers.

Table 4.10 Descriptive statistics of the cause of non-payment of loan timely

| Name of the causes | | Formal sources (N = 1158) | | | | Informal sources (N = 449) | | | Both |
|---|---|---|---|---|---|---|---|---|---|
| | HHs | Disagree | Neutral | Agree | HHS | Disagree | Neutral | Agree | Agree |
| Acute food problem | 726 | 15.8 | 20.5 | 63.6 | 349 | 25.2 | 5.7 | 69.1 | 65.4 |
| Medical treatment/medicine | 725 | 7.4 | 24.8 | 67.7 | 349 | 9.2 | 10.3 | 80.5 | 71.9 |
| Investment loss | 731 | 8.9 | 30.1 | 61.0 | 348 | 14.1 | 15.2 | 70.7 | 64.1 |
| Natural calamities | 733 | 15.4 | 16.6 | 67.9 | 349 | 19.5 | 9.5 | 71.1 | 68.9 |
| Insufficient of loan for investment | 721 | 12.1 | 25.9 | 62.0 | 349 | 15.8 | 21.2 | 63.0 | 62.3 |
| Duration of loan is short for return from investment | 724 | 7.5 | 27.2 | 65.3 | 349 | 14.6 | 26.4 | 59.0 | 63.3 |
| Installment period is very short | 723 | 5.8 | 13.3 | 80.9 | 349 | 14.6 | 13.2 | 72.2 | 78.1 |
| Rate of interest is very high | 729 | 14.4 | 15.0 | 70.6 | 349 | 5.2 | 2.0 | 92.8 | 77.8 |
| Renewal of loan is unavailable | 721 | 23.7 | 39.3 | 37.0 | 349 | 16.6 | 33.2 | 50.1 | 41.3 |
| Misappropriation of loan | 722 | 10.9 | 26.5 | 62.6 | 349 | 6.6 | 10.6 | 82.8 | 69.2 |
| Crop's failure | 725 | 20.8 | 33.4 | 45.8 | 349 | 23.8 | 13.8 | 62.5 | 51.2 |
| Expenditure for marriage of son/daughter etc. | 719 | 55.1 | 30.2 | 14.7 | 349 | 74.5 | 9.2 | 16.3 | 15.3 |
| Family problems and expenditure | 721 | 61.9 | 31.2 | 6.9 | 349 | 78.1 | 14.4 | 7.5 | 7.1 |
| Unexpected accident | 723 | 39.1 | 27.2 | 33.6 | 349 | 61.7 | 10.4 | 28.0 | 31.8 |

*Note:* HHs = Households

### 4.9.2 Inferential statistics on causes of non-payment of loans in time

In the previous section the descriptive statistics have been discussed. As there are fifteen causes of nonpayment of loans and to find out actual causes PCA is done. Based on Eigen value (more than 1.00) among the fourteen causes four factors is significantly mentionable for non-payment of loans in time as seen in the Scree plot given below in figure 4.2 (page 77).

Through the PCA with Varimax Orthogonal rotation method as in Table 4.11 it is clear that 53.52. % of total variation has been explained by the extracted factors with 0.715 KMO measure of sampling adequacy. It means the sample size is sufficient to explore the factors. Details of the factors or causes are given in table 4.11.

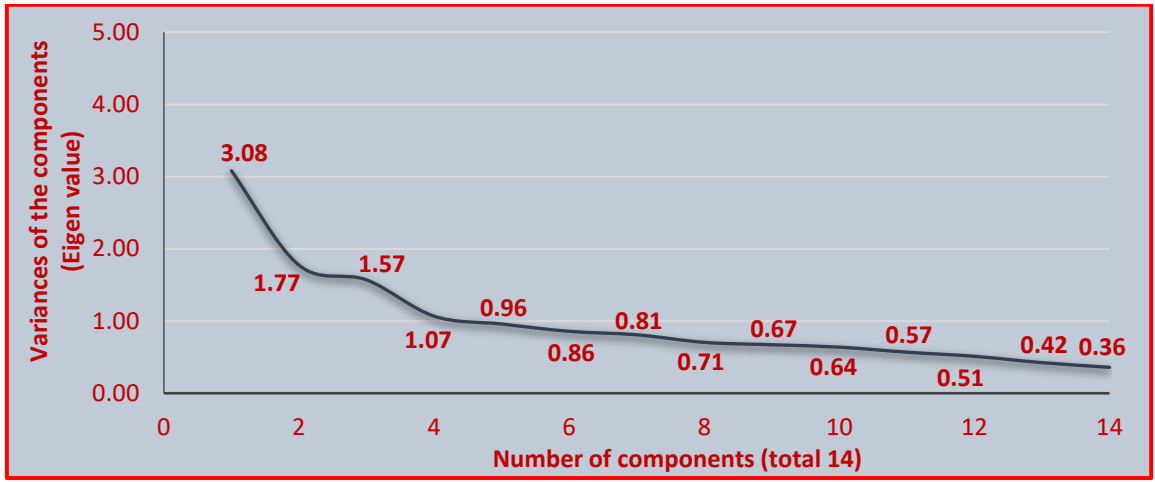

**Figure 4.2** Scree plot of factors for the causes of non-payment of loan

The first factor is composed of six factors with 22.2 (around 41% of 53.52%) percentage of total variation explained and the causes are associated to food, medicine and natural shocks and is composed of - (i) acute food problem; (ii) medical treatment/medicine; (iii) investment loss; (iv) natural calamities; (v) misappropriation of loan and (vi) crops' failure.

**Table 4.11** Major dimensional factors for the causes of scheduled non-payment of loan

| Serial | Causes of non-payment of loan timely | Rotated factor loadings (Varimax) | | | | Communalities |
|---|---|---|---|---|---|---|
| | | F1 | F2 | F3 | F4 | |
| a. | Acute food problem | 0.647 | | | | 0.487 |
| b. | Medical treatment/medicine | 0.640 | | | | 0.421 |
| c. | Investment loss | 0.702 | | | | 0.526 |
| d. | Natural calamities | 0.686 | | | | 0.586 |
| e. | Insufficient of loan for investment | | | 0.684 | | 0.553 |
| f. | Duration of loan is short for return from investment | | | 0.800 | | 0.666 |
| g. | Installment period is very short | | | | 0.609 | 0.650 |
| h. | Rate of interest is very high | | | -0.067 | 0.675 | 0.468 |
| i. | Renewal of loan is unavailable | | | | 0.478 | 0.438 |
| j. | Misappropriation of loan | 0.454 | | | | 0.387 |
| k. | Crop's failure | 0.641 | | | | 0.520 |
| l. | Expenditure for marriage of son/daughter etc. | | 0.767 | | | 0.616 |
| m. | Family legal problems and expenditure | | 0.794 | | | 0.636 |
| n. | Unexpected accident | | 0.673 | | | 0.538 |
| Percentage of variation explained | | 22.01 | 12.67 | 11.21 | 7.63 | |
| Total variation explains by the extracted factors | | | | | | 53.52 |
| Kaiser-Meyer-Olkin measure of sampling adequacy | | | | | | 0.715 |
| Bartlett's test of sphericity | | Chi-square = 2630.829; df = 91 & P-value < 0.001 | | | | |
| Extraction method | | Principal component analysis | | | | |

The second factor is connected to marriage and legal problems and is composed of three factors with 12.67 (around 24% of 53.52%) percentage of total variation explained and is included - (i) expenditure for marriage of son/daughter etc.; (ii) family legal problems and expenditure and (iii) unexpected accident. The third factor is related to terms of loans and investment loss and is composed of three factors with 11.21 (around 21% of 53.52%) percentage of total variation explained and the causes are (i) duration of loan is short for return from investment (ii) rate of interest is very high; and (iii) insufficient of loan for investment lastly the fourth factor is linked to cost of loan and is composed of three factors with 7.63% (around 14% of 53.52%) percentage of total variation explained and the causes are (i) installment period is very short; (ii) rate of interest is very high and (iii) renewable loan is unavailable.

### 4.10 Attitude of borrowers towards micro-credits

To measure the attitude of respondents towards micro-credits we have listed 16 variables measured in term of percentage of the respondents with respect to disagree, neutral and agree. The collected data is analyzed both in descriptive as well as inferential aspects through the subsections 4.10.1 and 4.10.2.

### 4.10.1 Descriptive analysis of attitude of borrowers towards micro-credits

Table 4.12 portrayed that among the sixteen dimensions in case of formal credits the top five agreed (positive) aspects are :(i) by micro-finance your food security has increased (54.3%); (ii) micro-finance is helping you in better access to healthcare (53.8%); (iii) micro-finance is helping you in better financial situation of your family (50.7%); (iv) Due to micro-finance, employment opportunities have been increased (45.6%) and (v) by micro-finance your income has increased (44%). In case of informal borrowers, the top five agreed (positive) attitude are :(i Local loans are easier to get than MFIs (76.8%); (ii) micro-finance is helping you in better access to healthcare (61.2%); (iii) by micro-finance your food security has increased (56.1%); (iv) micro-finance is helping you in better financial situation of your family (52.3%); and (v) Operational assistance received from MFIs was helpful to run the business (49.4%). Therefore, among the top five positive attitudes both formal and informal borrowers agreed that micro-creditss play positive roles in food security, healthcare and better financial position of the borrowers. In total the top five positive attitudes are: by micro-finance your food security has increased (54.3%); (ii) micro-finance is helping you in better access to healthcare (53.8%); (iii) micro-finance is helping you in better financial situation of your family (50.7%); (iv) Local loans are easier to get than MFIs (48.7%) and (v) Operational assistance received from MFIs was helpful to run the business (45.2%). In total top five disagreed (negative) attitudes are (i) cost of local loans is lower than MFIs (73.6); (ii) local lenders are friendly than MFIs (58.4%); (iii) terms and conditions of local loans are easier than MFIs (57.9); (iv) duration of credits is sufficient (56.4%); and (v) by micro-finance your savings has increased (56.1%). To identify the most important aspects of attitude the PCA is done in the following sub-section.

**Table 4.12** Distribution of the attitude of borrowers on micro-credits

| Statements of attitudes | Formal sources (N = 1158) | | | Informal sources (N = 449) | | | Both sources (N = 2340) | |
|---|---|---|---|---|---|---|---|---|
| | Disagree | Neutral | Agree | Disagree | Neutral | Agree | Disagree | Agree |
| The rate of interest of micro-credits is reasonable | 56.9 | 12.1 | 31.0 | 88.4 | 4.2 | 7.3 | 65.7 | 24.4 |
| Amount of credits is sufficient | 46.5 | 14.1 | 39.4 | 57.2 | 4.7 | 38.1 | 49.5 | 39.0 |
| Duration of credits is sufficient | 55.5 | 15.0 | 29.4 | 58.8 | 6.7 | 34.5 | 56.4 | 30.9 |
| Terms and conditions are not rigid | 41.4 | 33.4 | 25.2 | 63.9 | 12.7 | 23.4 | 47.7 | 24.7 |
| By micro-finance your food security has increased | 13.9 | 32.5 | 53.6 | 31.8 | 12.0 | 56.1 | 18.9 | 54.3 |
| By micro-finance your income has increased | 30.3 | 25.7 | 44.0 | 48.3 | 14.0 | 37.6 | 35.3 | 42.2 |
| By micro-finance your savings has increased | 49.9 | 22.9 | 27.2 | 71.9 | 13.1 | 14.9 | 56.1 | 23.8 |
| Micro-finance is helping you in better access to education | 25.5 | 40.2 | 34.4 | 38.3 | 26.7 | 35.0 | 29.1 | 34.5 |
| Micro-finance is helping you in better access to healthcare | 21.8 | 27.4 | 50.9 | 26.3 | 12.5 | 61.2 | 23.0 | 53.8 |
| Micro-finance is helping you in better financial situation of your family | 24.7 | 25.3 | 50.0 | 33.2 | 14.5 | 52.3 | 27.1 | 50.7 |
| Operational assistance received from MFIs was helpful to run the business | 21.3 | 35.1 | 43.6 | 28.7 | 21.8 | 49.4 | 23.4 | 45.2 |
| Due to micro-finance, employment opportunities have been increased | 27.5 | 26.9 | 45.6 | 47.7 | 16.3 | 36.1 | 33.1 | 42.9 |
| Local loans are easier to get than MFIs | 46.6 | 15.6 | 37.7 | 18.9 | 4.2 | 76.8 | 38.9 | 48.7 |
| Local lenders are friendly than MFIs | 63.0 | 16.6 | 20.5 | 46.8 | 5.6 | 47.7 | 58.4 | 28.1 |
| Cost of local loans is lower than MFIs | 74.9 | 16.8 | 8.4 | 70.4 | 6.0 | 23.6 | 73.6 | 12.6 |
| Terms and conditions of local loans are easier than MFIs | 59.3 | 17.8 | 22.9 | 54.1 | 8.9 | 37.0 | 57.9 | 26.8 |

*Note:* MFIs = Micro-finance institutions.

### 4.10.2 Factor analysis of attitudes towards micro-credits

Through the PCA with Varimax Orthogonal rotation method explained 57.38 % of total variation by the extracted factors with 0.747 Kaiser-Meyer-Olkin (KMO) measure of sampling adequacy. It means the sample size is sufficient to explore the factors. The PCA analysis showed that four attitudes are significant based on Eigen value (1.00 and above) as seen in the scree plot of major dimensional attitude of borrowers on micro-credits is given in figure 4.3.

The PCA as presented in table 4.13 showed that through the Varimax rotated factor loading sixteen dimensions converted into four factors explaining 57.38% of total variation through extraction. The first factor is associated with income and savings and has explained 20.38% of total variations with six dimensions (i) terms and conditions are not rigid (ii) by micro-finance your income has increased; (iii) by micro-finance your savings has increased; (iv) micro-finance is helping you in better financial situation of your family; (iv) operational assistance received from MFIs was helpful to run the business; (vi) due to micro-finance employment opportunities have been increased.

**Figure 4.3** Choice of the number of components among the attitude of borrowers on micro-credits programs in Bangladesh

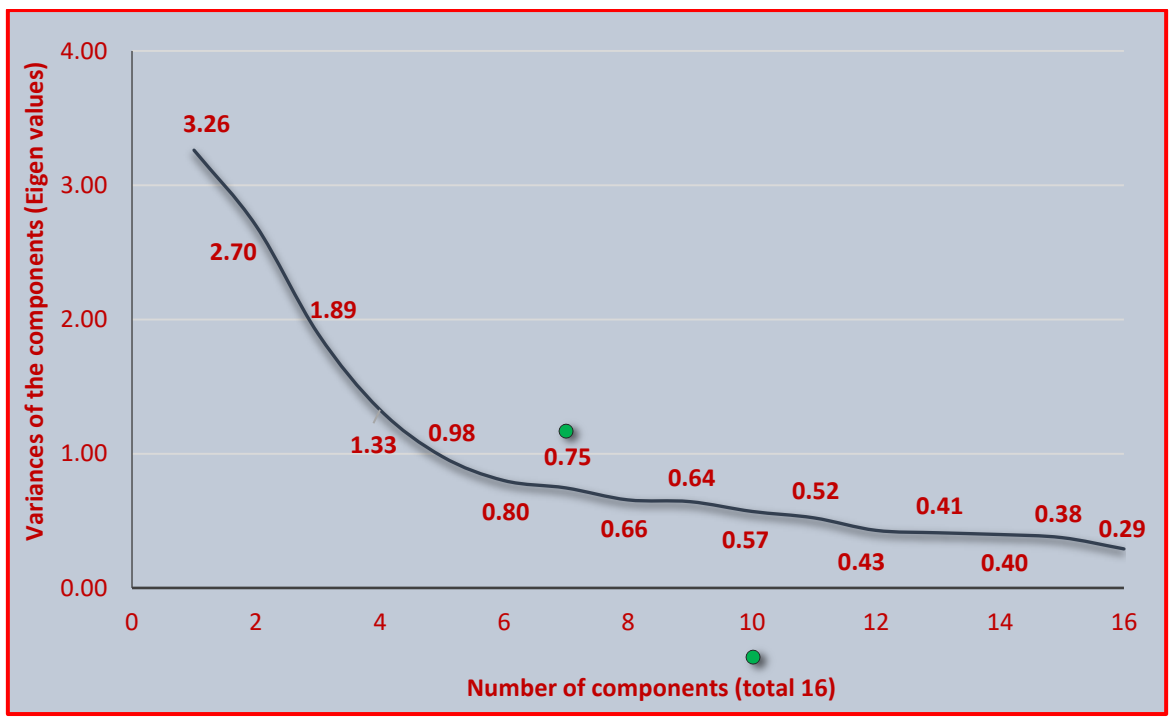

The second factor connected to terms and conditions of credit with explaining 16.86% of total variation and is composed of four dimensions: i) local loans are easier to get than MFIs; ii) local lenders are friendly than MFIs; iii) cost of local loans is lower than MFIs; iv) terms and conditions of local loans are easier than MFIs. The third factor regarded to cost of credits with explaining 11.84% of total variation is composed of three dimensions: i) rate of interest of micro-credits is reasonable; ii) amount of credits is sufficient; and iii) duration of credits is sufficient. The fourth related to food and health security with explaining 8.29% of total variation is composed of i) micro-finance your food security has increased; ii) micro-finance is helping you in better access to education; and iii) micro-finance is helping you in better access to healthcare.

Table 4.13 Major dimensional factors of the attitude of borrowers on micro-credits programs

| SL# | Attitude of borrowers on micro-credits | Rotated Factor Loadings (Varimax) | | | | Comunalities |
|---|---|---|---|---|---|---|
| | | F1 | F2 | F3 | F4 | |
| 01. | The rate of interest of micro-credits is reasonable | | | 0.694 | | 0.574 |
| 02. | Amount of credits is sufficient | | | 0.760 | | 0.633 |
| 03. | Duration of credits is sufficient | | | 0.809 | | 0.658 |
| 04. | Terms and conditions are not rigid | 0.446 | | | | 0.375 |
| 05. | By micro-finance your food security has increased | | | | 0.716 | 0.535 |
| 06. | By micro-finance your income has increased | 0.654 | | | | 0.564 |
| 07. | By micro-finance your savings has increased | 0.734 | | | | 0.567 |
| 08. | Micro-finance is helping you in better access to education | | | | 0.621 | 0.564 |
| 09. | Micro-finance is helping you in better access to healthcare | | | | 0.804 | 0.675 |
| 10. | Micro-finance is helping you in better financial situation of your family | 0.661 | | | | 0.495 |
| 11. | Operational assistance received from MFIs was helpful to run the business | 0.657 | | | | 0.464 |
| 12. | Due to micro-finance, employment opportunities have been increased | 0.682 | | | | 0.481 |
| 13. | Local loans are easier to get than MFIs | | 0.637 | | | 0.493 |
| 14. | Local lenders are friendly than MFIs | | 0.857 | | | 0.768 |
| 15. | Cost of local loans is lower than MFIs | | 0.792 | | | 0.667 |
| 16. | Terms and conditions of local loans are easier than MFIs | | 0.808 | | | 0.668 |
| Percentage of variation explained | | 20.38 | 16.86 | 11.84 | 8.29 | |
| Total Variation explain by the extracted factors | | | | | | 57.38 |
| Kaiser-Meyer-Olkin measure of sampling adequacy | | | | | | 0.747 |
| Bartlett's test of sphericity | | Chi-square = 7023.422; df = 120 & P-value < 0.001 | | | | |
| Extraction method | | Principal Component Analysis | | | | |

Note: F1 to F4 indicates four extracted factors; df = degrees of freedom; MFIs = Micro-finance Institutions.

**4.11 Conclusion**

This chapter has presented the profile of formal and informal micro-credits elaborately and consequences of micro-credits will be discussed in the following chapter.

# CHAPTER FIVE
# CONSEQUENCES OF MICRO-CREDITS

## 5.1 Introduction

In this chapter the role of micro-credits on poverty alleviation is analyzed based on the previous chapters. Specially does micro-credits help to come out from vicious cycle of poverty or not is tested in this chapter and also how does it help and what percentage of borrowers are coming out from vicious cycle of poverty through micro-credits are shown by the methods stated in chapter two. These are discussed below:

## 5.2 Role of micro-credits on graduation of selected socio-economic status of the borrowers

The role of micro-credit in graduating socio-economic status are- ensuring food security, social status and investment in education and healthcare.

### 5.2.1 Ensuring food security

Table 5.1 shows the self-rated food security condition through before-after and case-control comparison. The impact of micro-credits on the food security condition is administered though these comparisons.

**Table 5.1** Before-after comparison of prevalence of households' food insecurity status by case-control analysis in *Haor* regions of Bangladesh

| Status of food insecurity | Non-borrower HHs (%) (*N* = 733) | | | Borrower HHs (%) (*N* = 1607) | | | Both HHs (%) (*N* = 2340) | | |
|---|---|---|---|---|---|---|---|---|---|
| | 2016 | 2019 | P-value | 2016 | 2019 | P-value | 2016 | 2019 | P-value |
| Two meals in a day throughout year (Moderate food insecurity) | 14.5 | 12.4 | 0.199 | 19.0 | 14.5 | 0.001 | 17.6 | 13.8 | 0.001 |
| Some periods of hunger (Normal food insecurity) | 14.5 | 12.6 | 0.227 | 16.2 | 9.3 | <0.001 | 15.7 | 10.3 | <0.001 |
| Three meals a day throughout year (No food insecurity) | 70.5 | 75.0 | 0.061 | 63.3 | 76.2 | <0.001 | 65.6 | 75.9 | <0.001 |

*Note:* HHs = Households.

However, the households of *Haor* regions who received micro-credits benefit are found to be more food secure than the non-borrower households. The proportion of households with moderate food insecurity is significantly (p<0.01) reduced to 14.5% from 19.0% due to micro-credits program. This study also found that the incidence of normal food insecurity also significantly (p<0.01) reduced to 9.3% from 16.2%. On the other hand, such type of changes is also observed among non-borrower households but the decrement was found notably lower and statistically insignificant as well.

### 5.2.2 Social status

Table 5.2 shows the self-rated socio-economic condition of the study households by before-after and case-control comparison. The percentage households having extreme poor and moderately poor condition are estimated lower after receiving micro-credits. It is to be mention that the proportion also reduces to the non-borrower households but the decrement was statistically insignificant. However, the micro-credits have significant influence for reducing extreme poverty condition among credits receiving households.

**Table 5.2** Before-after comparison of households' social status in *Haor* regions of Bangladesh

| Social- status | Non-borrower HHs (%) (N = 733) | | | Borrower HHs (%) (N = 1607) | | | Both HHs (%) (N = 2340) | | |
|---|---|---|---|---|---|---|---|---|---|
| | 2016 | 2019 | P-value | 2016 | 2019 | P-value | 2016 | 2019 | P-value |
| Extremely poor | 22.4 | 16.9 | 0.012 | 23.1 | 13.6 | <0.001 | 22.9 | 14.6 | <0.001 |
| Moderately poor | 21.0 | 20.6 | 0.392 | 27.9 | 22.0 | <0.001 | 25.8 | 21.5 | 0.001 |
| Poor | 46.5 | 49.0 | 0.252 | 44.4 | 56.4 | <0.001 | 45.0 | 54.1 | <0.001 |
| Middle class | 8.2 | 11.5 | 0.042 | 4.5 | 8.0 | <0.001 | 5.6 | 9.1 | <0.001 |
| Rich | 1.9 | 2.0 | 0.395 | 0.1 | 0.1 | 0.399 | 0.7 | 0.7 | 0.399 |

*Note:* HHs = Households.

### 5.2.3 Investment on education and healthcare

Table 5.3 shows the self-rated condition on the changes of education and healthcare expenditure among the surveyed households. The percentage of households with the increase of educational expenditure are estimated higher in case of borrower households (63.8%) than that of non-borrower households (51.0%). This study found that the average educational expenditure increases significantly to Tk. 9893.7 from Tk. 8417.8 among borrower households and to Tk. 8742.3 from Tk. 7606.6 among non-borrower households.

In addition, the increment of healthcare expenditure also observed both case and control households. However, the before-after comparison among both type of households indicate that the increment of healthcare expenditure is also obvious in Bangladesh. The case-control comparison indicates that the healthcare expenditure increases more among borrower household (70.1%) than non-borrower household (60.3%). Besides, the before-after comparison indicates that the healthcare expenditures significantly increase to Tk. 9815.2 from Tk. 7730.9 among borrower households and also insignificantly increases to Tk. 7966.3 from Tk. 7404.2 among non-borrower households.

Table 5.3 Case-control comparison for the change in education and health expenditure in *Haor* regions of Bangladesh

| Indicators | Non-borrower ($N = 733$) | Borrower ($N = 1607$) | P-value | Overall ($N = 2340$) |
|---|---|---|---|---|
| **Education expenditure** | | | | |
| Increased (% of HHs) | 51.0 | 63.8 | <0.001 | 59.8 |
| No change (% of HHs) | 45.2 | 32.2 | <0.001 | 36.2 |
| Decreased (% of HHs) | 3.8 | 4.0 | 0.388 | 4.0 |
| Mean ± SD (Tk. in 2015) | 7606.6 ± 11624.6 | 8417.8 ± 10917.3 | 0.216 | 8181.1 ± 11130.2 |
| Mean ± SD (Tk. in 2018) | 8742.3 ± 11653.5 | 9893.7 ± 12092.6 | 0.076 | 9560.2 ± 11974.9 |
| P-value (2015 vs 2018) | 0.070 | <0.001 | | <0.001 |
| **Health expenditure** | | | | |
| Increased (% of HHs) | 60.3 | 70.1 | <0.001 | 67.0 |
| No change (% of HHs) | 30.0 | 20.4 | <0.001 | 23.4 |
| Decreased (% of HHs) | 9.7 | 9.5 | 0.394 | 9.6 |
| Mean ± SD (TK. in 2015) | 7404.2 ± 15200.8 | 7730.9 ± 8895.3 | 0.527 | 7629.1 ± 11244.1 |
| Mean ± SD (Tk. in 2018) | 7966.3 ± 9513.4 | 9815.2 ± 12145.1 | <0.001 | 9245.1 ± 11428.5 |
| P-value (2015 vs 2018) | 0.278 | <0.001 | | <0.001 |

*Note:* HHs = Households; SD = Standard Deviation.

## 5.3 Perceptions on micro-credits of borrowers

The perception of borrowers on micro-credits that means whether the micro-credits are blessing**s** or burden to the borrowers and if burden how overtake this is discussed below:

### 5.3.1 Major dimensional factors for not graduating from poverty

Table 5.4 presents the summary statistics of the perception on causes for not graduating from poverty among both case and control households. The principal cause was found as natural calamity among non-borrower households (75.3%) and pressure of the payment of loan among borrower households (83.9%). The second prime reason was loss of investment among non-borrower households (72%) and high interest rate among borrower households (73.1%).

**Table 5.4** Causes of not over-coming from vicious cycle of poverty of HHs s of *Haor* regions

| Causes of not overcoming from vicious cycle of poverty | Non-borrower HHs (%) (N = 733) | | | Borrower HHs (%) (N = 1607) | | | Both HHs (%) (N = 2340) | |
|---|---|---|---|---|---|---|---|---|
| | Disagree | Neutral | Agree | Disagree | Neutral | Agree | Disagree | Agree |
| Insufficient loan | 22.2 | 33.2 | 44.6 | 23.9 | 19.2 | 56.9 | 23.4 | 53.1 |
| Duration of loan is insufficient | 14.5 | 37.0 | 48.6 | 18.9 | 17.1 | 64.0 | 17.5 | 59.2 |
| High interest | 7.5 | 26.7 | 65.8 | 15.7 | 11.2 | 73.1 | 13.1 | 70.8 |
| Renewal of loan not get | 22.2 | 42.6 | 35.2 | 29.6 | 31.9 | 38.6 | 27.3 | 37.5 |
| Diversion of loan | 4.6 | 51.2 | 44.2 | 10.3 | 39.6 | 50.1 | 8.5 | 48.2 |
| Loss of investment | 5.5 | 22.5 | 72.0 | 8.6 | 27.2 | 64.2 | 7.6 | 66.7 |
| Natural calamity | 4.2 | 20.5 | 75.3 | 8.3 | 25.8 | 66.0 | 7.0 | 68.9 |
| Not getting loan in time | 7.6 | 24.8 | 67.5 | 19.7 | 18.4 | 62.0 | 15.9 | 63.7 |
| Pressure of loan payment | 8.7 | 26.3 | 64.9 | 6.2 | 9.9 | 83.9 | 7.0 | 77.9 |
| Number of dependent members are high | 11.9 | 21.1 | 67.0 | 12.9 | 22.0 | 65.1 | 12.6 | 65.7 |
| Income earning member is absent | 44.6 | 14.6 | 40.8 | 44.3 | 17.7 | 38.0 | 44.4 | 38.9 |
| Number of dependent incomes earning person is low | 14.7 | 24.6 | 60.7 | 14.6 | 24.6 | 60.7 | 14.7 | 60.7 |
| I have not taken loan | 24.3 | 14.6 | 61.1 | 84.6 | 11.0 | 4.4 | 65.7 | 22.1 |

*Note:* HHs = Households.

Figure 5.1 illustrates the scree plot of the extraction of key components for not graduating from poverty. The scree plot is a simple and graphical tool for correctly indicating the choice of the number components should be extracted under a theme.

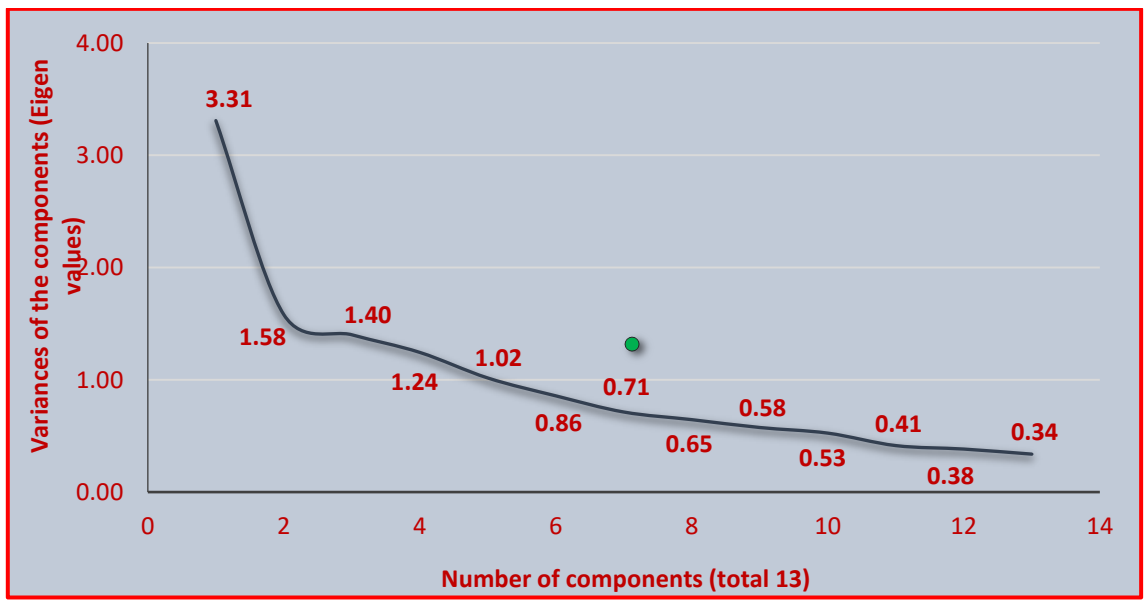

**Figure 5** Choice of the number of components among the causes of not over-coming from vicious cycle of poverty in Bangladesh

However, there should be five factors for not graduating from poverty. Mainly, first three factors account for more information than the fourth component however possess important and essential information. Therefore, the study assembles 13 unstructured reasons for not graduating from poverty to extract five dimensional factors using absolute magnitudes.

Table 5.5 shows the estimated loadings or contribution for assembling the source causes into five key factors using principal component analysis. In addition, the summary measure regarding the factor analysis is also presented for the validation of this procedure. The Bartlett's test of Sphericity is found highly significant and thus indicates that the determinant of the correlation matrix is significantly different from zero and thus the factor analysis is able to perform better for assembling key factors. The Kaiser-Meyer-Olkin measure of sampling adequacy is estimated at 0.711 which indicates that 71.1% of the variation of the unstructured sources of the cause of not graduating from poverty. In addition, the study found that about 65.77% of the total variation original sources or causes are explained by the extracted factors.

The extracted factors were created based on the loadings or the contribution of the original sets and the study mention the loading as influential if it is found greater than or equal to 0.3. The study used Varimax rotation for obtaining clear idea about the construction of factors. Using rotated factor loadings, first component explains 25.45% variation of the original set of causes and comprises of using i) Insufficient loan, ii) Insufficient duration of loan, iii) Not getting loan in time, iv) Absent of income earning member. Second component accounts for 12.16% variation of the original set and includes i) Higher number of dependent members, ii) Lower number of incomes earning members. In addition, third factor accounts for 10.78% variation and possess the reason of i) Loss of investment, ii) Natural calamity. Similarly, third factor explains almost 10% variation of the all reasons and includes i) Higher interest, ii) Pressure for loan payment, iii) Didn't receive loan. Finally, the last factor accounts for 7.82% variation and includes i) Didn't get the facility of loan renewal, ii) Diversion of loan.

Besides, the extracted factors are constructed as per the nature of the original set of reasons. Therefore, the first component can be named after 1) *imbalance between amount and time span of loan,* and the second factor should be termed as 2) *imbalance between dependent member and* *earning member*. Similarly, third factor can be defined as 3) *incidence of investment loss* and fourth factor can be termed as 4) *pressure created by the loan provider* and the last factor can be named after 5) *unable to utilize and diverse the loan*.

Table 5.5 Major dimensional factors of the causes of not over-coming from vicious cycle of poverty among the households of the *Haor* areas in Bangladesh

| SL# | Causes of not overcoming from vicious cycle of poverty | Rotated Factor Loadings (Varimax) | | | | | Comunalities |
|---|---|---|---|---|---|---|---|
| | | F1 | F2 | F3 | F4 | F5 | |
| 01. | Insufficient loan | 0.802 | | | | | 0.656 |
| 02. | Duration of loan is insufficient | 0.773 | | | | | 0.657 |
| 03. | High interest | | | | 0.527 | | 0.552 |
| 04. | Renewal of loan not get | | | | | 0.704 | 0.756 |
| 05. | Diversion of loan | | | | | 0.819 | 0.723 |
| 06. | Loss of investment | | | 0.828 | | | 0.711 |
| 07. | Natural calamity | | | 0.631 | | | 0.616 |
| 08. | Not getting loan in time | 0.505 | | | | | 0.489 |
| 09. | Pressure of loan payment | | | | 0.766 | | 0.700 |
| 10. | Number of dependent members are high | | 0.846 | | | | 0.748 |
| 11. | Income earning member is absent | 0.452 | | | | | 0.480 |
| 12. | Number of incomes earning person is low | | 0.871 | | | | 0.770 |
| 13. | I have not taken loan | | | | -0.716 | | 0.692 |
| Percentage of variation explained | | 25.45 | 12.16 | 10.78 | 9.57 | 7.82 | |
| Total variation explained by the extracted factors | | | | | | | 65.77 |
| Kaiser-Meyer-Olkin measure of sampling adequacy | | | | | | | 0.711 |
| Bartlett's test of sphericity | | Chi-square = 6601.464; df = 78 & P-value < 0.001 | | | | | |
| Extraction method | | Principal Component Analysis | | | | | |

*Note:* F1 to F5 indicates five extracted factors.

## 5.4 Impacts of micro-credits at the household level

Impacts of micro-credits on poverty of HHs are measured using Difference in Difference (DID) and before after comparison methods of income, expenditure and assets related indicators. These are as follows:

### 5.4.1 Comparison of income related indicators between 2016 and 2019

Appendix table 5.1 depicted that total income of borrowing HHs as well as non-borrowing HHs has been increased significantly over the period 2016-2019 and the income from agriculture, non-agriculture and labor sale have imbued this increase in total income in spite of insignificant

90 | P a g e

increase in income from business and donations. Finally, total income of borrowing HHs has been increased at higher rate than that of non-borrowing HHs over the period 2016-2019 significantly. This is permeated by the significant difference in increase of non-agricultural income and labor sale though the increase in agricultural income, business and donation are insignificant.

**5.4.2 Comparison of expenditure and investment between 2016 and 2019**

Appendix table 5.1 showed that total consumption expenditure with food and non-food expenditures of both borrowers and non–borrowers has been increased over the period 2016-2019 significantly. The increase in total consumption of borrowers and non-borrowers did not significantly differ though non-food expenditure differs significantly. Total investment expenditure of both borrowers and non –borrowers has been increased over the period 2016-2019 significantly. All the items of investment have been increased in case of borrowers except land purchase but in case of non- borrowers only land agriculture, family business, house repair and others investment are significant. Finally, the increase of total investment expenditure of borrowers has been significantly increased than that of non-borrowers and in item-wise all the investment items except education.

**5.4.3 Net impact of the micro-credits using difference-in-difference method**

The net impact of micro-credits is measured calculating Z-statistic which is found by deducting change in an economic indicator of borrowing HHs from non-borrowing HHs and multiplied by 100 as shown in appendix table 5.1. The table showed that total income of borrowers has been increased over non-borrowers (z=6.75) significantly. Among the components of income, non-agricultural income has been increased significantly on the other hand income from labor sale has been decreased significantly. The incomes from agriculture, business and donations have been increased but insignificantly. There are two heads of consumption expenditure viz. food and non- food and the non-food expenditure has been increased significantly but food expenditure has been decreased insignificantly while total food consumption has been increased (z=1.06) significantly of borrowers over non-borrowers. The investment expenditure of borrowers has been increased over the non-borrowers over the period 2016-2019 significantly. Among the heads of investment expenditure healthcare, poultry livestock and productive assets have been increased significantly over the period 2016-2019 of borrowers over non-borrowers while investment on education, agriculture, family business, house repair, land purchase and others investment have been increased insignificantly. Finally total expenditure, total investment and savings of borrowers have been increased significantly over the non-borrowers in 2019 based on 2016.

## 5.5 A comparative impact of formal and informal micro-credits

A comparative impact of formal and informal micro-credits is done in terms income, expenditure and investment as follows

### 5.5.1 Comparison of income related indicators in-between 2016 and 2019

Appendix table 5.2 showed that total income and its heads of income from agriculture, non-agriculture and labor sale have been increased significantly of formal borrowers over the study period while in case of informal sources total income and its heads of income from agriculture and labor sale have been increased significantly. The heads of income from business and donations in case of formal credits and incase of informal credits the heads of income from non-agriculture, business and donations have been increased but not increased significantly.

### 5.5.2 Comparison of expenditure and investment on different heads between 2016 and 2019

Appendix table 5.2 showed that total consumption expenditure with its heads of food and non-food consumption of both formal borrowers and informal borrowers have been increased over the period 2016-2019 significantly. In case of formal borrowers' total investment expenditure and its entire heads except land purchase and house repairing have been increased significantly over the period 2016-2019.

The increase in total consumption of borrowers and non-borrowers did not significantly differ though non-food expenditure differs significantly. Total investment expenditure of both borrowers and non –borrowers has been increased over the period 2016-2019 significantly. All the items of investment have been increased in case of borrowers except land purchase but in case of non- borrowers only land agriculture, family business, house repair and others investment are significant. Finally, the increase of total investment expenditure of borrowers has been significantly increased than that of non-borrowers and in item-wise all the investment items except education.

### 5.5.3 Net impact of the micro-credits using difference-in-difference method

The net impact on income and its heads showed that total income and its entire heads do not differ significantly among the formal and informal sources except donation income which has been decreased(Z=-20.6) of formal sources over informal sources significantly(p=.001). The data on total consumption showed that incase of formal borrowers the total consumption has two heads and total consumption and food consumption did not change significantly over informal borrowers (z=-13.68) but non-food expenditure has been decreased of formal borrowers than informal borrowers significantly ((p=<0.01). Total investment has been increased (z=18.53) of formal borrowers than that of informal significantly (p=<0.01). Among the items of investment, education, agriculture, poultry –livestock, productive assets and durable goods have been increased significantly of formal borrowers over informal borrowers won the other hand investment in health care, family business, house repair, land purchase and others investment have been decreased significantly. The total expenditure and savings have been increased of formal borrowers over informal over the period 2016-2019 significantly.

### 5.6 Conclusion

Consequences of micro-credits on borrowers has been analyzed in this chapter while the impact of micro-credits on poverty condition and efficiency of the programs will be analyzed in the next chapter.

# CHAPTER SIX
# IMPACT OF MICRO-CREDITS ON POVERTY CONDITION AND EFFICIENCY OF THE PROGRAM

## 6.1 Introduction

In this chapter we have analyzed the efficiency of micro-credit in poverty alleviation and which program is more efficient i.e., the formal or informal in poverty alleviation. With this aim, the discriminant analysis between formal and informal group is done.

## 6.2 Group means and standard deviation of the predictors of the discriminant analysis

Table 6.1 shows the measure of dispersion by groups to identify the variations among the predictors between formal and informal credits receiving status. In terms of borrowers' age at inclusion to micro-credits, credits receivers from informal source are older than the credits receivers from informal source.

**Table 6.1** Group means and standard deviation for group membership of micro-credits receivers

| List of predictors | Group means | | Group standard deviation | |
|---|---|---|---|---|
| | Formal | Informal | Formal | Informal |
| Gender of loan receiver | 0.64 | 0.87 | 0.48 | 0.34 |
| Age of loan receiver | 41.80 | 44.19 | 11.28 | 11.45 |
| Occupation of loan receiver | 2.73 | 2.38 | 1.20 | 1.21 |
| Earning status of loan receiver | 0.71 | 0.87 | 0.45 | 0.34 |
| Number of earning members | 1.59 | 1.81 | 1.00 | 1.05 |
| Dependency ratio | 0.75 | 0.81 | 0.58 | 0.73 |
| Regular income source | 1.48 | 1.65 | 0.50 | 0.48 |
| Asset index | 0.02 | -0.13 | 0.98 | 0.97 |
| Type of used toilet | 0.44 | 0.35 | 0.50 | 0.48 |
| Total amount of land | 67.29 | 80.75 | 112.39 | 112.20 |
| Total annual income | 105281.49 | 96233.18 | 58972.96 | 59387.91 |
| Amount of received loan | 34596.29 | 45474.61 | 40825.63 | 76125.59 |
| Installment type | 1.96 | 3.96 | 1.07 | 1.12 |
| Duration of loan | 11.86 | 21.25 | 3.51 | 22.80 |
| Interest rate | 17.77 | 26.00 | 7.55 | 19.12 |

The average number of earning members is slightly higher in informal credits receiver households than that of formal credits receivers. It appears from the group standard deviation that formal and informal credits receiving groups are more widely spreader in case of gender, earning status, dependency ratio, amount of received loan, duration of loan and interest rate. The variations between the groups were observed more in case of loan related factors. However, the average dependency ratio was estimated lower in formal credits receiving households while standard deviation was estimated higher than that of informal source of credits. The study also identifies that the mean difference between formal and informal credits groups in case of loan related characteristics is smaller but the standard deviation of these factors is relatively large. This finding initially indicates that loan related characteristics will contribute more to discriminate formal and informal source of loans.

The pooled within group correlation matrix are calculated to identify the linear relationship among predictors as the initial or descriptive measure of multicollinearity. The correlation the gender and occupation, gender and earning status, and occupation and earning status have moderate correlation among the pair of all predators (Table 6.2).

**Table 6.2** Pooled within-groups correlation matrix of the predictors

| List of covariates | 02. | 03. | 04. | 05. | 06. | 07. | 08. | 09. | 10. | 11. | 12. | 13. | 14. | 15. |
|---|---|---|---|---|---|---|---|---|---|---|---|---|---|---|
| 01. Gender | 0.18 | -0.70 | 0.68 | 0.01 | 0.02 | -0.05 | 0.06 | 0.01 | 0.07 | 0.02 | 0.04 | 0.11 | 0.04 | -0.05 |
| 02. Age | | -0.12 | 0.05 | 0.22 | -0.24 | -0.05 | 0.24 | 0.02 | 0.18 | 0.16 | 0.07 | 0.03 | 0.03 | -0.03 |
| 03. Occupation | | | -0.62 | -0.08 | -0.01 | 0.06 | -0.09 | -0.01 | -0.21 | 0.02 | -0.02 | -0.09 | -0.08 | 0.03 |
| 04. Earning status | | | | 0.25 | 0.03 | -0.10 | 0.06 | -0.06 | 0.06 | -0.01 | 0.02 | 0.09 | 0.05 | -0.03 |
| 05. No. of earner | | | | | -0.18 | -0.07 | 0.22 | -0.01 | 0.14 | 0.20 | 0.14 | 0.10 | 0.25 | 0.03 |
| 06. Dependency ratio | | | | | | 0.10 | -0.22 | -0.06 | 0.00 | -0.09 | 0.00 | 0.05 | 0.09 | 0.04 |
| 07. Regular income source | | | | | | | -0.15 | -0.15 | 0.07 | -0.08 | -0.06 | 0.11 | 0.11 | -0.05 |
| 08. Asset index | | | | | | | | 0.05 | 0.25 | 0.29 | 0.06 | -0.10 | -0.10 | -0.03 |
| 09. Toilet type | | | | | | | | | -0.03 | 0.09 | 0.07 | 0.08 | 0.10 | -0.04 |
| 10. Landholding | | | | | | | | | | 0.21 | 0.09 | 0.08 | 0.06 | -0.07 |
| 11. HH income | | | | | | | | | | | 0.23 | -0.04 | 0.10 | -0.07 |
| 12. Loan | | | | | | | | | | | | 0.17 | 0.30 | 0.17 |
| 13. Installment type | | | | | | | | | | | | | 0.23 | 0.21 |
| 14. Duration of loan | | | | | | | | | | | | | | 0.16 |
| 15. Interest rate | | | | | | | | | | | | | | |

However, the correlation matrix indicate very weak correlations among the predictors for rest set of pair of predictors and the coefficient of correlation is estimated below 0.3 for all set of pairs. Marginally, the multicollinearity is unlikely to be a problem for estimation as well as making conclusion for discriminating the formal and informal credits receiving likelihoods.

### 6.3 Significance of covariates for discriminating formal and informal credits source

In discriminant analysis, Wilk's lambda provides a test for assessing the null hypothesis that the vectors of means of the measurements (predictors) are the same in the two groups (formal and informal). The lambda coefficient is defined as the proportion of the total variance in the discriminant scores not explained by differences among the groups, here 55.6% (table 6.3).

**Table 6.3** Test of significance of function and equality of group means for each individual predictor

| Name of the predictor variables | Wilks' Lambda [a] | F-statistic | P-value |
|---|---|---|---|
| Test of function [b] | 0.556 | Chi-sq = 936.525 | <0.01 |
| Gender of loan receiver | 0.952 | 81.302 | <0.001 |
| Age of loan receiver | 0.991 | 14.409 | <0.001 |
| Occupation of loan receiver | 0.984 | 26.473 | <0.001 |
| Earning status of loan receiver | 0.973 | 43.798 | <0.001 |
| Number of earning members | 0.99 | 15.424 | <0.001 |
| Dependency ratio | 0.998 | 2.895 | 0.089 |
| Regular income source | 0.975 | 41.374 | <0.001 |
| Asset index | 0.995 | 8.392 | 0.004 |
| Type of used toilet | 0.994 | 10.094 | 0.002 |
| Total amount of land | 0.997 | 4.648 | 0.031 |
| Total annual income | 0.995 | 7.587 | 0.006 |
| Amount of received loan | 0.992 | 13.582 | <0.001 |
| Installment type | 0.592 | 1105.93 | <0.001 |
| Duration of loan | 0.897 | 185.231 | <0.001 |
| Interest rate | 0.913 | 153.155 | <0.001 |
| Model summary | | | |
| Eigen value | | 0.78 | |
| Canonical correlation | | 0.666 | |
| $R^2$ from canonical correlation | | 0.444 | |

[a] Wilks' $\lambda$ (U-statistic) and univariate F-ratio with 1 and 1605 degrees of freedom
[b] Wilks' $\lambda$ (U-statistic) and chi-square statistic tests the overall significance of function with 15 degrees of freedom

The formal test confirms that the sets of fifteen mean measurements differ significantly between the two sites (Chi-sq (15) = 936.525, p<0.01). The significance of the uni-variate F-ratios indicates that when the predictors are considered individually, is there any significantly difference between those who had been credited from formal source and those who had been credited from informal source. The study found that the mean dependency ratio significantly differentiates the formal credits and informal credits at 10% level of significance. Besides, average value of landholding indicate significantly differentiates the formal and informal credits at 5% level of significance. However, rests of the covariates also significantly differentiate the formal and informal credits receiving status at 1% or less than 1% level of significance. These tests provide validity to proceed in the next level of discriminant analysis. Because there are two groups, only one discriminant function can be estimated using all predictors. The eigenvalue associated with this function is estimated at .786, which accounts for 100 percent of the explained variance. The canonical correlation associated with this function is estimated at 0.666. The square of the canonical correlation in discriminant analysis demonstrates the interesting finding focusing of model fitting ability. However, the value of 0.444 of the $R^2$ from the estimated canonical correlation coefficient indicates that 44.4 percent of the variance in the dependent variable (credits type) is explained or accounted for by this model.

**6.4 Estimation and validation of two-group discriminant analysis**

The estimate of the coefficients for a certain variable depends on other variables included in the analysis. The sign of the coefficients indicates the effect of resulting variable on the formal and informal credits receiving status. The unstandardized coefficients measure absolute magnitude of the impact of each predictor on the discriminating power of the dependent variable. The larger estimates of the unstandardized coefficients identified the more powerful or influential predictor for discriminating the formal and informal credits types. The standardized coefficient is estimated because some of the predictor variables were in different scale or they are observed in different units. Besides the standardized coefficients are more prompt possessing the caveat of multicollinearity problem through implying the relative importance of each predictor for discriminating formal and informal credits receiving status. In addition, relatively large standardized coefficient demonstrates more discriminating power of the function as compared to the smaller standardized coefficients. These imply that the variable with relatively large estimates of the standardized coefficients is more important than the other coefficients. The coefficients of the model indicates that higher installment type (annual), more duration of loan, female loan receivers, earner loan receivers, households having regular income source, older loan receivers, households having more assets are more likely to receive the formal credits than that of informal credits (Table 6.4). However, smaller number of earning members in the borrower households, unhygienic toilet type, lower dependency ratio is more likely to receive the loan from formal source than that of informal source.

In addition, the canonical loadings or discriminant loadings are estimated to identify the magnitudes of the discriminating power of the predictors. These loadings measure the correlation between discriminating variables (study predictors) and the canonical discriminating functions which are presented at the structure matrix column in table 6.4.

Table 6.4 Estimates of two-group discriminant analysis for formal and informal credits

| Discriminating variables [a] | Unstandardized coefficients [b] | Standardized coefficients [c] | Structure Matrix [d] |
|---|---|---|---|
| Installment type | 0.802 | 0.867 | 0.930 |
| Duration of loan | 0.018 | 0.222 | 0.380 |
| Interest rate | 0.013 | 0.159 | 0.346 |
| Gender of loan receiver | 0.511 | 0.227 | 0.252 |
| Earning status of loan receiver | 0.015 | 0.006 | 0.185 |
| Regular income source | 0.109 | 0.054 | 0.180 |
| Occupation of loan receiver | 0.087 | 0.105 | -0.144 |
| Number of earning members | -0.028 | -0.028 | 0.110 |
| Age of loan receiver | 0.005 | 0.061 | 0.106 |
| Amount of received loan | 0.001 | -0.124 | 0.103 |
| Type of used toilet | -0.327 | -0.161 | -0.089 |
| Asset index | 0.051 | 0.050 | -0.081 |
| Total annual income | 0.001 | -0.036 | -0.077 |
| Total amount of land | 0.001 | -0.015 | 0.060 |
| Dependency ratio | -0.039 | -0.024 | 0.048 |
| Constant | -3.167 | - | - |

[a] Discriminating variable (predictors) are ordered by size of correlation within function to identify most contributor
[b] Coefficients are obtained from the original (unstandardized) set of predictors using canonical discriminant function
[c] Coefficients are obtained from the standardized set of predictors using canonical discriminant function
[d] Pooled within-groups correlations between discriminating variables and canonical discriminant functions, therefore called as canonical loadings or discriminant loadings

In short, the discriminating loadings identified the most important predictor among predictors for discriminating the borrowing source of micro-credits. Similarly, these estimates rank or order the predictors based on their discriminating power of the dependent variable. More precisely, installment types are the most important predictors in discriminating between formal and informal credits groups, followed by duration of loan, followed by interest rate, followed by gender of the loan receivers, followed by earning status of loan receivers, followed by regular income source in the households and so on. On the contrary, dependency ratio, household possession of land, households' total income is the least contributory predictor, however significant, for discriminating the borrowing source of micro-credits.

In discriminant analysis, the validation of the model is grabbed using cross-validation approaches. Generally, the discriminant analysis model splits the whole data sets into two equal parts in which first one is termed as estimation sample or analysis sample and rest one is termed as holdout or validation sample.

Table 6.5 Validation of the two-group discriminant analysis model

| | | | Prior probabilities based on groups (informal = 0.279, formal = 0.721) | | |
|---|---|---|---|---|---|
| No cross-validation [a] | | Type of credits | Predicted | | Total |
| | | | Informal | Formal | |
| | Observed | Informal | 240 | 209 | 449 |
| | | Formal | 24 | 1134 | 1158 |
| Leave one out cross-validation [b] | | Type of credits | Predicted | | Total |
| | | | Informal | Formal | |
| | Observed | Informal | 239 | 210 | 449 |
| | | Formal | 25 | 1133 | 1158 |

[a] Hit ratio of 0.855 measure that 85.5% of original grouped cases correctly classified
[b] Hit ratio of 0.854 measure that 85.4% of cross-validated (leave one out) grouped cases correctly classified

On the other hand, leave one out cross-validation approach is also examined to estimate the accuracy of correct classification rate by splitting the samples equivalent to its size (n-term). Each time one sample is reserved for validation sample while the rest sample are used as the estimation sample and this process is repeated n-times holding each sample as validation sample. It is to be noted that the prior distribution is assumed equivalent to the sample estimate for each group of the dependent variable rather than 0.5 for each group. Finally, the validation of the discriminant analysis is documented using the percentage of correct classification which is termed as hit ratio in discriminant analysis. However, there is no substantial improvement in the leave one out cross-validation approach as compared to the no cross-validation which is presented in table 6.5. In addition, no cross-validation approach estimates that almost 86% of the original grouped cases such as formal and informal source of credits are correctly classified by the discriminant analysis model.

**6.5 Estimation of Fisher's linear discriminant function**

Final, this study estimates the parameters of the Fisher's linear discriminant function for discriminating the formal and informal source of credits by the study variables. This study mainly examined three types of predictors for discriminating formal and informal source. These are: individual level factors or the characteristics of the borrowers (loan receivers), their household characteristics, and finally micro-credits characteristics. The estimated Fisher's linear discrimination function is as follows:

$$D = -8.12 + 1.02 \times \text{Sex} + 0.01 \times \text{Age} + 0.17 \times \text{Occupation} + 0.03 \times \text{Earning status} - 0.06 \times \text{Number of earner} - 0.08 \times \text{Dependency ratio} + 0.22 \times \text{Household's regular income source} + 0.10 \times \text{Asset index} - 0.65 \times \text{Toilet type} + 0.01 \times \text{Landholding} + 0.01 \times \text{Annual income} + 0.01 \times \text{Amount of loan} + 1.59 \times \text{Installment type} + 0.04 \times \text{Duration of loan} + 0.03 \times \text{Interest rate}$$

This function is used to predict or classify a new individual with the respective characteristics who are supposed to borrow from formal source or informal source. The average of the group means commonly known as group centroids and the score or value obtain from Fisher's linear discriminant function are used and compared to allocate new members into a group. The average of the functions at group centroid is estimated at 0.44 (Table 6.6). Therefore, it is obvious that new members with discriminant scores above 0.44 supposed to borrow micro-credits form the informal sources otherwise he/she would be preferred to borrow from formal source.

**Table 6.6** Estimates of Fisher's linear discriminant function by two-group discriminant analysis

| Predictors | Informal | Formal | Fisher's discriminant function coefficients |
|---|---|---|---|
| (Constant) | -41.02 | -32.90 | -8.12 |
| Gender of loan receiver | 7.95 | 6.94 | 1.02 |
| Age of loan receiver | 0.38 | 0.37 | 0.01 |
| Occupation of loan receiver | 6.33 | 6.16 | 0.17 |
| Earning status of loan receiver | 10.00 | 9.97 | 0.03 |
| Number of earning members | 0.41 | 0.46 | -0.06 |
| Dependency ratio | 2.97 | 3.04 | -0.08 |
| Regular income source | 7.31 | 7.10 | 0.22 |
| Asset index | -0.35 | -0.45 | 0.10 |
| Type of used toilet | 3.05 | 3.70 | -0.65 |
| Total amount of land | 0.01 | 0.01 | <0.01 |
| Total annual income | <0.01 | <0.01 | <0.01 |
| Amount of received loan | <0.01 | <0.01 | <0.01 |
| Installment type | 2.18 | 0.58 | 1.59 |
| Duration of loan | 0.01 | -0.03 | 0.04 |
| Interest rate | 0.19 | 0.17 | 0.03 |
| General classification rule | | | |
| Functions at group centroids | 1.43 | -0.56 | 0.44 |

## 6.6 Conclusion

The impact of micro-credits on poverty condition and efficiency of the programs has been analyzed in this chapter and the next chapter has attempted to present the findings from the qualitative surveys are presented in the seventh chapter.

# CHAPTER SEVEN
# FINDINGS FROM THE QUALITATIVE SURVEYS

**7.1 Introduction**

As per the plan of the study, 30 Key Informant Interviews (KIIs) and 30 In-depth Interviews (IDIs) on the research topic "Vicious Cycle of Poverty in *Haor* Region of Bangladesh: Impact of Formal and Informal Credit" were conducted in different locations of the *Haor* region of Bangladesh. The aim of these qualitative surveys was to get in-depth understanding of the credit-trap and its impact on poverty situation of the concerned households.

The key informants were the community leaders (preferably Chairman of Union Parishad), delegates from civil society, and other stakeholders including the NGO delegates working in *Haor* areas. On the other hand, the participants of in-depth interviews were the selected victims of poverty cycle, local money-lenders and delegates of existing MFIs working in the *Haor* region. The principal investigator and co-investigator have mainly conducted the key informant interviews and in-depth interviews. Besides, some of the in-depth interviews were done by research assistants. All the interviews were recorded through personal digital assistant (PDA) at the time of survey. The responses have been transcribed to extract conclusive comments on the issues. Following strategies have been followed in conducting the KIIs/IDIs:

At first, the consent of the key informants and participants of in-depth interviews were sought by briefing the objectives of research project and aims of qualitative surveys. The interviews/discussions were conducted if consent were found duly. If consent was not found in any case, the interview was postponed for the particular case (key informants/participants). The key informant interviews and in-depth interviews were performed covering the following issues:

The Issues are as follows
- (1) Availability of the micro-credits in the community;
- (2) Interest rate of micro-credits in the community;
- (3) Purposes receiving micro-credits;
- (4) Causes of not getting formal and informal micro-credits;

(5)  Heads of utilization heads of micro-credits;

(6) Attitudes of borrowers on formal and informal micro-credits programs;

(7) Causes of ineffectiveness of micro-credits for poverty reduction perspective;

(8) Causes of remaining poor after using micro-credits; and

(9) Suggestions for maximization of   benefits from micro-credits

## 7.2 Findings

The key findings of the quantitative household survey have been shared with the key informants to make the discussion easier as well as to know their views on the quantitative results. Most of the key informants agreed that the findings are very much consistent with prevailing condition of micro-credits in *Haor* region. However, some of them raised question about the impacts of micro-credits. They argued that there is no straightforward positive impact of micro-credits on poverty condition of the households. Some of them have mentioned that the increase of income and food expenditure is not the good indicator of measuring poverty alleviation. They have given emphasis on qualitative views of the victims/respondents rather than quantitative data

**Issue#01: Availability of the micro-credits in the community**

The key informants/participants informed that micro-credit is available in each and every corner of the *Haor* region. However, the sources of micro-credits, in terms of formal and informal, are vary in different locations. In the early days, only informal sources (like *Mahajan's* loan) of micro-credits were available in *Haor* region. However, the pattern has shifted as several government and non-government organizations have come forward with micro-credit facility in the *Haor* region. Now-a-days formal sources of micro-credits are dominating over informal sources. Most of the formal sources are mainly for SMEs purpose. A loan to meet the family consumption expenditure is rare from formal sources.  The *Haor* people still depends on informal sources of micro-credits for covering the emergency situation like treatment of household members, covering the cost of food items during severe food shortages. The key informants/participants viewed that the exiting micro-credit system of formal sources is not user friendly to the credit receivers. The necessary documents along with the provision of collateral and the loan re-payment system are regarded as the main hindrances for user friendly micro-credit system from formal sources.

**Issue#02: Interest rate of micro-credits in the community**

The key informants/participants informed that the interest rate of micro-credits vary remarkably between formal and informal sources. The interest rate of micro-credits is overwhelmingly higher for informal sources like *Mahajan's* credit. Among other terms and conditions, the duration is very limited for informal sources of micro-credits. The key informants and participants of IDIs informed that the interest rate of micro-credits from informal sources is sometimes rise to 60% in some season, particularly before harvesting the crops. On the other hand, the interest rate of micro-credits is remains constant throughout the year for institutional sources like banks, NGOs etc.

**Issue#03: Purposes of receiving micro-credits**

Regarding reasons of receiving micro-credits, the key informants and participants of IDIs informed that covering the food expenditure, repairmen of living house, investment in agriculture farming, investment in purchasing livestock, and expenditure for treatment of household members are the main reasons for receiving micro-credits. Some informants also mentioned purchasing motorcycle and expenditure of education of children are the reasons for receiving micro-credits. But repayment of earlier loan is one of the prime reasons of micro-credits. As a result, most of the credit receivers could not come out from credit cycle. Ultimately, they failed to come out from poverty status and remain in the vicious cycle of poverty.

**Issue#04: Causes of not getting formal and informal micro-credits**

The participants said that they do not know about some micro-credits programs, nepotism, misappropriation of credits, lack of entry fee/bribe and bureaucratic problem are the main causes of not getting micro-credits from formal sources. The discussant told that high rate of interest, social status, non-availability collateral, relationship, lobbing, non-cooperation of local members budget limitation are the main causes of not getting micro-credits from in-formal sources.

**Issue#05: Heads of utilization of micro-credits**

The key informants and participants have informed that the micro-credits are mostly utilized to meet-up the food expenditure, repairing the living house, investment in agriculture for harvesting crop along with loan repayment. Due to the threat of losing the collateral, most of the credit-receivers used to get another loan to pay the earlier credits. A few of the credit-receivers can come out from this sort of credit-trap unless asset depletion.

**Issue#06: Attitudes of borrowers on formal and informal micro-credits programs**

The attitudes towards formal and informal micro- credits are explored by the discussants that the rate of interest is very high, credits period is insufficient and terms are rigid.

**Issue#07: Causes of ineffectiveness of micro-credits for poverty reduction perspective**

Since most of the micro-credits are utilized in non-productive heads and consequently the credit-receivers cannot pay the installment in time. When credit-receivers fails to pay the installment in time, the credit amount becomes bigger and bigger by adding the interest with the principal amount as time goes. Ultimately, most of the credit-receivers cannot come out from credit cycle unless sale of any valuable property. The economic condition of the concerned households becomes poorer due to the asset depletion. The key informants and participants have informed that lack of household-level policy is the main reason for this sort of asset depletion. In addition, lack of skills on SMEs to carry out a productive economic activity is also vital reason for ineffectiveness of micro-credits. Thus, the lack of policy, lack of skills, and investment in non-productive heads appeared as the hindrances for effectiveness of the micro-credits on the perspective of poverty reduction.

**Issue#08: Causes of remaining poor after using micro-credits**

The participants explored the causes of not come out from vicious cycle of poverty in spite of using loans due to pressure of repayment of loans, loss of investment, not renewal of loans in need, short duration of loans diversion of loans from main purposes, absent of income earning member in the family.

**Issue#09: Suggestions for maximization of benefits from micro-credits**

The analysis of the threadbare discussions of the key informants and participants, following suggestions could be made for getting maximum benefits from micro-credits:

- ➢ In general, the micro-credits from informal sources should be stopped. Instead, a new social safety nets program can be introduced for the *Haor* people to help them during food shortages.
- ➢ The most of the participants suggested that for breaking vicious cycle of poverty by micro-credits the duration of loans should be at least five year and the volume of loans must be minimum 500,000 and repayment should at not be less than monthly. The rate of interest should not be more than 5%.
- ➢ The micro-credits from formal sources should be encouraged for the households whose members acquired capability of start any economic activity through skill-development program.
- ➢ The government should undertake a plan of action to review the existing micro-credit programs of *Haor* region to make the programs user friendly in terms of required documents, interest rate and duration. The zero-interest micro-credit system can be introduced for vulnerable households.
- ➢ Skill-development training programs should be strengthening covering different investment areas for the adult members of credit-receiver households.
- ➢ The key informants also argued for more research to find out different pathways for eradication of poverty of *Haor* people in order to achieve the targets of SDGs.

# CHAPTER EIGHT
# SUMMARY OF FINDINGS AND POLICY IMPLICATIONS

**Summary of findings**: After analyzing the data the following findings are found:

- Socio-demographic characteristics of the respondents and study population;
- Housing conditions and facilities of housing;
- Landholdings of households;
- Possession of assets by households;
- Reasons of not getting micro-credit by eligible non-borrowers;
- Sources and types of micro-credits;
- Purpose of micro-credits;
- Volume of both formal and informal micro-credits;
- Terms and conditions of micro-credits;
- Payment status of loans;
- Expenditure and investment pattern of micro-credits;
- Income of both borrowers and non-borrower households;
- Expenditure of both borrowers and non-borrower households;
- Causes of non-payment of loans in time;
- Attitude of borrowers towards micro-credits;
- Role of micro-credits on graduation of selected socio-economic status;
- Major dimensional factors of not graduating from poverty from micro-credit;
- Impact of micro-credit at the household level;
- Comparative impact between formal and informal micro-credit;
- Efficiency of the programs;
- Findings from the qualitative surveys.

**Socio-economic profile of the respondents and study population:** The real data of the depicted out of total 2340 respondents 69% were borrowers (treatment group) and 31% were non-borrowers (control group). In terms of relationship of respondents with household heads lion portion (81%) respondents were the household heads. In respect of gender 76.8% was male and 23.2 was female and in respect of age 62% were between 31 and 50. Among the respondents 92.2% was married and 5% was widowed and in regard of literacy level more than half of the respondent's attended primary. In respect occupation of respondents 20% engaged in agriculture, 23% in labor selling, 13% in off-farm activities, 19% in business/service and 18.9% in household work. In respect of income earning nearly half are found as full-time income earner and about 29% were part-time and 21% of the respondents had no earning source. The distribution of household population in respect of sex 51.17% was male and 48.83% female and in respect of age 35.2% male and 37.9% female are up to 15 years, 26.7% male and 28.3% female between 16-30 years and 25.5% male 25.3% female are between 31-50 years 6.8% male, 4.9% female are in between 51-60 age and 5.7% male and 3.7% female are above 60. The marital status of the population 16 and above showed that the lion part of population is married (66.4% for male and 73% for female) and female population is being getting early married than their male counterpart. Near half of the study population are found to attend primary level education and only about 2% completed graduate and the illiteracy rate was higher of women than men. In respect of occupation male aged between 16 years and 60 years showed that 8.9% farming, 13.5%day laborer, 8.4% off-firm activities 13.7% service/business 11% student, 34.6% household work and 9.9% others while 71% women in household work including vegetable cultivation, poultry rearing, weaving mattress, stitch *khatha*, fire-wood collection, feeding cattle, helping in agricultural work *etc*. The distribution of earning of household members aged 16 and above showed that 28% were full-time earner and about 22% were part-time earner and about one-quarter male and three-quarters female are found to no work at the survey point.

**Housing conditions and facilities of housing:** The ownership pattern of housing showed that about 95% of the households owns house for living and 20% of the households uses one room, 42% of uses 2 rooms and 10% four and more rooms for living. Most of the houses (86%) are made of made of tin shed roof with different kinds of fence and only 4.3% is made of straw roof with bamboo/muddy fence. The quality of houses was found inferior for borrower households in comparison to non-borrower households. About 39% of the households does not own separate kitchen room for cooking while the main sources of cooking fuel were straw/leaf/husk (43.6%) and wood/kerosene (30.1%) and 74% of the households used these. In respect of utilities facilities 87% of the area has electricity connection while 81% of the surveyed households have access to this electricity and 19% uses lamp of kerosene oil. The data on sanitation facilities revealed that 91.6% owns toilet, 15% used hygienic latrine with water-preventing capacity 28% used pit latrine without water preventing capacity and over half of the households used completely non-hygienic toilet that include *katcha* and open field/space.

**Landholdings of households:** The study showed that 8.5% of the surveyed households had no homestead land, 79.2% owned homestead land between 1 to 15 decimals and only 12.3% owned homestead land more than 15 decimals. Among the surveyed households, over two-thirds have no agricultural land and 7% owned agricultural land between 1 to 15 decimals, 8% owned agricultural land between 16 to 50 decimals, and only 17% over 50 decimals. Among the land-owning households, the average cultivable agricultural land is 105.85 decimals for non-borrower households and 96.96 decimals for borrower households. It is found that 27% of the households had some sort of leased land and the average size of this kind of land is 102.66 decimals. About 94% of study households possessed pond and average size of the pond land is 8.65 decimals.

**Possession of assets by households**: The study revealed that 14.8% household owned boats 1.2% owned deep tube well, 20.6% owned tube well and 8.1% owned bicycle, 16.2% owned television, 6.2% owned drawing room furniture, 28.5% owned watch 69.7% owned electric fan and 90% owned mobile phone. The possession of productive assets revealed that 664 borrowers HHs and 311 of non-borrowers HHs possessed cultivating instruments on average 3.27 and 3.05 respectively while livestock possessed by 694 HHs of borrowers and 287 HHs of non -borrowers The average possession of cultivating instruments and livestock is 3.20 and 2.49 respectively.

**Reasons of not getting micro-credit by eligible non-borrowers**: Fifteen causes were given to the eligible but not receivers of micro-credits and through PCA with    Varimax rotation technique we reduced major four dimensional factors and factor one related to credit misappropriation and biasness, factor two related to non-cooperation of local authority, factor three related to bureaucracy and budget limitation and forth factor is related to ignorance and corruption.

**Sources and types of micro-credits**: The study revealed that in terms of loan receivers 1595 (99.3%) HHs borrowed cash loan and only 12 (0.7%) in kinds and in respect of sources 1158

HHs (72.1%) borrowed from formal sources and 449 HHs (27.1%) from informal sources. There are two types of formal sources viz. (i) government (Banks/Co-operatives; and (ii) nongovernment (MFI/NGO/Insurance) and 88 HHs (5.5%) borrowed from the former and the 1070 HHs (66.6%) from the later. There are three informal sources viz., (i) local money lender (mohajan/private samittee); (ii) non-interest loan (relatives/ friends/ neighbors); and (iii) more than one sources and the data revealed that 393 HHs (24.5%) taken loans from the first, 53 HHs (3.3%) from the second and only 3 HHs (0.2%) from the third. Again, the informal sources have been divided into interest bearing and non-interest bearing and 408 HHs (90.9%) have taken from the former and HHs 41(9.1%) from the later.

**Purpose of micro-credits:** To identify the purpose of borrowing seventeen purposes were given to borrowers through pilot survey and using PCA the seventeen purposes reduced to five factors as factor- one related to daily life and livelihoods, factor-two related to adaptability of natural shocks and farming, factor three related to cropping and raring cattle the elements, factor four related to business and marriage and factor five related to sending family member to abroad.

**Volume of both formal and informal micro-credits:** The data revealed that minimum amount of all types of loans is Tk. 2000, maximum amount of formal loan, informal loan, in total are Tk. 550,000 and Tk. 11, 00,000 and 11, 00,000 respectively and mean is Tk. 34596 for formal and 45475 for informal and 37636 in total. Among the informal loans maximum amount of interest bearing and non-interest bearing are Tk. 1100000 and Tk. 150000 respectively. The mean of interest bearing and non-interest bearing are loans are Tk. 47696 and Tk. 233656 respectively.

**Terms and Conditions of micro-credits:** In respect of interest rate for formal credits it varies between 1% and more than 25% and 50% of HHs (525 out of 1158) borrowed at 11% to 15% and 211HHsat 16% to 20% while in case of in formal loans more than 41% HHs (185) get at more than at 25%, and 25% HHs (106) borrowed at 1% 10%. Regarding the type of installment in case of formal credit more than 53% HHs (619) borrowed weekly basis while about 50% of informal borrowers on monthly basis. In respect of total installment more than 52% more than 24 installments while 50% informal borrowers on 2 to 12 installments. Regarding the duration of loans about 95% (1096) formal borrowers and 71 % (317) informal borrowed for one year and most of the credits of both formal (94%) and informal (94%) are non-collateral.

**Payment status of loans:** The payment status of loans showed that 97% (1113) formal and 55% (245) of informal borrowers paid totally in time in which 1108 paid principal and 1092 paid interest under formal credit and in case of informal 144 paid principal and 233 paid interests. In overall 1358 (86%) borrowers paid totally in time in which 1252 principal and 1345 interest.

**Expenditure and investment pattern of micro-credits:** The respondents were given a list of 24 expenditures and investments items of their loans and among those 14 items were marked by them. The items marked by 10% or above borrowers in case of formal credits are, family enterprise (15.76%) followed by agricultural loan (15.09%), food consumption (15.5%), others (13%) and payment of loan (12%). The investment and expenditure items of formal borrowers 10% and above are food consumption (23.24%), others (16.86%), payment of loan (12.97%) and health care (11.68%) and in total of both food consumption (17.34%), others (14.08%), family enterprise (13.60%), agricultural inputs (13.03%) and payment of loan (12.29%).

**Income of both borrowers and non-borrower households:** The study showed that annual income of majority borrowers 1053 (66%) depend on income from labor sale following from agriculture 830 (56%), non-agriculture 83 (52%), business 400 (25%) and donation 208(13%). Similarly, the annual income of majority non-borrowers 464(63%) comes from labor sale following agriculture 333(45%), non-agriculture 331 (45%), business 197(27%) and donations 81(11%).

**Expenditures of both borrowers and non-borrower households:** Data analysis depicted that both the borrowers and non-borrowers HHs incurred expenditure for consumption in respect of both food and non-food but borrowers HHs incurred 80% for food and 20% for non-food while non-borrowers HHs 84% for food and 16% for non-food and there is significant difference in respect non- food consumption as well as total consumption between borrowers and non-borrowers HHs. There are twelve items of investment expenditure and the items above 5% of total expenditure in case of borrowers' HHs are agriculture (20%), healthcare (20%) family business (16%), house repair (9%) and poultry /livestock (7%) while in respect of non – borrowers the investment items above 5% of in terms of percentage of total expenditure are healthcare (27%), agriculture (23%), education (18%), family business 9% and house repair (9%). There is significant difference between borrowers and non-borrowers in the items of education, healthcare, poultry/livestock, productive assets, durable goods, house repair, others investment and also in total investment expenditure as well as total expenditure (consumption plus investment).

**Causes of non-payment of loan in time:** The borrowers were given a list of fourteen possible causes of non- payment of loan in time and PCA reduced the fourteen causes into four factors. The first factor is related to food and medicine, the second factor is related to marriage and legal problems of family, the third factor is related to terms of loan and investment lastly the fourth factor related to cost and unavailability of loan in time.

**Attitude of borrowers towards micro-credits:** The borrowers were given sixteen dimensions of attitudes related to micro-credits and after statistical analysis and PCA these dimensions are reduced into major four factors that micro-credits help in increasing the income and savings firstly, secondly terms and conditions of loans are rigid, third factor related to cost of loans is high finally the fourth factor is that micro-credits boost up food and health.

**Role of micro-credits on graduation of selected socio-economic status**: The role of micro-credits is measured in terms of food security, socio-economic status and investment on education and healthcare.

The study showed that the households of *Haor* regions who received micro-credits benefit are found in more food security than the non-borrower households. The proportion of households with moderate food insecurity is significantly ($p<0.01$) reduced to 14.5% from 19.0% due to micro-credits program. This study also found that the incidence of normal food insecurity also significantly ($p<0.01$) reduced to 9.3% from 16.2%. On the other hand, such type of changes is also observed among non-borrower households but the decrement was found notably lower and statistically insignificant as well.

The percentage households having extreme poor and moderately poor condition are estimated lower after receiving micro-credits. It is noticeable that the proportion also reduces to the non-borrower households but the decrement was statistically insignificant. However, the micro-credits have significant influence for reducing extreme poverty condition among credits receiving households.

The data on the percentage of households with the increase of educational expenditure are estimated higher in case of borrower households (63.8%) than that of non-borrower households (51.0%). This study found that the average educational expenditure increases significantly among borrower households as well as among non-borrower households. In addition, the increment of healthcare expenditure also observed both case and control households. The case-control comparison indicates that the healthcare expenditure increases more among borrower household (70.1%) than non-borrower household (60.3%). Besides, the before-after comparison indicates that the healthcare expenditures among borrower households but insignificant among non-borrower households.

**Major dimensional factors for not graduating from poverty by micro-credits:** To find out the causes of not graduating from poverty in spite of using micro-credits among both case and control households' fourteen dimensions were listed to the respondents. The principal cause was found as natural calamity among non-borrower households (75.3%) and pressure of the payment of loan among borrower households (83.9%). The second prime reason was loss of investment among non-borrower households (72%) and high interest rate among borrower households (73.1%). The PCA reduced these dimensions into five factors- (i) imbalance between amount and time span of loan; (ii) imbalance between dependent member and earning member; (iii) incidence of investment loss; (iv) pressure created by the loan provider and (v) unable to utilize and diverse the loan.

**Impact of micro-credit at the household level**: Impact of micro-credits at household level is measured by comparing between 2016 and 2019 that means before and after comparison and DID method of income related indicators and expenditure and investment. The study showed that total income of borrowing HHs as well as non-borrowing HHs has been increased significantly and the income from agriculture, non-agriculture and labor sale have imbued this increase in total income in spite of insignificant increase in income from business and donations. Finally, total income of borrowing HHs has been increased at higher rate than that of non-borrowing HHs significantly. This is permeated by the significant difference in increase of non-agricultural income and labor sale though the increase in agricultural income, business and donation are insignificant. Analysis of data showed that total consumption expenditure with food and non-food expenditures of both borrowers and non–borrowers has been increased significantly. The increase in total consumption of borrowers and non-borrowers did not significantly differ though non-food expenditure differs significantly. Total investment expenditure of both borrowers and non –borrowers has been increased significantly. All the items of investment have been increased in case of borrowers except land purchase but in case of non- borrowers only land agriculture, family business, house repair and others investment are significant. Finally, the increase of total investment expenditure of borrowers has been significantly increased than that of non-borrowers and in item-wise all the investment items except education.

The relevant data showed that total income of borrowers has been increased over non-borrowers (z=6.75) significantly. Among the components of income, non-agricultural income has been increased significantly on the other hand income from labor sale has been decreased significantly. The incomes from agriculture, business and donations have been increased but insignificantly. The non-food expenditure has been increased significantly but food expenditure has been decreased insignificantly while total food consumption has been increased (z=1.06) significantly of borrowers over non-borrowers. The investment expenditure of borrowers has been increased over the non-borrowers significantly. Among the heads of investment expenditure healthcare, poultry livestock and productive assets have been increased significantly of the borrowers over non-borrowers while investment on education, agriculture, family business, house repair, land purchase and others investment have been increased insignificantly. Finally total expenditure, total investment and savings of borrowers have been increased significantly over the non-borrowers.

**Comparative impact of formal and informal micro-credits:** A comparative impact of formal and informal micro-credits is done in terms income, expenditure and investment as follows:

The data on income related showed that total income and its heads of income from agriculture, non-agriculture and labor sale have been increased significantly of formal borrowers over the study period while in case of informal credits total income and its heads of income from agriculture and labor sale have been increased significantly. The heads of income from business and donations in case of formal credits and incase of informal credits the heads of income from non-agriculture, business and donations have been increased insignificantly.

Data showed that total consumption expenditure with its heads of food and non-food consumption of both formal borrowers and informal borrowers have been increased significantly. In case of formal borrowers' total investment expenditure and its entire heads except land purchase and house repairing have been increased significantly. The increase in total consumption of borrowers and non-borrowers did not significantly differ though non-food expenditure differs significantly. Total investment expenditure of both borrowers and non-borrowers has been increased significantly. All the items of investment have been increased in case of borrowers except land purchase but in case of non- borrowers only land agriculture, family business, house repair and others investment are significant. Finally, the increase of total investment expenditure of borrowers has been significantly increased than that of non-borrowers and in item-wise all the investment items except education.

The net impact on income and its heads showed that total income and its entire heads do not differ significantly among the formal and informal sources except donation income which has been decreased ($z=-20.6$) of formal sources over informal sources significantly ($p=.001$).The data on total consumption showed that incase of formal borrowers the total consumption has two heads and total consumption and food consumption did not change significantly over informal borrowers ($z=-13.68$) but non-food expenditure has been decreased of formal borrowers than informal borrowers significantly ($p=<0.01$). Total investment has been increased ($z=18.53$) of formal borrowers than that of informal significantly ($p=<0.01$) among the items of investment, education, agriculture, poultry–livestock, productive assets and durable goods have been increased significantly of formal borrowers over informal borrowers on the other hand investment in health care, family business, house repair, land purchase and others investment have been decreased significantly. The total expenditure and savings have been increased of formal borrowers over informal significantly.

**Efficiency of the programs:** The discriminant analysis between formal and informal borrowers was done in respect of selected fifteen predictors (variables) to assess the efficiency of the programs. The analysis showed that there is significant discriminant between the two groups of borrowers in respect all predicators. Based on the estimates of the discriminant analysis such as un-standardized coefficients, standardized coefficients and structure matrix we found the magnetite of discriminating power of the predictors. The top most ten magnitude predictors are types of installments followed by duration of loan, interest rate, gender, earning status, regularity of income, occupation, number of earning members, age and amount of loan. We have established the Fisher's linear discriminant function for discriminating the formal and informal source of credits by the fifteen stated variables above to predict or classify a new individual with the respective characteristics who are supposed to borrow from formal source or informal source. The average of the functions at group centroid is estimated at 0.44. Therefore, it is obvious that new members with discriminant scores above 0.44 supposed to borrow from the informal sources, otherwise he/she would be preferred to borrow from formal sources.

**Findings from the qualitative surveys:** The key informants/participants informed that micro-credit is available in each and every corner of the *Haor* region. Though there are numerous formal sources yet, the *Haor* people still depends on informal sources of micro-credits for covering the emergency situation like treatment of household members, covering the cost of food items during severe food shortages. The interest rate of micro-credits from informal sources is sometimes rose to 60% in some season, particularly before harvesting the crops. On the other hand, the interest rate of micro-credits is remains constant throughout the year for institutional sources like banks, NGOs etc. The main purpose of getting loans is for food expenditure, repairmen of living house, investment in agriculture farming, investment in purchasing livestock, and expenditure for treatment of household members are the main reasons for receiving micro-credits. The causes of not getting formal and informal micro-credits are they do not know about some micro-credit programs, nepotism, misappropriation of credits, lack of entry fee/bribe and bureaucratic problem. The heads of utilization of micro-credits are to meet-up the food expenditure, repairing the living house, investment in agriculture for harvesting crop along with loan repayment. The borrowers said that the rate of interest is very high, credits period is insufficient and terms are rigid. The causes of ineffectiveness of micro-credits for poverty reduction perspective are the lack of policy, lack of skills, and investment in non-productive heads appeared as the hindrances for effectiveness of the micro-credits on the perspective of poverty reduction. The main suggestions for maximization of benefits from micro-credits are that in general, the micro-credits from informal sources should be stopped, the volume of loan should be increased at lower rate, the micro-credits from formal sources should be encouraged for the households whose members acquired capability of start any economic activity through skill-development program. The government could undertake a plan of action to review the existing micro-credit programs. Skill-development training programs should be strengthening covering different investment areas for the adult members of credit-receiver households and finally new research may be undertaken to find out different pathways for eradication of poverty of *Haor* people in order to achieve the targets of SDGs.

# Annex-1: Survey Questionnaire

## Vicious Cycle of Poverty in Haor Region of Bangladesh: Impact of Formal and Informal Credits

**ID** ☐ ☐ ☐ ☐ ☐ ☐ ☐

### HOUSEHOLD INTERVIEW SCHEDULE

| | |
|---|---|
| District | |
| Upazila | |
| Union | |
| Mouza/Village | |
| PSU number | |
| Household number | |
| | |
| Name of household head | |
| Name of the Father/husband/wife of household head | |
| Mobile number | |
| Status of NGO membership (Yes = 1, No = 2) | |
| Religion [Muslim-1, Hindu-2, Buddhist-3, Christian-4] | |
| **Type of Micro-credits:** Government (Banks/ co-operatives) =1, Nongovernment (MFI/NGO/Insurance) =2, Local Money Lender (Mohajan/Private Samittee =3, Non-interest Loan (Relatives/friends/ neighbors/ Land Owner) =4; More than one sources=5 | |
| **Household type** [Borrower = 1, Non-borrower = 2] | |
| Respondent name | |
| Relation with household head (if the respond is not head) | |
| Interviewer name | |
| Date of interview | |
| Name of data entry operator (DEO) | |

| **Household Type (Put√)** | ☐ Borrower | ☐ Old Beneficiary HH |
|---|---|---|

**Section-1: Profile of Household Members of the Respondent (for all)**

|      | 1 | 2 | 3 | 4 | 5 | 6 | 7 | 8 | 9 |
|------|---|---|---|---|---|---|---|---|---|
| Sl. No. | Name of household members (Write head's name first) | Relationship with HH (code)[2] | Age (in full years) | sex (Male = 1, Female =2) | Main Occupation Age>=16) (Code)[5] | Education (Age>7) code[6] | Marital Status (age≥15) (Code)[7] | Borrower status (Code)[8] | Work Condition (Age>=16) (Code)[9] |
| 01 | | | | | | | | | |
| 02 | | | | | | | | | |
| 03 | | | | | | | | | |
| 04 | | | | | | | | | |
| 05 | | | | | | | | | |
| 06 | | | | | | | | | |
| 07 | | | | | | | | | |
| 08 | | | | | | | | | |
| 09 | | | | | | | | | |
| 10 | | | | | | | | | |

**2. Relationship code:** Household head = 1, Husband/Wife = 2, Son/Daughter = 3, Brother/Sister = 4, Father/Mother = 5, Father-in-law/Mother-in-law = 6, Son-in-law/Daughter-in-law = 7, Grandson/daughter = 8, Relative = 9, Non-relative = 10, Servant = 11, Other (Specify) …….........

**5. Occupation code:** Farming (own land or leased land) = 1, Agri-labour = 3, Non-agri-labour (Driver/Rickshaw /Van puller/boatman)=4,Service/Self Employed/Entrepreneur/Business/Tuition=5,Student=6, Unable to work (Old person, Disabled)=7, House-wife=8, Unemployed=9, Other(Specify)=10

**6. Education Status:** Years of schooling (*i.e.*, if class 5 = 5, if class six = 6)

**7. Marital status code:** Married = 1, Unmarried = 2, Widow = 3, Separated/Divorced = 4.

**8. Borrower's Status:** 1 year but less than 2 years=1; 2 years but less than 3 years=2; 3 year but less than 4 year =3; 4 years but less than 5 years = 4; 5 year and above = 5

**9. Income Earning Condition:** Full time work = 1, Part time work = 2, No Work = 3, Otherwise = 4.

**Section-2: Basic Household Information (for all)**

| | |
|---|---|
| 2.1. Do you live in your own house?    (Yes=1, No= 2) | |
| 2.2. How many rooms in your house? | |
| 2.3. Does kitchen room is separated from bed room in your own house**?** (Yes=1, No= 2) | |
| 2.4. Do you have regular source of income? (Yes=1, No= 2) | |
| 2.5. Housing Information: | |

| Sl | | |
|---|---|---|
| 1 | **Type of main house** [Straw roof and bamboo wall = 1; Straw roof and muddy wall = 2; Tin shed roof and muddy wall = 3; Tin shed roof and wall and muddy floor = 4; Tin shed building =5; Pucca building = 6; Slum =7; Others = 8] | |
| 2 | **Source of cooking fuel of your households** [Wood = 1; Gas = 2; Kerosene = 3; Straw/Leaf/Husk = 4; Electricity =5; Cow dung = 6; Jute stick =7; others = 5] | |
| 3 | **Main source of drinking water of your household** [Supply water = 1; Tube-well = 2; Pond = 3; Well = 3; Rain water = 5; River/Canal = 6] | |
| 4 | **Electricity coverage in the area:**         (Yes=1            No= 2) | |
| 5 | **Electricity coverage in the   household**: (Yes=1            No= 2) | |
| 6 | **Ownership of Toilet:**                 (Yes=1            No= 2) | |
| 7 | **Type of Toilet used by HH members:** [*Pucca* toilet (water resistant) = 1;   *Pucca* toilet (not water resistant) = 2;   *Katcha* toilet = 3;   Open field/Others = 4] | |

**Section-3: Possessions of Assets (for all)**

| Sl. No. | Items | Quantity | Sl. No. | Items | Quantity |
|---|---|---|---|---|---|
| 1 | Radio | | 11 | Furniture of drawing room | |
| 2 | Television | | 12 | Bicycle | |
| 3 | Mobile Phone | | 13 | Boat | |
| 4 | Electric Fan | | 14 | Tube-well | |
| 5 | Almirah | | 15 | Deep tube-well | |
| 6 | Cookeries | | 16 | Clock | |
| 7 | Cutleries | | 17 | Shelf | |
| 8 | Table chair | | 18 | *Alna* | |
| 9 | Bed, Cushion | | 19 | Motor cycle | |
| 10 | *Choki* (cot) | | 20 | Others (specify) | |

**Section-4. Household Productive Assets (for all)**

| Sl. No. | Productive Assets | Yes=1 | No=2 | |
|---|---|---|---|---|
| 1 | Cultivable land | 1 | 2 | Amount in............................decimals |
| 2 | Cultivable land: leased or cropped share | 1 | 2 | Amount in............................decimals |
| 3 | Homestead land | 1 | 2 | Amount in............................decimals |
| 4 | cultivation Instruments | 1 | 2 | Amount in............................decimals |
| 5 | Livestock | 1 | 2 | Amount in............................Number |
| 6 | Rickshaw/Van | 1 | 2 | Amount in............................Number |
| 7 | Auto Rickshaw | 1 | 2 | Amount in............................Number |
| 8 | Sprayer | 1 | 2 | Amount in............................Number |
| 9 | Fishing Net | 1 | 2 | Amount in …........................Number |
| 10 | Bee Box | 1 | 2 | Amount in............................Number |
| 11 | Sewing Machine | 1 | 2 | Amount in …........................Number |
| 12 | Motor (Engine) | 1 | 2 | Amount in …........................Number |
| 13 | Family Business | 1 | 2 | Amount in …........................Number |
| 14 | Others | 1 | 2 | Amount in …........................Number |
| | Others (if any) | | | |
| 16 | | | | |
| 17 | | | | |

**Section 5: Knowledge of household about Micro-credits Information (for all)**

| | Question | Response |
|---|---|---|
| | 501. Do you or any member of your HHs have/has Knowledge about Micro-credits benefit? (Yes=1, No= 2) | |
| | 502. Do you or any member of your HHs have/has try to get micro-credits benefit? (Yes=1, No= 2) | |
| | 503. For which type of credits do you take first attempt for getting Micro-credits benefit (use code given below) | |
| | **Codes of Micro credits Programs:**<br>Government (Banks/ co-operatives) =1,<br>Nongovernment (MFI/NGO/Insurance) =2,<br>Local Money Lender (Mohajan/Private Samittee =3,<br>Non-interest Loan (Relatives/friends/ neighbors/ Land Owner) =4;<br>More than one sources=5 | |

| | **504. Please specify the level of response on the causes of not including into micro-credits program (for non-borrowers only)** | |
|---|---|---|
| a. | **Bureaucratic complexity** (Strongly disagree=1; Disagree=2; no comment=3; Neutral =4; Strongly Agree=5.) | |
| b. | **Limitation of budget** (according to selector): (Strongly disagree=1; Disagree=2; no comment=3; Neutral =4; Strongly Agree=5.) | |
| c. | **Couldn't provide bribe or entry fee:** (Strongly disagree=1; Disagree=2; no comment=3; Neutral =4; Strongly Agree=5.) | |
| d. | **No political exposure:** (Strongly disagree=1; Disagree=2; no comment=3; Neutral =4; Strongly Agree=5.) | |
| e. | **Didn't have any idea about such program:** (Strongly disagree=1; Disagree=2; no comment=3; Neutral =4; Strongly Agree=5.) | |
| f. | **Nepotism:** (Strongly disagree=1; Disagree=2; no comment=3; Neutral =4; Strongly Agree=5.) | |
| g. | **Non-cooperation from public delegate of Micro-credits institution:** (Strongly disagree=1; Disagree=2; no comment=3; Neutral =4; Strongly Agree=5.) | |
| h. | **Non-cooperation from local lenders:** (Strongly disagree=1; Disagree=2; no comment=3; Neutral =4; Strongly Agree=5.) | |
| i. | **Non-availability of NID:** (Strongly disagree=1; Disagree=2; no comment=3; Neutral =4; Strongly Agree=5.) | |
| j. | **Lack of networking or lobbying:** (Strongly disagree=1; Disagree=2; no comment=3; Neutral =4; Strongly Agree=5.) | |
| k. | **Distance from Micro-credits from the village:** (Strongly disagree=1; Disagree=2; no comment=3; Neutral =4; Strongly Agree=5.) | |
| l. | **No Micro-credits in the area:** (Strongly disagree=1; Disagree=2; no comment=3; Neutral =4; Strongly Agree=5.) | |
| m. | **Non availability of collateral:** (Strongly disagree=1; Disagree=2; no comment=3; Neutral =4; Strongly Agree=5.) | |
| n. | **Misappropriation of credits:** (Strongly disagree=1; Disagree=2; no comment=3; Neutral =4; Strongly Agree=5.) | |
| o. | **Biasness:** (Strongly disagree=1; Disagree=2; no comment=3; Neutral =4; Strongly Agree=5.) | |
| p. | Others: (specify)…………… | |
| | 505. Ask for help to get Micro-credits benefit:<br>(UP Chairman=1; UP Member=2; UP Office=3; NGO=4;<br>Bank officer-5; Relatives/Neighbors/Friends/ others=6) | |
| | 506. Do you ask for money for giving Micro-credits benefit? (Yes=1, No= 2) | |
| | 507. Does micro-credits help to remove poverty? (Yes=1, No= 2) | |

**Section-6: Profile of micro-credits received (Borrowers only)**

| 1 | 2 | 3 | | 4 | 5 | 6 | | | | | 7 | | | 8 | |
|---|---|---|---|---|---|---|---|---|---|---|---|---|---|---|---|
| Sl. No | Sources code[2] | Date of loan | | Amount of loan (in Taka) | Type of loan code[5] | Terms of loan | | | | | Payment of loan | | | Unpaid loan | |
| | | Year | Month | | | Interest Rate (%) | Installment type code[6] | Total installments | Duration (in month) | collateral code[6a] | Principal | Interest | total | Principal | Interest |
| | | | | | | | | | | | | | | | |
| | | | | | | | | | | | | | | | |
| | | | | | | | | | | | | | | | |

**2. Sources of credits:** Government (Banks/ co-operatives) =1,
Nongovernment (MFI/NGO/Insurance) =2,
Local Money Lender (Mohajan /Private Samittee =3.
Non-interest Loan (Relatives/friends/ neighbors/ Land Owner) =4
More than one sources =5
**5. Type of loan:** Cash=1; food items =2; Both cash and food items=3; others=4
**6. Installment type:** Weekly=1; Biweekly=2; Monthly =3; Quarterly =4; Annually=5
**6a. Collateral:** If not collateral =0; If collateral =1

**609: Expenditure and Investment Pattern of Micro-credits**

| SL# | Expenditure Heads | Amount | % | SL# | Expenditure Heads | Amount | % |
|---|---|---|---|---|---|---|---|
| 1. | Consumption on food | | | 13. | Boat for rent | | |
| 2. | Clothing and others essentials | | | 14. | Payment of loan | | |
| 3. | Purchasing agricultural inputs | | | 15. | Family enterprises | | |
| 4. | Purchasing durable goods | | | 16. | Health care | | |
| 5. | Housing improvements | | | 17. | Human capital (Education, training) | | |
| 6. | Purchasing land/Lease | | | 18. | Savings (Bank, MIF, NGO) | | |
| 7. | Purchasing animals | | | 19. | Sending son to abroad | | |
| 8. | Natural calamities | | | 20. | Release mortgaged land | | |
| 9. | Fishing net | | | 21. | Mortgage land | | |
| 10. | Bee box | | | 22. | Daughter's/Son's marriage | | |
| 11. | Sewing machine | | | 23. | Fisheries pond lease | | |
| 12. | Motor (engine) | | | | Others: | | |

**Section- 7: Impact of Micro-credits on the households for both borrower and non-borrower**

|      | **Indicators** | **Status** ||
|------|----------------|-------------|---------------|
|      |                | Survey Point | Before loan or 3 years back |
| 7.1  | Food expenditure (annual) | | |
| 7.2  | Non-food consumption expenditure (annual) | | |
| **7.3** | **Consumption Expenditure (Annual) (7.1 + 7.2)** | | |
| 7.4  | Education (annual) | | |
| 7.5  | Medical (annual) | | |
| 7.6  | Training (annual) | | |
| 7.7  | Agriculture (annual) | | |
| 7.8  | Purchasing productive equipment (annual) | | |
| 7.9  | Purchasing durable assets (annual) | | |
| 7.10 | House repair/development (annual) | | |
| 7.11 | Land purchase (annual) | | |
| 7.12 | Family enterprise (annual) | | |
| 7.13 | Livestock and Poultry rearing (annual) | | |
| 7.14 | Other investment | | |
| **7.15** | **Total Investment Expenditure (Annual) (Total of 7.4 to 7.14)** | | |
| **7.16** | **Savings (annual) (in banks, MFI, etc.)** | | |
| **7.17** | **Total Expenditure and Savings (Annual) (7.3 + 7.15 + 7.16)** | | |
| 7.18 | On-farm (Agricultural) income (annual) | | |
| 7.19 | Off-farm (Non-agricultural) income (annual) | | |
| 7.20 | Labor sale income (annual) | | |
| 7.21 | Business (annual) | | |
| 7.22 | Donation/Begging (annual) | | |
| **7.23** | **Debt (annual)** | | |
| **7.24** | **Total Income and Debt (Total of 7.18 to 7.23)** | | |
| 7.25 | **Food security status** (Some periods of hunger during the year (Normal Food Insecurity) =1, Two meals a day throughout year (Moderate food insecurity) =2, Three meals a day throughout year (No Food Insecurity=3)) | | |
| 7.26 | **Socio-economic status** (Very poor=1, Moderately poor=2, Poor=3, Middle class=4, Rich=5) | | |
| 7.27 | Total Land (Homestead, Cultivable, Pond) *in Decimals* | | |
| 7.28 | **Type of House** [Straw roof and bamboo wall = 1; Straw roof and muddy wall = 2; Tin shed roof and muddy wall = 3; Tin shed roof and wall and muddy floor = 4; Tin shed building =5; Pucca building = 6; Slum =7; Others = 8] | | |
| 7.29 | Status of Electricity Connection (Yes=1, No=2) | | |
| 7.30 | Types of Latrines [*Pucca* toilet (water resistant) = 1; Pucca toilet (not water resistant) = 2; *Katcha* toilet = 3; Open field/Others = 4] | | |
| 7.31 | Ownership of Television (Yes=1, No=0) | | |
| 7.32 | Number of Fan | | |
| 7.33 | Number of Mobile Phone | | |
| 7.34 | Status of Education Expenditure (Increased=1, Decreased=2, No Change=0) | | |
| 7.35 | Status of Health Expenditure (Increased=1, Decreased=2, No Change=0) | | |

**Section -8. Purpose of Loan: Please put tick () mark(s) for which you have taken loan (Borrower only)**

| SL# | Purpose | Code | | SL# | Purpose | code |
|---|---|---|---|---|---|---|
| a. | Purchasing food items | 1 | | k. | Purchasing of livelihood equipment | 12 |
| b. | Crop production | 2 | | l. | payment of loan | 13 |
| c. | Rearing cattle/poultry | 3 | | m. | Repairing cost of houses | 14 |
| d. | Sending family member to abroad | 4 | | n. | Health care expenditure | 15 |
| e. | Trade/Business/Industry | 6 | | o. | Education | 16 |
| f. | Fish farming/ Fishing | 7 | | p. | Others: | 17 |
| g. | Daughter/son's marriage | 8 | | | | |
| h. | Constructing housing | 9 | | | | |
| i. | Tackling shocks of natural calamities | 10 | | | | |
| j. | Tackling shocks of sudden death of HH head | 11 | | | | |

**Section -9: Causes of nonpayment of loan timely**

| SL# | Causes of nonpayment of loan timely | Strongly agree=5 | Agree =4 | Neutral =3 | Disagree =2 | Strongly disagree=1 |
|---|---|---|---|---|---|---|
| 1. | Acute food problem | | | | | |
| 2. | Medical treatment/medicine | | | | | |
| 3. | Investment loss | | | | | |
| 4. | Natural calamities | | | | | |
| 5. | Insufficient of loan for investment | | | | | |
| 6. | Duration of loan is short for return from investment | | | | | |
| 7. | Installment period is very short | | | | | |
| 8. | Rate of interest is very high | | | | | |
| 9. | Renewal of loan is unavailable | | | | | |
| 10. | Misappropriation of loan | | | | | |
| 11. | Crop's failure | | | | | |
| 12. | Expenditure for marriage of son/daughter etc. | | | | | |
| 13. | Family legal problems and expenditure | | | | | |
| 14. | Unexpected accident | | | | | |

**Section -10. Attitude of Borrowers' on Micro Credits (Borrower Only)**

| Sl No. | Question Statement | Strongly agree=5 | Agree =4 | Neutral =3 | Disagree =2 | Strongly disagree=1 |
|---|---|---|---|---|---|---|
| 1 | The rate of interest of micro credits is reasonable | | | | | |
| 2 | Amount of credits is sufficient | | | | | |
| 3 | Duration of credits is sufficient | | | | | |
| 4 | Terms and conditions are not rigid | | | | | |
| 5 | By microfinance your food security has increased | | | | | |
| 6 | By microfinance your income has increased | | | | | |
| 7 | By microfinance your savings has increased | | | | | |
| 8 | Microfinance is helping you in better access to education | | | | | |
| 9 | Microfinance is helping you in better access to healthcare | | | | | |
| 10 | Microfinance is helping you in better financial situation of your family | | | | | |
| 11 | Operational assistance received from MFIs was helpful to run the business | | | | | |
| 12 | Due to microfinance, employment opportunities have been increased | | | | | |
| 13 | Local loans are easier to get than MFIs | | | | | |
| 14 | Local lenders are friendly than MFIs | | | | | |
| 15 | Cost of Local loans is lower than MFIs | | | | | |
| 16 | Terms and conditions local loans are easier than MFIs | | | | | |

**Section -11. Causes of not over-coming from vicious cycle of poverty (for all)**

| SL# | Causes | Strongly agree=5 | Agree =4 | Neutral =3 | Disagree =2 | Strongly disagree=1 |
|---|---|---|---|---|---|---|
| 1. | Insufficient loan | | | | | |
| 2. | Duration of loan is insufficient | | | | | |
| 3. | High Interest | | | | | |
| 4. | Renewal of loan not get | | | | | |
| 5. | Diversion of loan | | | | | |
| 6. | Loss of investment | | | | | |
| 7. | Natural calamity | | | | | |
| 8. | Not getting loan in time | | | | | |
| 9. | Pressure of loan payment | | | | | |
| 10. | Number of dependent members is high | | | | | |
| 11. | Income earning member is absent | | | | | |
| 12. | No. of income earning member is low in terms of dependent person | | | | | |
| 13. | I have not taken loan | | | | | |

Study Conducting by:

**Professor Dr. Md. Nazrul Islam**
Department of Business Administration
and Member, SUST Research Centre
Shahjalal University of Science & Technology (SUST)
Contact# 01712817424
E-mail: dnislam69@gmail.com

# *Appendices*

**Table 3.1** Profile of respondent by micro-credits receiving status in the *Haor* regions

| Characteristics | Borrower (N = 1607) | Non-borrower (N = 733) | Z-statistic | P-value | Both (N = 2340) |
|---|---|---|---|---|---|
| **Relation with household head** | | | | | |
| Head | 73.8 | 98.1 | -14.015 | <0.001 | 81.4 |
| Husband/wife | 23.7 | 1.8 | 13.130 | <0.001 | 16.8 |
| Son/daughter | 1.0 | 0.0 | 2.717 | 0.010 | 0.7 |
| Father/mother | 1.2 | 0.1 | 2.680 | 0.011 | 0.9 |
| Others | 0.3 | 0.0 | 1.484 | 0.133 | 0.1 |
| **Sex of respondent** | | | | | |
| Male | 70.6 | 90.3 | -10.467 | <0.001 | 76.8 |
| Female | 29.4 | 9.7 | 10.467 | <0.001 | 23.2 |
| **Age of respondent** | | | | | |
| 16-30 | 17.9 | 14.6 | 1.977 | 0.056 | 16.8 |
| 31-50 | 61.9 | 62.5 | -0.277 | 0.384 | 62.1 |
| 51-60 | 13.9 | 14.7 | -0.515 | 0.349 | 14.2 |
| Above 60 | 6.3 | 8.2 | -1.682 | 0.097 | 6.9 |
| Mean ± SD (in years) | 42.47 ± 11.38 | 43.57 ± 11.60 | -2.140 | 0.040 | 42.81 ± 11.46 |
| **Marital status** | | | | | |
| Married | 92.7 | 91.1 | 1.339 | 0.163 | 92.2 |
| Unmarried | 2.4 | 2.9 | -0.711 | 0.310 | 2.6 |
| Widow | 4.7 | 5.7 | -1.028 | 0.235 | 5.0 |
| Separated/Divorced | 0.1 | 0.3 | -1.114 | 0.215 | 0.2 |
| **Educational status** | | | | | |
| No education | 27.7 | 24.3 | 1.726 | 0.090 | 26.6 |
| 1-5 years of schooling | 53.9 | 52.3 | 0.720 | 0.308 | 53.4 |
| 6-9 years of schooling | 12.4 | 13.6 | -0.807 | 0.288 | 12.8 |
| SSC / HSC | 5.6 | 8.7 | -2.807 | 0.008 | 6.6 |
| Graduate and above | 0.4 | 1.1 | -2.002 | 0.054 | 0.6 |
| Mean ± SD (in years) | 3.76 ± 3.13 | 4.19 ± 3.42 | -2.896 | 0.006 | 3.89 ± 3.23 |
| **Occupation** | | | | | |
| Farming | 19.7 | 19.8 | -0.056 | 0.398 | 19.7 |
| Day laborer | 22.5 | 23.3 | -0.428 | 0.364 | 22.7 |
| Off-farm activities | 11.1 | 17.7 | -4.379 | <0.001 | 13.2 |
| Service/business | 16.2 | 24.6 | -4.821 | <0.001 | 18.8 |
| Student | 0.2 | 0.0 | 1.212 | 0.191 | 0.2 |
| Household work | 24.9 | 5.7 | 11.006 | <0.001 | 18.9 |
| Others | 5.4 | 8.9 | -3.186 | 0.002 | 6.5 |
| **Income earner** | | | | | |
| Full time | 47.6 | 54.7 | -3.186 | 0.002 | 49.8 |
| Part time | 27.9 | 32.7 | -2.364 | 0.024 | 29.4 |
| No work | 24.5 | 12.6 | 6.581 | <0.001 | 20.7 |
| **Disability status** | | | | | |
| Yes | 2.9 | 4.4 | -1.865 | 0.070 | 3.3 |
| No | 97.1 | 95.6 | 1.865 | 0.070 | 96.7 |

*Note*: SSC = Secondary School Certificate; HSC = Higher Secondary Certificate; SD = Standard Deviation

**Table 3.2** Distribution of household population by gender in the *Haor* regions of Bangladesh

| Characteristics | Male (%) (N = 5950) | Female (%) (N = 5678) | Z-statistic | P-value | Both (%) (N = 11628) |
|---|---|---|---|---|---|
| **Age group** | | | | | |
| 0-15 | 35.2 | 37.9 | -3.023 | 0.004 | 36.5 |
| 16-30 | 26.7 | 28.3 | -1.932 | 0.062 | 27.5 |
| 31-50 | 25.5 | 25.3 | 0.248 | 0.387 | 25.4 |
| 51-60 | 6.8 | 4.9 | 4.356 | <0.001 | 5.9 |
| Above 60 | 5.7 | 3.7 | 5.082 | <0.001 | 4.7 |
| Mean ± SD | 27.52 ± 18.92 | 25.38 ± 17.50 | 6.335 | <0.001 | 26.47 ± 18.27 |
| **Marital status (Age above 15 years)** | | | | | |
| Married | 66.4 | 73.0 | -6.156 | <0.001 | 69.5 |
| Unmarried | 31.7 | 18.4 | 13.124 | <0.001 | 25.4 |
| Widow | 1.8 | 8.1 | -12.637 | <0.001 | 4.8 |
| Separated/divorced | 0.1 | 0.5 | -3.187 | 0.002 | 0.3 |
| Total (n) | 3856 | 3528 | | | 7384 |
| **Educational status (Age above 6 years)** | | | | | |
| No education | 17.3 | 21.1 | -4.904 | <0.001 | 19.1 |
| 1-5 years schooling | 49.8 | 48.8 | 1.015 | 0.238 | 49.3 |
| 6-9 years schooling | 19.2 | 18.5 | 0.908 | 0.264 | 18.9 |
| SSC / HSC | 11.4 | 10.1 | 2.129 | 0.041 | 10.8 |
| Above HSC | 2.3 | 1.5 | 2.965 | 0.005 | 1.9 |
| Mean ± SD | 4.91 ± 3.67 | 4.50 ± 3.56 | 5.760 | <0.001 | 4.71 ± 3.62 |
| Total (n) | 5328 | 4992 | | | 10320 |
| **Occupation (Age 16-60 years)** | | | | | |
| Farming | 17.0 | 0.3 | 24.245 | <0.001 | 8.9 |
| Day laborer | 24.7 | 1.7 | 27.783 | <0.001 | 13.6 |
| Off-farm activities | 15.8 | 0.6 | 22.618 | <0.001 | 8.4 |
| Service/business | 23.0 | 3.8 | 23.088 | <0.001 | 13.7 |
| Student | 10.9 | 11.2 | -0.395 | 0.369 | 11.0 |
| Household work | 0.5 | 70.7 | -60.976 | <0.001 | 34.6 |
| Others | 8.3 | 11.6 | -4.565 | <0.001 | 9.9 |
| Total (n) | 3515 | 3319 | | | 6834 |
| **Income earner (Age above 15 years)** | | | | | |
| Full time | 45.6 | 8.6 | 35.401 | <0.001 | 27.9 |
| Part time | 28.7 | 14.3 | 14.965 | <0.001 | 21.8 |
| No work | 25.4 | 76.5 | -43.867 | <0.001 | 49.8 |
| Otherwise | 0.4 | 0.6 | -1.222 | 0.189 | 0.5 |
| Total (n) | 3856 | 3528 | | | 7384 |
| **Disability status** | | | | | |
| Yes | 4.0 | 4.7 | -1.851 | 0.072 | 4.4 |
| No | 96.0 | 95.3 | | | 95.6 |

*Note:* SD=Standard Deviation; SSC=Secondary School Certificate; HSC=Higher Secondary School Certificate

**Table 3.3** Housing condition and sanitation facilities in the *Haor* regions of Bangladesh

| Household characteristics | Targeted variables | Borrowing status (%) | | | Household type by loan (%) | | | Overall (2340) |
|---|---|---|---|---|---|---|---|---|
| | | Non-borrower (733) | Borrower (1607) | P-value | Formal (1158) | Informal (449) | P-value | |
| **House ownership** | Yes | 94.1 | 94.8 | 0.314 | 94.5 | 95.5 | 0.288 | 94.6 |
| | No | 5.9 | 5.2 | 0.314 | 5.5 | 4.5 | 0.288 | 5.4 |
| **Number of Rooms** | One rooms | 22.6 | 18.2 | 0.018 | 17.0 | 21.4 | 0.049 | 19.6 |
| | Two rooms | 40.5 | 43.1 | 0.199 | 44.0 | 41.0 | 0.220 | 42.3 |
| | Three rooms | 27.3 | 28.3 | 0.352 | 28.5 | 27.6 | 0.374 | 27.9 |
| | Four or more rooms | 9.5 | 10.4 | 0.319 | 10.5 | 10.0 | 0.382 | 10.1 |
| **Mean person per room** | | **2.38** | **2.52** | **0.011** | **2.48** | **2.63** | **0.057** | **2.48** |
| **Separate Kitchen** | Yes | 59.5 | 62.4 | 0.163 | 56.7 | 76.8 | <0.001 | 61.5 |
| | No | 40.5 | 37.6 | 0.163 | 43.3 | 23.2 | <0.001 | 38.5 |
| **Type of main house** | Straw roof & Bamboo/muddy wall | 3.8 | 4.5 | 0.295 | 5.8 | 1.1 | <0.001 | 4.3 |
| | Tin shed roof & Tin/ muddy wall | 84.3 | 87.1 | 0.076 | 86.3 | 89.1 | 0.129 | 86.2 |
| | Tin shed building | 7.1 | 7.0 | 0.397 | 7.3 | 6.2 | 0.295 | 7.1 |
| | *Pucca* building | 4.5 | 1.4 | <0.001 | 0.5 | 3.6 | <0.001 | 2.4 |
| | Others | 0.3 | 0.1 | 0.215 | 0.1 | 0.0 | 0.319 | 0.1 |
| **Source of cooking fuel** | Wood/kerosene | 30.3 | 30.1 | 0.397 | 36.0 | 14.7 | <0.001 | 30.1 |
| | Gas | 7.8 | 3.8 | <0.001 | 3.5 | 4.5 | 0.256 | 5.0 |
| | Straw/husk/jute stick | 39.0 | 45.7 | 0.004 | 42.9 | 53.0 | 0.001 | 43.6 |
| | Cow dung | 22.8 | 20.2 | 0.143 | 17.4 | 27.4 | <0.001 | 21.0 |
| | Others | 0.1 | 0.2 | 0.344 | 0.2 | 0.4 | 0.310 | 0.2 |
| **Source of drinking water** | Supply water | 0.4 | 1.3 | 0.053 | 1.5 | 0.9 | 0.256 | 1.0 |
| | Tube-well | 95.8 | 93.2 | 0.019 | 92.0 | 96.2 | 0.004 | 94.0 |
| | Others | 3.8 | 5.5 | 0.085 | 6.6 | 2.9 | 0.006 | 5.0 |
| **Electricity coverage in the village** | Yes | 85.4 | 86.9 | 0.246 | 85.5 | 90.6 | 0.010 | 86.5 |
| | No | 14.6 | 13.1 | 0.246 | 14.5 | 9.4 | 0.010 | 13.5 |
| **Electricity coverage in the house** | Yes | 80.8 | 80.7 | 0.398 | 80.8 | 80.4 | 0.392 | 80.8 |
| | No | 19.2 | 19.3 | 0.398 | 19.2 | 19.6 | 0.392 | 19.2 |
| **Ownership of Toilet** | Yes | 91.1 | 91.8 | 0.340 | 92.1 | 91.3 | 0.347 | 91.6 |
| | No | 8.9 | 8.2 | 0.340 | 7.9 | 8.7 | 0.347 | 8.4 |
| **Type of Sanitation** | *Pucca* toilet (water seal) | 20.5 | 12.2 | <0.001 | 10.5 | 16.5 | 0.002 | 14.8 |
| | *Pucca* toilet | 25.9 | 29.2 | 0.103 | 33.3 | 18.7 | <0.001 | 28.2 |
| | *Katcha* toilet | 52.5 | 54.7 | 0.244 | 52.0 | 61.7 | 0.001 | 54.0 |
| | Open field/others | 1.1 | 3.9 | <0.001 | 4.1 | 3.1 | 0.257 | 3.0 |

*Note:* **Formal source:** Govt. bank or Govt. co-operatives, and Nongovernment (MFI/NGO/Insurance).
**Informal source:** Local Money Lender (Mohajan/Private samitte) & Non-interest Loan (Relatives/Land Owner)

**Table 3.4** Landholdings pattern of the study households in the *Haor* regions of Bangladesh

| Types of land | Classification and statistics | Micro-credits type | | | Household type | | | Overall (2340) |
|---|---|---|---|---|---|---|---|---|
| | | Non-borrower (733) | Borrower (1607) | P-value | Formal (1158) | Informal (449) | P-value | |
| Homestead Land | No land (%) | 5.9 | 9.6 | 0.005 | 11.8 | 4.0 | <0.001 | 8.5 |
| | 1 to 15 decimals (%) | 81.2 | 78.3 | 0.110 | 77.0 | 81.7 | 0.049 | 79.2 |
| | 16-50 decimals (%) | 9.8 | 9.3 | 0.371 | 8.8 | 10.7 | 0.200 | 9.5 |
| | 50+ decimals (%) | 3.1 | 2.7 | 0.345 | 2.3 | 3.6 | 0.139 | 2.8 |
| | Mean (in decimal) | 10.58 | 10.83 | 0.385 | 10.96 | 10.52 | 0.362 | 10.75 |
| | SD (in decimal) | 21.96 | 19.99 | | 21.31 | 16.48 | | 20.64 |
| | Median (in decimal) | 5.00 | 5.00 | | 5.00 | 5.00 | | 5.00 |
| | IQR (in decimal) | 7.00 | 8.00 | | 7.00 | 9.00 | | 7.00 |
| Cultivable own land | Landless (%) | 66.0 | 68.8 | 0.161 | 68.8 | 68.6 | 0.398 | 67.9 |
| | 1 to 15 decimals (%) | 8.2 | 6.0 | 0.057 | 6.5 | 4.9 | 0.193 | 6.7 |
| | 16-50 decimals (%) | 8.0 | 8.2 | 0.394 | 8.3 | 8.0 | 0.391 | 8.2 |
| | 50+ decimals (%) | 17.7 | 17.0 | 0.366 | 16.4 | 18.5 | 0.241 | 17.2 |
| | Mean (in decimal) | 105.85 | 96.60 | 0.148 | 102.80 | 80.74 | <0.001 | 99.67 |
| | SD (in decimal) | 155.49 | 127.41 | | 139.72 | 86.86 | | 137.33 |
| | Median (in decimal) | 56.00 | 60.00 | | 60.00 | 60.00 | | 60.0 |
| | IQR (in decimal) | 103.00 | 100.00 | | 100.00 | 65.00 | | 100.00 |
| Cultivable leased-in or sharecropped land | No land (%) | 80.8 | 70.1 | <0.001 | 71.0 | 67.7 | 0.172 | 73.4 |
| | 1 to 15 decimals (%) | 4.1 | 3.9 | 0.389 | 4.7 | 1.8 | 0.010 | 3.9 |
| | 16-50 decimals (%) | 3.4 | 7.3 | <0.001 | 8.8 | 3.6 | 0.001 | 6.1 |
| | 50+ decimals (%) | 11.7 | 18.7 | <0.001 | 15.5 | 26.9 | <0.001 | 16.5 |
| | Mean (in decimal) | 98.53 | 103.88 | 0.188 | 88.16 | 140.29 | <0.001 | 102.66 |
| | SD (in decimal) | 100.23 | 92.76 | | 85.03 | 99.75 | | 94.45 |
| | Median (in decimal) | 66.00 | 90.00 | | 60.00 | 120.00 | | 90.00 |
| | IQR (in decimal) | 121.00 | 120.00 | | 95.00 | 120.00 | | 120.00 |
| Pond land | Don't have (%) | 92.4 | 94.5 | 0.058 | 95.2 | 92.7 | 0.057 | 93.8 |
| | 1 to 15 decimals (%) | 6.8 | 4.4 | 0.021 | 3.7 | 6.2 | 0.036 | 5.2 |
| | 16-50 decimals (%) | 0.8 | 1.1 | 0.318 | 1.0 | 1.1 | 0.393 | 1.0 |
| | Mean (in decimal) | 6.83 | 9.79 | <0.001 | 11.19 | 7.42 | <0.001 | 8.65 |
| | SD (in decimal) | 7.32 | 11.11 | | 12.69 | 7.31 | | 9.89 |
| | Median (in decimal) | 5.00 | 5.00 | | 5.00 | 5.00 | | 5.00 |
| | IQR (in decimal) | 7.30 | 7.00 | | 11.80 | 7.00 | | 7.00 |
| Total land | Don't have (%) | 4.6 | 5.6 | 0.241 | 7.3 | 1.3 | <0.001 | 5.3 |
| | 1 to 15 decimals (%) | 49.2 | 40.5 | <0.001 | 40.1 | 41.6 | 0.343 | 43.2 |
| | 16-50 decimals (%) | 17.5 | 18.6 | 0.325 | 19.9 | 15.1 | 0.034 | 18.2 |
| | 50+ decimals (%) | 28.6 | 35.3 | 0.002 | 32.7 | 41.9 | 0.001 | 33.2 |
| | Mean (in decimal) | 68.58 | 75.84 | 0.178 | 73.14 | 82.40 | 0.135 | 73.55 |
| | SD (in decimal) | 134.09 | 114.46 | | 115.17 | 112.58 | | 121.01 |
| | Median (in decimal) | 15.00 | 25.00 | | 25.00 | 27.00 | | 21.00 |
| | IQR (in decimal) | 61.00 | 95.50 | | 87.00 | 115.00 | | 88.00 |

*Note:* SD=Standard Deviation; IQR= Inter Quartile Range

**Table 3.5** Possession of durable assets by the study households in *Haor* regions

| Name of household assets | Micro-credits type (%) | | | Household type (%) | | | Overall |
|---|---|---|---|---|---|---|---|
| | Non-borrower (733) | Borrower (1607) | P-value | Formal (1158) | Informal (449) | P-value | (2340) |
| Radio | 2.9 | 1.4 | 0.052 | 1.6 | 0.9 | 0.225 | 1.9 |
| Television | 18.8 | 14.9 | 0.064 | 18.2 | 6.5 | <0.001 | 16.2 |
| Mobile phone | 85.8 | 91.7 | 0.001 | 91.5 | 92.2 | 0.360 | 89.9 |
| Electric fan | 71.2 | 69.1 | 0.285 | 73.4 | 57.9 | <0.001 | 69.7 |
| Almirah | 37.8 | 34.0 | 0.143 | 37.3 | 25.4 | <0.001 | 35.2 |
| Cookeries | 96.3 | 89.4 | <0.001 | 86.4 | 96.9 | <0.001 | 91.5 |
| Cutleries | 96.3 | 89.4 | <0.001 | 86.3 | 97.3 | <0.001 | 91.5 |
| Table/chair | 78.9 | 81.1 | 0.242 | 82.1 | 78.6 | 0.109 | 80.4 |
| Bed | 92.6 | 93.5 | 0.324 | 94.0 | 92.0 | 0.139 | 93.2 |
| *Chauki* | 94.5 | 93.5 | 0.302 | 92.7 | 95.8 | 0.030 | 93.8 |
| Drawing room furniture | 6.0 | 6.2 | 0.394 | 5.0 | 9.4 | 0.002 | 6.2 |
| Bicycle | 10.6 | 7.0 | 0.023 | 6.9 | 7.1 | 0.395 | 8.1 |
| Boat | 11.6 | 16.2 | 0.027 | 16.8 | 14.9 | 0.260 | 14.8 |
| Shallow tube-well | 19.9 | 20.9 | 0.361 | 20.9 | 20.9 | 0.399 | 20.6 |
| Deep tube-well | 0.5 | 1.6 | 0.086 | 1.1 | 2.7 | 0.026 | 1.2 |
| Watch | 25.5 | 29.9 | 0.086 | 31.1 | 26.7 | 0.089 | 28.5 |
| Shelf | 29.6 | 28.4 | 0.356 | 27.9 | 29.6 | 0.317 | 28.8 |
| *Alna* | 58.3 | 59.9 | 0.336 | 59.1 | 61.9 | 0.235 | 59.4 |
| Motorcycle | 0.0 | 0.4 | 0.162 | 0.5 | 0.0 | 0.129 | 0.3 |
| Others | 4.9 | 5.4 | 0.368 | 6.5 | 2.4 | 0.002 | 5.2 |

*Note:* **Formal source:** Govt. bank or Govt. co-operatives, and Nongovernment (MFI/NGO/Insurance).
**Informal source:** Local Money Lender (Mohajan/Private samitte) & Non-interest Loan (Relatives/Land Owner)

**Table 3.6** Possession of durable assets by the study households in *Haor* regions of Bangladesh

| Name of assets | Borrower (N = 1607) | | Non-borrower (N = 733) | | P-value | Overall (N = 2340) | |
|---|---|---|---|---|---|---|---|
| | HHs | Mean ± SD | HHs | Mean ± SD | | HHs | Mean ± SD |
| Cultivating equipment | 664 | 3.27 ± 2.27 | 311 | 3.05 ± 2.05 | 0.146 | 975 | 3.20 ± 2.20 |
| Livestock | 694 | 2.39 ± 1.51 | 287 | 2.73 ± 1.82 | 0.003 | 981 | 2.49 ± 1.61 |
| Rickshaw/van | 038 | 1.32 ± 0.66 | 015 | 1.40 ± 1.06 | 0.741 | 053 | 1.34 ± 0.78 |
| Auto rickshaw | 020 | 1.10 ± 0.45 | 016 | 1.00 ± 0.00 | 0.382 | 036 | 1.06 ± 0.33 |
| Sprayer | 029 | 1.17 ± 0.38 | 014 | 1.29 ± 0.61 | 0.433 | 043 | 1.21 ± 0.47 |
| Fishing net | 336 | 1.62 ± 0.94 | 127 | 1.74 ± 1.03 | 0.233 | 463 | 1.65 ± 0.96 |
| Bee box | 014 | 1.00 ± 0.00 | 002 | 1.50 ± 0.71 | 0.004 | 016 | 1.06 ± 0.25 |
| Sewing machine | 089 | 1.07 ± 0.25 | 035 | 1.06 ± 0.24 | 0.840 | 124 | 1.06 ± 0.25 |
| Motor (engine) | 041 | 1.17 ± 0.50 | 017 | 1.06 ± 0.24 | 0.392 | 058 | 1.14 ± 0.44 |
| Family business | 135 | 59377.78 | 057 | 104000.00 | 0.023 | 192 | 72625.00 |
| Others | 024 | 4.71 ± 8.09 | 020 | 2.30 ± 2.70 | 0.210 | 044 | 3.61 ± 6.30 |

*Note:* HHs = Households; SD = Standard Deviation

**Table 3.7** Knowledge and perception on micro-credits benefits of respondents in *Haor* regions

| Statement of knowledge and/or perception | Number (*N* = 2340) | Percentage of total | Total in percentage |
|---|---|---|---|
| **Knowledge about micro-credits benefit** | | | |
|     Yes | 2212 | 94.5 | 100 |
|     No | 128 | 5.5 | |
| **Trying for getting micro-credits benefit** | | | |
|     Yes | 2008 | 85.8 | 100 |
|     No | 332 | 14.2 | |
| **First attempt for getting micro-credits benefit with type** | | | |
|     Government (*Banks/Co-operatives*) | 96 | 4.1 | |
|     Nongovernment (*MFI/NGO/Insurance*) | 1289 | 55.1 | |
|     Local money lender (*Mohajan/Private Samittee*) | 374 | 16.0 | 85.8 |
|     Non-interest loan (*Relatives/friends/neighbors*) | 56 | 2.4 | |
|     More than one sources | 193 | 8.2 | |
| **Second attempt for getting Micro-credits benefit with type** | | | |
|     Government (*Banks/Co-operatives*) | 21 | 0.9 | |
|     Nongovernment (*MFI/NGO/Insurance*) | 38 | 1.6 | |
|     Local money lender (*Mohajan/Private Samittee*) | 128 | 5.5 | 8.6 |
|     Non-interest loan (*Relatives/friends/neighbors*) | 11 | 0.5 | |
|     More than one sources | 3 | 0.1 | |
| **Communication for getting Micro-credits benefit** | | | |
|     UP chairman | 12 | 0.5 | |
|     UP member | 20 | 0.9 | |
|     UP office | 804 | 34.4 | 58.60 |
|     Government officer | 112 | 4.8 | |
|     Relatives/neighbors/friends | 343 | 14.7 | |
|     NGO | 78 | 3.3 | |
| **Ask for money for giving Micro-credits benefit** | | | |
|     Yes | 91 | 3.9 | 59.00 |
|     No | 1290 | 55.1 | |
| **Micro-creditss help to remove poverty** | | | |
|     Yes | 1413 | 60.4 | 66.10 |
|     No | 134 | 5.7 | |

*Note:* MFI=Microfinance Institution; NGO=Nongovernment Organization; UP=Union Parishad.

**Table 3.8** Reasons of not getting micro-credits benefits of eligible borrowers' of *Haor* regions

| Name of the reasons | Household (N = 733) | Strongly disagree | Disagree | No comment | Agree | Strongly agree |
|---|---|---|---|---|---|---|
| Bureaucratic complexity | 579 | 2.3 | 23.6 | 26.9 | 12.6 | 13.6 |
| Limitation of budget (according to selector) | 577 | 11.7 | 21.8 | 25.9 | 11.6 | 7.6 |
| Couldn't provide bribe or entry fee | 575 | 12.7 | 37.8 | 17.3 | 6.7 | 4.0 |
| No political exposure | 577 | 10.6 | 18.4 | 24.0 | 16.9 | 8.7 |
| Didn't have any idea about such program | 578 | 19.4 | 31.9 | 7.4 | 15.3 | 4.9 |
| Nepotism | 561 | 9.0 | 15.1 | 25.5 | 15.7 | 11.2 |
| Non-cooperation from public delegate | 576 | 2.5 | 26.3 | 13.1 | 25.9 | 10.8 |
| Non-cooperation from local lenders | 571 | 2.0 | 18.7 | 17.3 | 28.6 | 11.2 |
| Non-availability of NID | 573 | 39.4 | 25.4 | 9.4 | 2.7 | 1.2 |
| Lack of networking or lobbying | 574 | 9.4 | 11.3 | 19.2 | 29.2 | 9.1 |
| Distance from Micro-credits from the village | 569 | 9.0 | 35.6 | 6.8 | 21.6 | 4.6 |
| No Micro-credits in the area | 442 | 25.5 | 28.2 | 3.7 | 1.1 | 1.8 |
| Non availability of collateral | 388 | 2.6 | 19.6 | 11.9 | 13.8 | 5.0 |
| Misappropriation of credits | 384 | 2.3 | 14.3 | 12.0 | 17.7 | 6.0 |
| Biasness | 348 | 1.5 | 19.2 | 22.8 | 2.0 | 1.9 |
| Others | 207 | 1.1 | 1.9 | 16.4 | 2.0 | 4.9 |

*Note:* NID = National Identity.

**Appendix table 5.1** Impact of micro-credits benefits on the households' of *Haor* areas by before-after borrowers and non- borrowers as well as case-control.

| Indicator of economic status | Non-borrowing households | | | | | | Borrowing households | | | | | | Diff-in-diff Z=y-x | p-value (x vs. y) |
|---|---|---|---|---|---|---|---|---|---|---|---|---|---|---|
| | 2016 | | 2019 | | t-statistic (p-value) | $x = \frac{b-a}{a} \times 100$ | 2016 | | 2019 | | t-statistic (p-value) | $y = \frac{d-c}{c} \times 100$ | | |
| | HHs | (a) | HHs | (b) | | | HHs | (c) | HHs | (d) | | | | |
| **Income Sources** | | | | | | | | | | | | | | |
| Agricultural | 318 | 40116.4 | 333 | 45803.6 | -1.84(0.066) | 14.18 | 813 | 38105.9 | 900 | 44654.2 | -4.53(<0.01) | 17.18 | 3.01 | 0.188 |
| Non-agricultural | 295 | 51237.3 | 331 | 55763.8 | -0.85(0.394) | 8.83 | 665 | 37516.6 | 830 | 43169.3 | -2.42(0.016) | 15.07 | 6.23 | 0.012 |
| Labor Sale | 456 | 44855.7 | 464 | 54409.5 | -5.69(<0.01) | 21.30 | 1002 | 48847.3 | 1053 | 56527.9 | -4.65(<0.01) | 15.72 | -5.58 | 0.014 |
| Business | 179 | 68607.0 | 197 | 74070.1 | -0.93(0.352) | 7.96 | 344 | 62094.2 | 400 | 68855.3 | -1.71(0.088) | 10.89 | 2.93 | 0.226 |
| Donation/Begging | 83 | 9342.2 | 81 | 11041.9 | -0.74(0.463) | 18.19 | 156 | 7854.5 | 208 | 9804.3 | -1.66(0.098) | 24.82 | 6.63 | 0.202 |
| **Total Income** | **732** | **83855.5** | **733** | **101558.8** | **-5.40(<0.01)** | **21.11** | **1586** | **80365.3** | **1607** | **102753.4** | **-11.57(<0.01)** | **27.86** | **6.75** | **0.001** |
| **Expenditure heads** | | | | | | | | | | | | | | |
| Food | 733 | 46479.8 | 733 | 54700.3 | -6.72(<0.01) | 17.69 | 1607 | 47632.1 | 1607 | 56030.3 | -10.85(<0.01) | 17.63 | -0.05 | 0.399 |
| Non-food | 733 | 8738.4 | 733 | 11470.7 | -7.24(<0.01) | 31.27 | 1607 | 9903.0 | 1607 | 13528.0 | -12.38(<0.01) | 36.61 | 5.34 | 0.017 |
| **Total consumption** | **733** | **55218.6** | **733** | **66171.0** | **-8.01(<0.01)** | **19.83** | **1607** | **57535.1** | **1607** | **69558.3** | **-13.83(<0.01)** | **20.90** | **1.06** | **0.334** |
| Education | 407 | 7917.4 | 481 | 9015.2 | -1.37(0.171) | 13.87 | 997 | 8567.5 | 1190 | 10193.0 | -3.24(0.001) | 18.97 | 5.11 | 0.030 |
| Healthcare | 688 | 7404.2 | 686 | 7966.3 | -0.82(0.412) | 7.59 | 1519 | 7731.0 | 1539 | 9815.1 | -5.41(<0.01) | 26.96 | 19.37 | <0.01 |
| Agriculture | 276 | 16273.6 | 279 | 19157.7 | -1.99(0.047) | 17.72 | 742 | 15280.1 | 811 | 19016.8 | -6.31(<0.01) | 24.45 | 6.73 | 0.030 |
| Poultry-livestock | 201 | 4380.6 | 202 | 5018.3 | -1.10(0.273) | 14.56 | 445 | 4925.1 | 541 | 9604.3 | -8.10(<0.01) | 95.01 | 80.45 | <0.01 |
| Family business | 72 | 18416.7 | 79 | 22810.1 | -2.29(0.023) | 23.86 | 150 | 17952.0 | 220 | 25086.4 | -4.65(<0.01) | 39.74 | 15.89 | 0.027 |
| Productive asset | 41 | 5682.9 | 54 | 6157.4 | -0.35(0.724) | 8.35 | 136 | 7157.4 | 233 | 11639.9 | -3.30(0.001) | 62.63 | 54.28 | <0.01 |
| Durable goods | 76 | 4802.6 | 84 | 3756.0 | 1.07(0.268) | -21.79 | 91 | 5238.5 | 146 | 8630.8 | -2.66(0.008) | 64.76 | 86.55 | <0.01 |
| House repair | 225 | 9487.1 | 235 | 7005.1 | 2.11(0.035) | -26.16 | 499 | 8420.4 | 574 | 11401.7 | -2.84((0.005) | 35.41 | 61.57 | 0.019 |
| Land purchase | 10 | 15000.0 | 13 | 25192.3 | -1.26(0.223) | 67.95 | 21 | 23333.3 | 38 | 29710.5 | -0.78(0.439) | 27.33 | -40.62 | 0.039 |
| Other investment | 125 | 7917.4 | 158 | 11063.3 | -2.40(0.017) | 39.73 | 356 | 10411.5 | 561 | 16049.2 | -6.73(<0.01) | 54.15 | 14.41 | 0.009 |
| **Total investment** | **727** | **25864.7** | **730** | **30591.2** | **-3.06(0.002)** | **18.27** | **1589** | **29128.0** | **1606** | **46091.0** | **-15.54(<0.01)** | **58.24** | **39.96** | **<0.01** |
| **Total expenditure** | **733** | **80871.5** | **733** | **96637.0** | **-6.48(<0.01)** | **19.49** | **1606** | **86325.2** | **1607** | **115620.6** | **-18.21(<0.01)** | **33.94** | **14.44** | **<0.01** |
| Debt | 192 | 13791.7 | 158 | 12018.4 | 1.03(0.303) | -12.86 | 830 | 20689.2 | 1253 | 28671.2 | -5.51(<0.01) | 38.58 | 51.44 | <0.01 |
| Savings | 118 | 18827.1 | 111 | 19438.3 | -0.19(0.846) | 3.25 | 308 | 4440.4 | 507 | 3965.4 | 1.37(0.170) | -10.70 | -13.94 | 0.020 |

*Note:* HH = Households;
a = Average value of non-beneficiary HHs in 2016; b = Average value of non-beneficiary HHs in 2019; x = Impact of non-borrower HHs from before-after comparison;
c = Average value of beneficiary HHs in 2016; d = Average value of beneficiary HHs in 2019; y = Impact of borrower HHs from before-after comparison;
z = Net impact of micro-creditss through case-control as well as before-after comparison using diff-in-diff method.

**Appendix table 5.2** A comparative impact of micro-credits benefits of formal and informal borrowers in the *Haor* regions of Bangladesh by before-after as well as case-control method

| Indicator of Economic Status | Informal sources | | | | | | Formal sources | | | | | | Diff-in-diff Z=y-x | p-value (x vs y) |
|---|---|---|---|---|---|---|---|---|---|---|---|---|---|---|
| | 2016 | | 2019 | | t-statistic (p-value) | $x=\frac{b-a}{a}\times 100$ | 2016 | | 2019 | | t-statistic (p-value) | $y=\frac{d-c}{c}\times 100$ | | |
| | HHs | (a) | HHs | (b) | | | HHs | (c) | HHs | (d) | | | | |
| **Income Sources** | | | | | | | | | | | | | | |
| Agricultural | 221 | 33511.8 | 245 | 40303.7 | -2.70(0.007) | 20.27 | 592 | 39821.0 | 655 | 46281.5 | -3.71(<0.01) | 16.22 | -4.04 | 0.108 |
| Non-agricultural | 174 | 36086.5 | 216 | 41351.4 | -0.96(0.336) | 14.59 | 491 | 38023.5 | 614 | 43808.9 | -2.31(0.021) | 15.22 | 0.63 | 0.386 |
| Labor Sale | 340 | 47110.3 | 348 | 55379.3 | -3.80(<0.01) | 17.55 | 662 | 49739.4 | 705 | 57094.9 | -3.30(0.001) | 14.79 | -2.76 | 0.162 |
| Business | 63 | 56317.5 | 74 | 60270.3 | -0.67(0.506) | 7.02 | 281 | 63389.3 | 326 | 70804.1 | -1.59(0.112) | 11.70 | 4.68 | 0.096 |
| Donation/Begging | 54 | 6672.2 | 72 | 9311.1 | -1.80(0.074) | 39.55 | 102 | 8480.4 | 136 | 10065.4 | -0.98(0.330) | 18.69 | -20.86 | 0.001 |
| **Total Income** | 447 | 75192.3 | 449 | 96233.2 | -5.74(<0.01) | 27.98 | 1139 | 82395.4 | 1158 | 105281.5 | -10.07(<0.01) | 27.78 | -0.21 | 0.397 |
| **Expenditure heads** | | | | | | | | | | | | | | |
| Food | 449 | 46731.9 | 449 | 54261.5 | -5.14(<0.01) | 16.11 | 1158 | 47981.2 | 1158 | 56716.2 | -9.58(<0.01) | 18.21 | 2.09 | 0.185 |
| Non-food | 449 | 8447.0 | 449 | 12418.9 | -7.55(<0.01) | 47.02 | 1158 | 10467.6 | 1158 | 13958.0 | -10.00(<0.01) | 33.34 | -13.68 | <0.01 |
| **Total consumption** | 449 | 55178.8 | 449 | 66680.4 | -7.11(<0.01) | 20.84 | 1158 | 58448.7 | 1158 | 70674.2 | -11.89(<0.01) | 20.92 | 0.07 | 0.399 |
| Education | 275 | 8104.7 | 298 | 9283.6 | -1.30(0.193) | 14.55 | 722 | 8743.8 | 892 | 10496.9 | -2.92(0.004) | 20.05 | 5.50 | 0.022 |
| Healthcare | 438 | 8379.5 | 436 | 10845.2 | -3.56(<0.01) | 29.43 | 1081 | 7468.2 | 1103 | 9408.0 | -4.20(<0.01) | 25.97 | -3.45 | 0.095 |
| Agriculture | 221 | 16156.6 | 222 | 18959.9 | -2.44(0.015) | 17.35 | 521 | 14908.3 | 589 | 19038.2 | -5.99(<0.01) | 27.70 | 10.35 | 0.001 |
| Poultry-Livestock | 194 | 3996.4 | 206 | 5496.1 | -2.74(0.006) | 37.53 | 251 | 5642.8 | 335 | 12130.5 | -7.52(<0.01) | 114.97 | 77.45 | <0.01 |
| Family business | 21 | 14038.1 | 31 | 22580.7 | -2.53(0.015) | 60.85 | 129 | 18589.2 | 189 | 25497.4 | -4.08(<0.01) | 37.16 | -23.69 | 0.002 |
| Productive asset | 60 | 5888.3 | 80 | 8137.5 | -1.36(0.178) | 38.20 | 76 | 8159.2 | 153 | 13471.2 | -2.74(0.007) | 65.10 | 26.91 | 0.004 |
| Durable goods | 26 | 6096.2 | 40 | 8975.0 | -0.94(0.353) | 47.22 | 65 | 4895.4 | 106 | 8500.9 | -2.73(0.007) | 73.65 | 26.43 | 0.001 |
| House repair | 178 | 7906.7 | 208 | 12846.2 | -2.45(0.015) | 62.47 | 321 | 8705.3 | 366 | 10580.9 | -1.58(0.115) | 21.55 | -40.93 | <0.01 |
| Land purchase | 4 | 12500.0 | 11 | 34090.9 | -0.93(0.370) | 172.73 | 17 | 25882.4 | 27 | 27925.9 | -0.25(0.807) | 7.90 | -164.8 | <0.01 |
| Other investment | 99 | 7681.8 | 130 | 13179.2 | -3.22(0.001) | 71.56 | 257 | 11463.0 | 431 | 16914.9 | -5.71(<0.01) | 47.56 | -24.00 | <0.01 |
| **Total investment** | 445 | 29658.2 | 449 | 42999.2 | -6.14(<0.01) | 44.98 | 1144 | 28921.7 | 1157 | 47290.8 | -14.60(<0.01) | 63.51 | 18.53 | <0.01 |
| **Total expenditure** | 448 | 84527.2 | 449 | 109679.6 | -7.76(<0.01) | 29.76 | 1158 | 87020.8 | 1158 | 117924.2 | -16.78(<0.01) | 35.51 | 5.76 | 0.010 |
| Debt | 265 | 20720.8 | 323 | 24791.0 | -1.66(0.098) | 19.64 | 565 | 20674.4 | 930 | 30018.9 | -5.25(<0.01) | 45.20 | 25.56 | <0.01 |
| Savings | 31 | 5312.9 | 29 | 4372.4 | 0.82(0.416) | -17.70 | 277 | 4342.7 | 478 | 3940.7 | 1.10(0.270) | -9.26 | 8.45 | 0.021 |

*Note:* HHs = Households;
a=Average value from informal sources in 2016; b=Average value from informal sources in 2019; x=Impact of credits from informal sources using before-after comparison;
c=Average value from informal sources in 2016; b=Average value from informal sources in 2019; x=Impact of credits from informal sources using before-after comparison;
z=Net impact of formal credits over informal credits through case-control as well as before-after comparison using diff-in-diff method